\pdfoutput=1
\documentclass[useAMS,fleqn,usenatbib]{mnras}

\usepackage[T1]{fontenc}
\usepackage{ae,aecompl}

\usepackage{amsmath}
\usepackage{amsopn}
\usepackage[british]{babel}
\usepackage[varg]{txfonts}
\usepackage{biblio} 
\usepackage{natbib}
\usepackage{color}

\usepackage{graphicx}
\usepackage{xpatch}
\makeatletter
\xpatchcmd\Gin@setfile{\expandafter\strip@prefix\meaning\@tempa}{\makebox[\Gin@req@width]{Image place holder (draft mode)}}{}{}
\makeatother

\definecolor{orange}{rgb}{1.0,0.5,0.}

\def\MDM{\ifmmode{\>M_{\textnormal{\sc dm}}}\else{$$M_{\textnormal{\sc dm}}}\fi}

\def\XH{\ifmmode{\>X_{\textnormal{\sc h}}} \else{$X_{\textnormal{\sc h}}$}\fi}
\def\nH{\ifmmode{\>n_{\textnormal{\sc h}}} \else{$n_{\textnormal{\sc h}}$}\fi}

\def\maspyr{\ifmmode{\>\textnormal{mas~yr}^{-1}}\else{mas~yr$^{-1}$}\fi}

\def\mG{\ifmmode{\>\mu\mathrm{G}}\else{$\mu$G}\fi}
\def\erg{\ifmmode{\> {\rm erg}}\else{erg}\fi}
\def\keV{\ifmmode{\> {\rm keV}}\else{keV}\fi}

\def\deg{\ifmmode{\>^{\circ}}\else{$^{\circ}$}\fi}
\def\onedeg{\ifmmode{\>1^{\circ}}\else{$1^{\circ}$}\fi}

\def\xvir{\ifmmode{\>\!x_{vir}}\else{$x_{vir}$}\fi}
\def\Mvir{\ifmmode{\>\!M_{vir} }\else{$M_{vir} $}\fi}
\def\rvir{\ifmmode{\>\!r_{vir}}\else{$r_{vir}$}\fi}
\def\vvir{\ifmmode{\>\!v_{vir}}\else{$v_{vir}$}\fi}
\def\Vvir{\ifmmode{\>\!V_{vir} }\else{$V_{vir} $}\fi}

\def\tratio{\ifmmode{\>\tau}\else{$\tau$}\fi}

\def\rms{\ifmmode{\>r_{\textnormal{\sc ms}}}\else{$r_{\textnormal{\sc ms}}$}\fi}

\def\Mpc{\ifmmode{\>\!{\rm Mpc}} \else{Mpc}\fi}
\def\kpc{\ifmmode{\>\!{\rm kpc}} \else{kpc}\fi}
\def\pc{\ifmmode{\>\!{\rm pc}} \else{pc}\fi}

\def\Gyr{\ifmmode{\>\!{\rm Gyr}} \else{Gyr}\fi}
\def\Myr{\ifmmode{\>\!{\rm Myr}} \else{Myr}\fi}
\def\yr{\ifmmode{\>\!{\rm yr}} \else{yr}\fi}
\def\pyr{\ifmmode{\>\!{\rm yr}^{-1}}\else{yr $^{-1}$} \fi}
\def\s{\ifmmode{\>\!{\rm s}}\else{s}\fi}
\def\ps{\ifmmode{\>\!{\rm s}^{-1}}\else{s$^{-1}$}\fi}
\def\Hz{\ifmmode{\>\!{\rm Hz}}\else{Hz}\fi}

\def\kms{\ifmmode{\>\!{\rm km\,s}^{-1}}\else{km~s$^{-1}$}\fi}

\def\K{\ifmmode{\>\!{\rm K}}\else{K}\fi}

\def\sr{\ifmmode{\>\!{\rm sr}}\else{sr}\fi}
\def\psr{\ifmmode{\>\!{\rm sr}^{-1}}\else{sr$^{-1}$}\fi}
\def\arcs{\ifmmode{\>\!{\rm arcsec}}\else{arcsec}\fi}
\def\parcs{\ifmmode{\>\!{\rm arcsec}^{-1}}\else{arcsec${-1}$}\fi}
\def\parcss{\ifmmode{\>\!{\rm arcsec}^{-2}}\else{arcsec${-2}$}\fi}

\def\cm{\ifmmode{\>\!{\rm cm}}\else{cm}\fi}
\def\cc{\ifmmode{\>\!{\rm cm}^{3}}\else{cm$^{3}$}\fi}
\def\sqc{\ifmmode{\>\!{\rm cm}^{2}}\else{cm$^{2}$}\fi}
\def\pcc{\ifmmode{\>\!{\rm cm}^{-3}}\else{cm$^{-3}$}\fi}
\def\psc{\ifmmode{\>\!{\rm cm}^{-2}}\else{cm$^{-2}$}\fi}

\def\g{\ifmmode{\>\!{\rm g}}\else{g}\fi}
\def\Msun{\ifmmode{\>\!{\rm M}_{\odot}}\else{M$_{\odot}$}\fi}
\def\hMsun{\ifmmode{\> h^{-1}{\rm M}_{\odot}}\else{$h^{-1}$M$_{\odot}$}\fi}

\def\Zsun{\ifmmode{\>\!{\rm Z}_{\odot}}\else{Z$_{\odot}$}\fi}

\def\rayl{\ifmmode{\>\!{\rm R}}\else{R}\fi}
\def\mR{\ifmmode{\>\!{\rm mR}}\else{mR}\fi}

\renewcommand{\ion}[2]{\hbox{#1\,{\sc #2}}}

\def\lya{\ifmmode{\>\!{\rm Ly}\alpha}\else{Ly$\alpha$}\fi}

\def\Ha{\ifmmode{\>\!{\rm H}\alpha}\else{H$\alpha$}\fi}
\def\Hb{\ifmmode{\>\!{\rm H}\beta}\else{H$\beta$}\fi}

\def\HI{\ifmmode{\> \textnormal{\ion{H}{i}}} \else{\ion{H}{i}}\fi}
\def\HII{\ifmmode{\> \textnormal{\ion{H}{ii}}} \else{\ion{H}{ii}}\fi}
\def\CIV{\ifmmode{\> \textnormal{\ion{C}{iv}}} \else{\ion{C}{iv}}\fi}
\def\SiIV{\ifmmode{\> \textnormal{\ion{S}{iv}}} \else{\ion{Si}{iv}}\fi}

\def\NHI{\ifmmode{\> {\rm N}_{\HI}} \else{N$_{\HI}$}\fi}
\def\MHI{\ifmmode{\> {\rm M}_{ \HI}} \else{M$_{\HI}$}\fi}

\def\mua{\ifmmode{\>\mu_{ \textnormal{\Ha}}}\else{$\mu_{ \textnormal{\Ha}}$}\fi}
\def\alphabha{\ifmmode{\>\alpha_{B}^{(\textnormal{\Ha})}}\else{$\alpha_{B}^{(\textnormal{\Ha})}$}\fi}

\newcommand{\myemail}{tepper@physics.usyd.edu.au}
\newcommand{\ramses}{{\sc ramses}}
\newcommand{\dice}{{\sc dice}}

\defcitealias{pla14b}{\textcolor{black}{Planck Collaboration} 2014}
\defcitealias{ast13a}{\textcolor{black}{Astropy Collaboration} 2013}

\title[ Gas loss in Sgr ]{ The Sagittarius dwarf galaxy: Where did all the gas go? }

\author[T.~Tepper-Garc\'\i{}a and J.~Bland-Hawthorn]{%
Thor Tepper-Garc\'\i{}a\thanks{\myemail}$^{,1}$ and Joss Bland-Hawthorn$^{1,2,3}$
\\
$^1$Sydney Institute for Astronomy, School of Physics, University of Sydney, NSW 2006, Australia\\
$^2$Centre of Excellence for All Sky Astrophysics in Three Dimensions (ASTRO-3D), Australia\\
$^3$Miller Professor, Miller Institute, University of California Berkeley, Berkeley, CA 94720, USA\\
}

\date{Accepted ---. Received ---; in original form ---}

\pubyear{\date{year}}

\begin{document}
\label{firstpage}
\pagerange{\pageref{firstpage}--\pageref{lastpage}}
\maketitle

\pdfminorversion=5
\begin{abstract}
The remarkable 1994 discovery of the Sagittarius dwarf galaxy (Sgr) revealed that, together with the Magellanic Clouds, there are at least three major dwarf galaxies, each with a total mass of order $10^{10} - 10^{11}$ \Msun, falling onto the Galaxy in the present epoch. Beyond a Galactic radius of 300 kpc, dwarfs tend to retain their gas. At roughly 50 kpc, the Magellanic Clouds have experienced substantial gas stripping as evidenced by the Magellanic Stream which extends from them. Since Sgr experienced star formation long after it fell into the Galaxy, it is interesting to explore just how and when this dwarf lost its gas. To date, there has been no definitive detection of an associated gas component. We revisit recent simulations of the stellar and dark matter components of Sgr but, for the first time, include gas that is initially bound to the infalling galaxy. We find that the gas stripping was 30 - 50\% complete at its first disc crossing $\sim 2.7$ Gyr ago, then entirely stripped at its last disc crossing $\sim 1$ Gyr ago. Our timeline is consistent with the last substantial burst of star formation in Sgr which occurred about the time of the last disc crossing. We discuss the consequences of gas stripping and conclude that the vast majority of the stripped gas was fully settled onto the Galaxy by $\sim 300$ Myr ago. It is {\it highly unlikely} that any of the high- or intermediate-velocity clouds have a direct association with the Sgr dwarf.

\end{abstract}


\section{Introduction} \label{sec:intro}

The Sagittarius dwarf galaxy \citep[Sgr; ][]{iba94a} is devoid of neutral atomic gas (\HI) down to the detection limit \citep[][]{kor94a,bur99b}, as are generally other satellite systems of the Galaxy \citep[and Andromeda;][]{grc09a,spe14b}. Two notable exceptions are the Magellanic Clouds (MCs): the Large Magellanic Cloud (LMC), and the Small Magellanic Cloud \citep[SMC; q.v.][]{mcc12a}. New mass determinations now render the precursors of Sgr \citep[][]{muc17a,gib17a} comparable to the LMC \citep[][]{jet16a}, with combined dark-matter (DM) and baryonic masses of up to $\sim 10^{11}$ \Msun. Why have the Magellanic Clouds retained much of their gas while Sgr has apparently lost it all?

A key difference between Sgr and the LMC is their orbital history. The LMC is likely on its first approach to the Galaxy \citep[][]{shu92a,byr94a,bes07a}, currently at an estimated distance from the Galactic centre of $d \approx 50$ kpc \citep[][]{pie13b}. Models of the evolution of Sgr from infall to its present location at $d \approx 16 - 20$ kpc \citep[or $D \approx 25 - 29$ kpc from the Sun; q.v.][]{kun09a} agree that Sgr has undergone already several pericentric passages \citep[e.g.][]{fel06a,law10a,nie12a}.\footnote{For a brief overview of the extensive literature prior to 2010 see \citet[][]{mye10a}. For a recent review on Sgr, including numerical models, see \citet[][]{law16a}.} The LMC began to experience significant tidal gas stripping assisted by ram pressure due to the tenuous corona of the Galaxy \citep[$\nH \sim 10^{-4} ~\pcc$ at $r \approx 50$ kpc; e.g.][]{tep15a} only recently \citep[$\sim 1.5$ Gyr;][]{gug14a}, as attested by the prominent Magellanic Stream that extends from the MCs over nearly 200\deg\ across the sky \citep[][]{mat74a,nid10a}.
In contrast, most dynamical models place the Sgr precursor on an orbit around the Galaxy for no less than $\sim 3$ Gyr \citep[e.g.][]{gom15a,die17b}, and indicate that it has likely collided with the (outer) Galactic disc {\em at least} once in the past \citep[][]{pur11a}. As a result of the prolonged interaction with the Galaxy, the Sgr precursor has experienced significant tidal disruption, leaving behind a dense, prolate core and multiple extended stellar streams along its orbit out to $\sim 100$ kpc \citep[][]{bel14a}.

In spite of its observed lack of gas today, Sgr must have had a significant gas reservoir prior to falling onto the Galaxy and along much of its orbit within the Galactic virial radius. The age-metallicity trends associated with the stellar populations reveal that Sgr experienced multiple episodes of star formation, with a significant drop in its efficiency already $\sim 2$ Gyr ago \citep[][]{muc17a}, terminating altogether about 1 Gyr ago \citep[][]{sie07a}. These studies conclude that an event of considerable gas loss within that epoch is required in order to explain the observed chemical pattern of the stellar population in the nucleus of Sgr. Any realistic {\em gas-dynamical} model of Sgr therefore needs to explain both its tidal disruption and its gas loss within the timeframe bracketed by this event and its infall onto the Galaxy.

The aim of this paper is to introduce -- for the first time in the literature -- a model for the infall of Sgr where gas, in addition to stars and dark matter, is considered. In essence, we show that it is possible to account for the present-day stellar kinematic properties of Sgr and its debris, as well as its observed lack of gas, starting from a DM dominated, gas-bearing precursor with a spheroidal stellar component that experiences a long-term interaction the Galaxy. Our main working hypothesis --supported by the observational constraints discussed above -- is that Sgr did not exhaust its initial gas content entirely as a result of its star forming events, but rather retained some up until perhaps only recently, part of which may have allowed the formation of its youngest generation of metal rich stars \citep[see e.g.][]{nic15a}. Thus, we assume that stellar feedback within the dwarf does not play an important role in the dynamic evolution of its gas. We assume further that Sgr's stellar component was in place before infalling onto the Galaxy. In other words, we ignore for now the process of star formation and its associated feedback. In consequence, we do not attempt to reproduce the star formation history and, for that matter, the details of the stellar population (age, metallicity, spatial distribution) of Sgr. Our focus is rather on the fate of the gas initially bound to the dwarf. We argue that the key process behind the total gas removal is the interaction with the Galactic gas disc. We explore whether any of this gas has a direct association to the \HI\ high- or intermediate velocity clouds \citep[HVCs, IVCs;][]{wak91b} around the Galaxy, as has been claimed before \citep[][]{bla98a,put04a}.

Where applicable, we assume a flat, dark-energy- and matter (baryonic and cold dark-matter; CDM) dominated Universe, with a cosmology defined by the simplified set of parameter values $h = 0.7$, $\Omega_m = 0.3$, and $\Omega_{\Lambda} = 0.7$.

\section{A minimal model for the infall of Sgr} \label{sec:mod}

We simulate the infall of the Sgr precursor onto the Galaxy along a prescribed orbit taken from previous work \citep[][see below]{die17a}. In doing so, we ignore the presence of other nearby systems, in particular, we neglect the perturbation induced by the LMC on the Galaxy--Sgr interaction which may affect the distribution of tidal debris \citep[][]{gom15a}. This is of no concern for now as we do not attempt to model the precise location of the Sgr core and its streams. Most importantly, we do not expect this simplification to affect in any significant way the evolution of the gas initially bound to the dwarf, which is the main focus of this study.

Our simulation framework is similar to the one adopted in our previous work on the Smith Cloud \citep[][]{tep18a}. Therefore, in what follows we shall omit some details and refer the reader to the latter reference when necessary.

\begin{table}
\begin{center}
\caption{Relevant model parameters (initial values).  Column headers are as follows: $M_t$ := total mass ($10^{9}$ \Msun); $r_s$ := scalelength (kpc); $r_{tr}$ :=  truncation radius (kpc); $N_p$ := particle number ($10^{5}$); $Z$ := gas metallicity ($\Zsun$).}
\label{tab:comp}
\begin{tabular}{lcccccc}
\hline
\hline
 				& Profile	& $M_t$ 		&  $r_s$ 		& $r_{tr}$ 	& $N_p$  & $Z$ ~\\
\hline
Galaxy\\
\hline
DM halo$^{\,a}$	& H			& $10^3$		& 38.4		& 250	& 	5	&	--	~\\
Disc$^{\,b}$		& MN		& 46			& 5.0	$^{d}$	& 20		&	4.6	&	--	~\\
Bulge			& H			& 9			& 0.7			& 4		&	5	&	--	~\\
Corona		 	& H			& 18			& 38.1		& 250	&	50	&	0.3	~\\
Gas disc			& Exp$^{\,c}$	& 14			& 7.0	$^{e}$	& 60		&	100	&	0.3	~\\
\hline
Sgr precursor\\
\hline
DM subhalo$^{\,f}$	& H			& 10			& 1.9			& 25		&	1	&	--	~\\
Bulge			& H			& 0.4			& 0.85		& 2.5		&	1	&	--	~\\
Gas halo			& H			& 0.6			& 1.7			& 5		&	1	&	0.2	~\\
\end{tabular}
\end{center}
\begin{list}{}{}
\item {\em Notes}. H := \citet{her90a} profile; MN := \citet{miy75a} profile; Exp := Radial exponential profile\\
$^{a\,}$Mass enclosed within $r_{tr}$ is $9.7\times10^{11}$ \Msun.\\
$^{b\,}$The stellar metallicity is ignored as it is of no relevance for our study.\\
$^{c\,}$In vertical hydrostatic equilibrium initially at $T = 10^4 ~\K$\\
$^{d\,}$Scaleheight set to 0.5 kpc.\\
$^{e\,}$Scaleheight set by vertical hydrostatic equilibrium (`flaring' disc).\\
$^{f\,}$Mass enclosed within $r_{tr}$ is $\sim 7\times10^{9}$ \Msun.\\
 \end{list}
\end{table}

Our Galaxy model consists of `live', collisionless components (a DM host halo, a stellar disc, and a stellar bulge), and gaseous components (a slowly {\em spinning} corona with a peak velocity $\sim 70$ \kms\ at $R \approx 30$ kpc, and a centrifugally supported, vertically stratified gas disc) with properties broadly consistent with observational constraints \citep[see][]{tep18a}. Given its success in reproducing the phase-space properties of the Sgr tidal tails, we adopt the properties of the Galactic collisionless components from \citet[][their Table 1]{die17a}, with one difference: We adopt a {\em lighter} stellar disc with a \citet[][]{miy75a} profile (rather than exponential); the mass difference is put into a gas disc. As we show later, the presence of a gas disc in the Galaxy is essential. The parameter values characterising the corona and the gas disc are taken from our previous work, with some minor differences \citep[cf. Table \ref{tab:comp} and][their Sec.~2]{tep18a}.

Like the Galaxy model, the Sgr precursor consists of live collisionless components and a gaseous component (also summarised in Table \ref{tab:comp}). The collisionless components (DM subhalo and stellar bulge) are set according to the precursor model of \citet[][]{die17a}. The necessity for a DM halo around the baryonic component of the precursor was recognised early on \citep[][]{iba98a}. The DM subhalo is modelled as a (cored) \citet[][]{her90a} sphere with a virial mass of $10^{10}$ \Msun, at the lower end of the plausible range of precursor masses \citep[e.g.][]{nie10a}, and its extension is limited to a spherical radius of 25 kpc, roughly corresponding to the tidal radius of a $\sim 10^{10}$ \Msun\ object at $d = 125$ kpc -- the initial distance of the precursor; see below -- from the centre of a $\sim 10^{12}$ \Msun\ host. The stellar bulge initially follows a \citet[][]{her90a} profile as well,\footnote{The choice of profile matters, as dwarf systems may be more or less prone to stripping depending on whether initial profile is cored or cuspy \citep[][]{pen08a}.} with a total stellar mass of $\sim 4 \times 10^8$ \Msun\ and a scale radius of 0.85 kpc. The initial maximum velocity dispersion of the bulge is $\sigma_b \sim 25$ \kms, slightly higher than observed \citep[][]{lok10a} in order to account for its evolution \citep[i.e. decrease;][]{pen08a}.

In our model, the Sgr precursor includes an extended, initially spherical, gas halo embedded within the DM subhalo, in thermostatic equilibrium with the precursor's total gravitational potential, with temperatures in the range $\sim 10^4 - 10^5$ K. The temperatures at the high-end make up to some extent for the absence of stellar feedback -- in particular heating -- in our model (see below). We set the ratio of total baryonic mass (stellar bulge and gas halo) to DM mass to 1:10, with a mass ratio of gas-to-stars of 3:2. Our adopted baryons-to-DM mass ratio is well below the cosmic baryon fraction $f_b \approx 0.16$ \citepalias[][]{pla14b}, but is adequate for systems with a dynamical mass $\sim 10^{10}$ \Msun\ \citep[][]{mcg10a}. Even if the Sgr precursor initially had a baryon fraction close to the cosmic value, it is very likely that some of its gas (or even stars) would have gotten stripped prior to being accreted by the Galaxy \citep[i.e. due to `group preprocessing';][]{wet15b}. Our choice of gas-to-star mass ratio is based on the results of simulations of dwarf galaxies in isolation which suggest that DM haloes with masses $\sim 10^{10}$ \Msun\ contain half of its baryons in the form of gas at $z = 0$ \citep[][]{gon14a}. A gas-to-stars mass ratio of order unity is also consistent with the distribution of baryons among stars and (neutral) gas in low-mass galaxies \citep[$M_{\star} \lesssim 4 \times 10^ 8$ \Msun;][]{bra15a}, in particular of dwarf irregulars in the Local Group \citep[][]{mcc12a}. We choose to adopt a slightly higher gas fraction, corresponding to an initial total gas mass of $\sim 6 \times 10^8$ \Msun, thus {\em favouring} the presence of gas within the Sgr precursor.

In order to assess the dependence of our result to the choice of total baryon fraction, we consider an extreme case: a gas-rich precursor with a total (gas and stars) baryon to DM mass fraction equal to $f_b$, with a gas-to-star mass ratio of 5:3. In this case the DM subhalo and stellar bulge are marginally more massive ($\sigma_b \sim 30$ \kms), and both the bulge and the gas halo slightly more extended. In what follows, for ease of discussion we shall refer to the latter as the `gas-rich' model, and to the former as the `reference' model.

Note that we do not include a stellar disc in the dwarf as \citet[][]{die17a} do, and instead put the corresponding mass into a gas halo, as explained above. Our ignoring the presence of a stellar (or gas) disc in the Sgr precursor follows from the absence of residual rotation in the Sgr core \citep[][]{pen11b} which would be observable if the Sgr precursor had been a disc galaxy \citep[][but see \citealt{lok10a}]{pen10b}. Gas-bearing, even gas-rich dwarf galaxies with no (substantial) disc component are not rare \citep[][]{bli00a}. Others with clear rotation signatures have a circular velocity comparable to their internal velocity dispersion \citep[e.g. Leo P;][]{ber14b}, thus making the disc component kinematically less relevant.

The initial conditions of the Galaxy and the Sgr precursor (i.e. particle positions, velocities, and additionally internal energies for the gas components) are constructed following the approach developed by \citet[][]{spr05c} as implemented in the \dice\ code \citep[][]{per14c}.\footnote{See footnote \ref{foo:code}.} This approach relies on the {\em local Maxwellian approximation} \citep[][]{her93b} to calculate particle velocities, which may not be fully adequate to model $N$-body systems in strict dynamic equilibrium, in particular dwarf systems \citep[][]{kaz04a}. This is, however, of no concern in our case since: i) we are not interested in the onset of instabilities; and ii) the infall distance of the Sgr precursor is large enough that the dwarf evolves into a self-consistent configuration before tidal effects become important.\footnote{We refer the reader to \citet[][]{tep18a} for more details on the setup of the initial conditions.}

Following \citet[][]{die17a}, the barycentre of the Sgr precursor is placed at $\vec{r}_0 = \left(125, \, 0, \, 0 \right)$ kpc (with respect to the initial geometric centre of the Galaxy), and given an initial velocity\footnote{ More specifically, the velocity vector has a magnitude $v_0 = 72.6 ~\kms$, and is directed along the $xz$-plane at an angle $\theta = 80.8^{\circ}$ with respect to the negative $x$-axis. See Sec.~\ref{sec:res} for a definition of our adopted coordinate frame. } $\vec{v}_0 \approx \left( -10, \, 0, \, 70 \right) ~\kms$. We note that the initial angular momentum defined by our adopted orbit (and our adopted Galaxy virial mass) for an infall scenario from $z \approx 0.35$ (see Sec.~\ref{sec:sdm}) is essentially identical to the initial angular momentum required for a more massive Sgr precursor \citep[e.g. $6 \times 10^{10}$ \Msun; ][]{gib17a} to approximately match the present-day Sgr phase-space coordinates \citep[][]{die17b}. Also, our chosen orbital parameter values are consistent with Sgr's orbit-mass degeneracy \citep[][]{jia00a}.

It is important to note that \citet[][]{die17a} neglect the effect of gas drag \citep[][see also \citealt{nic15a}]{ben97a} when calculating the orbital initial conditions for the Sgr precursor by integrating its equation of motion within the Galactic potential backwards from its observed location at the present epoch. Therefore, we should expect some differences in the orbital evolution between their model and ours.

We calculate the time evolution of the composite system representing the Galaxy and the Sgr precursor with their collisionless and gaseous components by solving the Vlasov-Poisson and Euler equations with the adaptive mesh refinement, $N$-body, gravito-hydrodynamics code \ramses\ \citep[version 3.0 of the code last described by][]{tey02a}.\footnote{Our setup files are freely available upon request to the corresponding author (TTG). \label{foo:code}} All gaseous components are assumed to consists of a monoatomic, ideal gas, with an hydrogen fraction (by mass) $\XH = 0.76$. And the contribution of {\em all} components to the overall gravitational field is taken into account at all times.

We ignore in our simulation star formation and feedback of any kind. The only microphysical process we consider is radiative cooling by hydrogen, helium, and heavy elements \citep[cf.][their Sec.~4.2]{tep18a}. The relevant parameters in this respect are the gas metallicity and $\XH$. The Sgr halo gas metallicity is initially set to $Z = 0.2$ \Zsun, corresponding to lowest metallicity of stars of age $\sim 4$ Gyr \citep[][]{maj03a}, the approximate infall epoch of Sgr in our model (see Sec.~\ref{sec:sdm}). For reference, estimates of the metallicity of its youngest ($\lesssim 1$ Gyr) stellar population range from $\sim 0.4$ \Zsun\ to $\sim 4$ \Zsun\ \citep[][]{maj03a,sie07a}. Because we are ignoring the formation of stars and their metal yields, the halo gas metallicity may only change (increase) due to the interaction with the Galactic corona or the Galactic gas disc, which are both initially set at 0.3 \Zsun\ \citep[][]{mil15a,hou00a}. Such a relatively low value is chosen to avoid excessive cooling at the Galactic disc-halo interface, thus compensating for the absence of heating via stellar feedback in our models.

It is worth noting that a higher initial gas metallicity within the dwarf would promote gas cooling, which could potentially reduce gas stripping by forcing gas deeper into the dwarf's potential well \citep[e.g.][but see \citealt{ton09a} for the case of more massive galaxies]{may06a}. This effect may be further suppressed by our neglecting the chemical enrichment of the gas. On the other hand, we are also ignoring stellar energy injection into the gas within the dwarf in the form of kinetic energy and ionising radiation, which would unavoidably heat the gas within the dwarf. This is compensated to some extent by the initial temperature of the gas halo within the dwarf ($\sim 10^4 - 10^5$ K) as mentioned above. Internal heating is believed to assist gas stripping via ram pressure off low-mass ($\lesssim 10^9$ \Msun) galaxies \citep[e.g.][]{nic11a}. But simulations that self-consistently take into account the effect of radiative gas cooling, star formation, stellar feedback, gas enrichment and even ultraviolet radiation heating indicate that ram pressure stripping is unlikely to be relevant in galaxies with total masses $\gtrsim 5\times 10^9$ \Msun\ \citep[e.g.][]{saw12a}, as is the Sgr precursor in our model. Thus we argue that a (relatively low) initial gas metallicity of the dwarf's halo on the one hand, and the absence of stellar feedback in the form of metal yields and energy on the other, pose reasonable assumptions to model the stripping of the gas associated to Sgr.

\begin{figure*}
\centering
\includegraphics[width=0.33\textwidth]{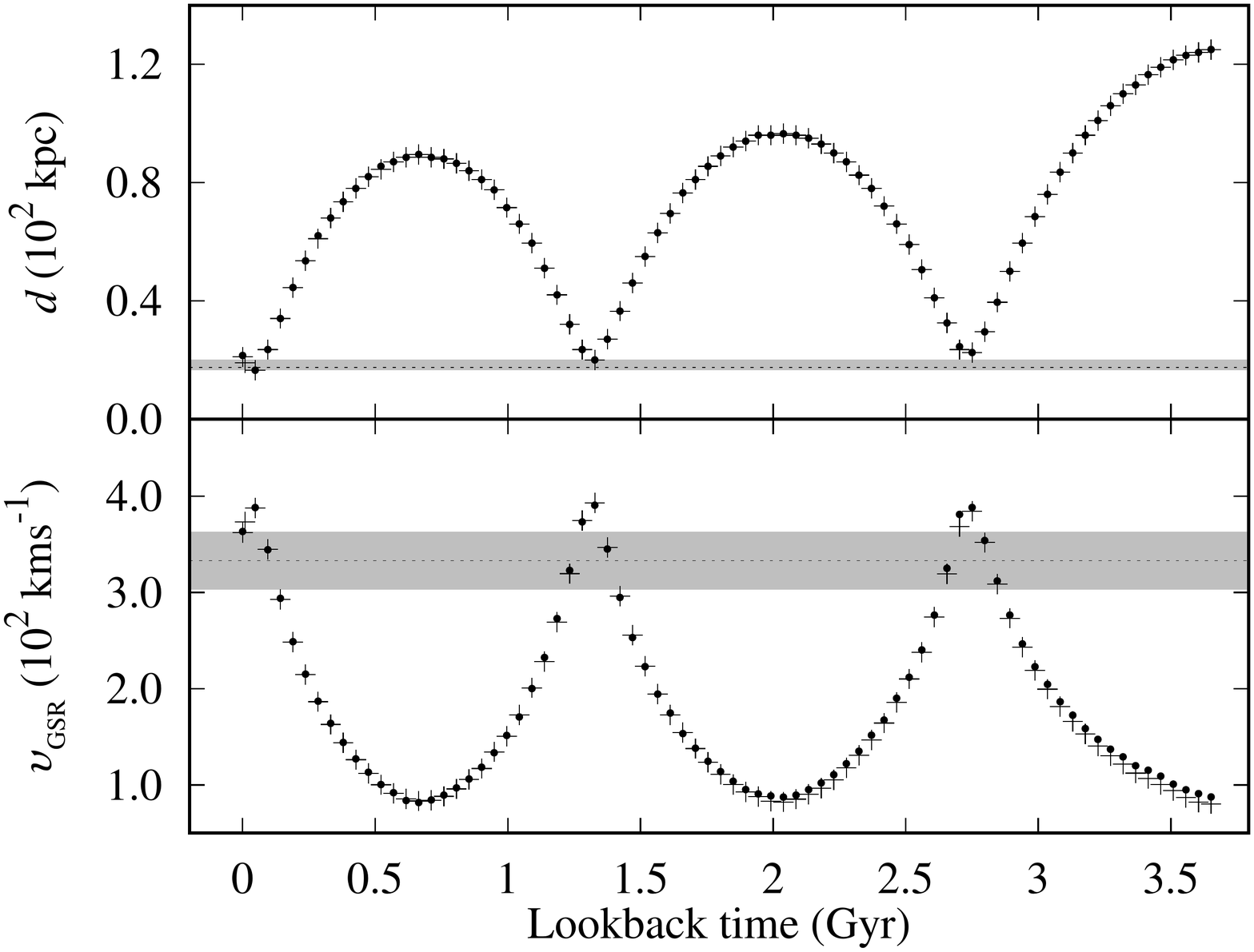}
\includegraphics[width=0.33\textwidth]{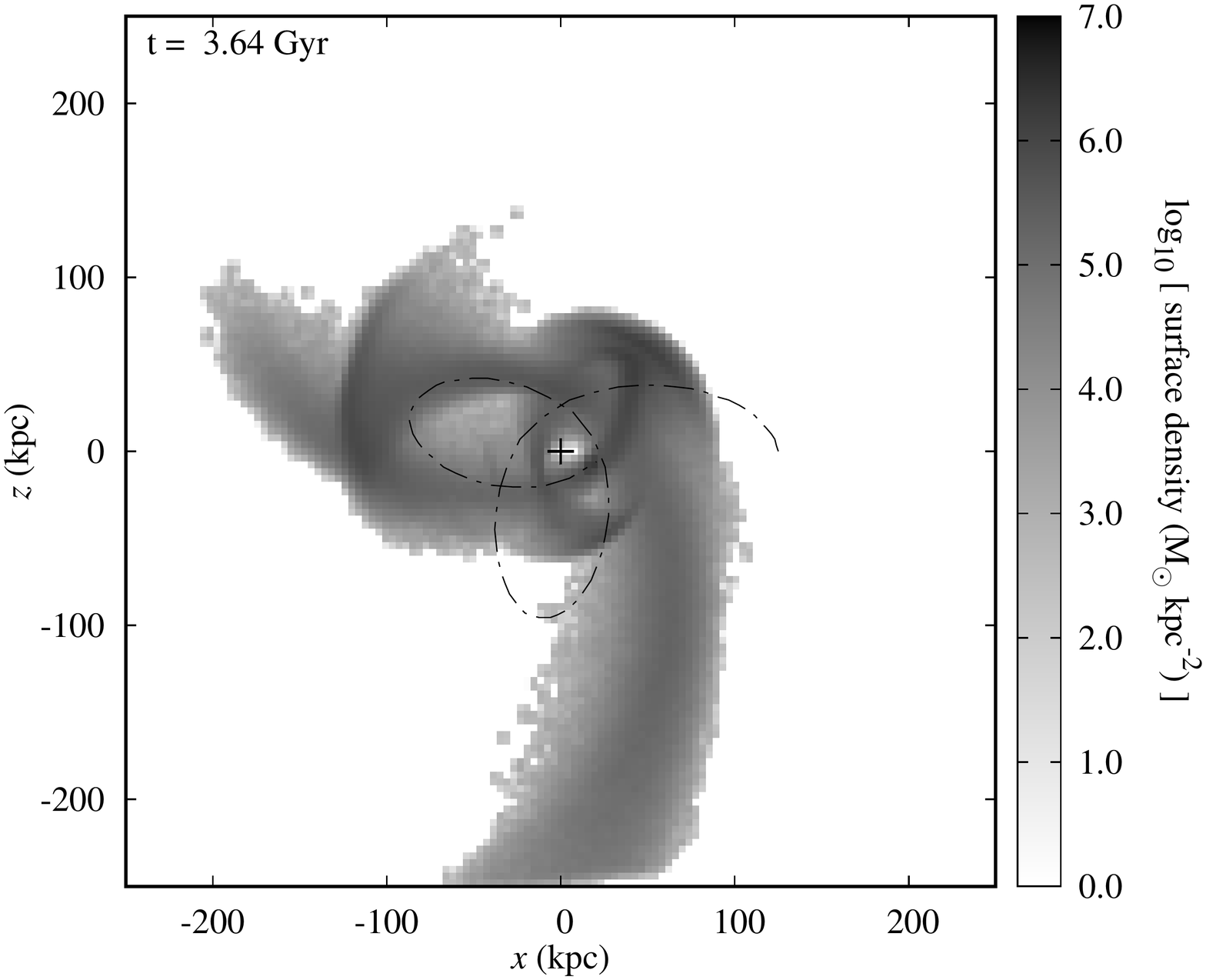}
\includegraphics[width=0.33\textwidth]{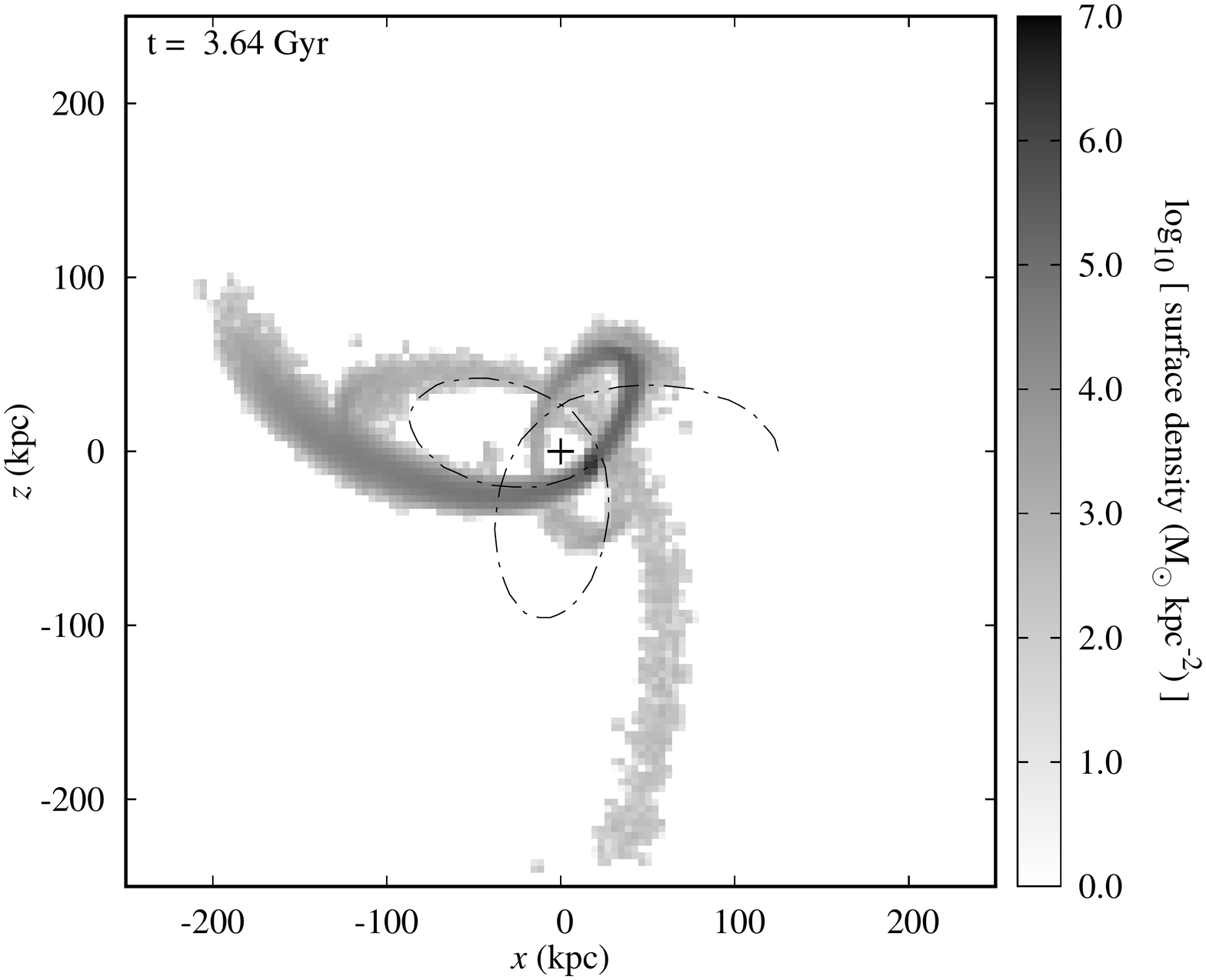}
\caption[ Model C Sgr core kinematics; DM and star distribution at snap = 385  ]{ Left: Kinematics of the model Sgr core from infall to the present epoch. Dots (plus signs) correspond to the DM (stellar) component. The top (bottom) sub-panel displays the Galactocentric distance (Galactocentric speed) along its orbit. The horizontal dashed line and hatched area corresponds to the range of distance determinations (top) and mean total speed (bottom) of the Sgr core and its corresponding uncertainty, as compiled by \citet[][]{kun09a} and estimated by \citet[][]{die17a}, respectively. The effect of dynamical friction in reducing the apo- and pericentric distances is apparent. Centre / Right: Present-day configuration (i.e. at a lookback time of $\tau = 0$ Gyr, or $t \approx 3.64$ Gyr since infall) of the distribution of DM (centre) and stars (right) of the Sgr model projected along a vector perpendicular to the orbital plane of Sgr, coincident with an edge-on view of the Galactic disc (not shown). The cross flags the position of the Galactic Centre. The dot-dashed curve indicates to the precursor's orbit, from infall to its current location. Clearly, the DM and stellar cores track each other well, both spatially and kinematically.}
\label{fig:kinC}
\end{figure*}

A final important parameter of our simulation is the limiting spatial resolution. As we are being guided by the $N$-body models by \citet[][]{die17a}, and results from $N$-body experiments, in particular the disruption of haloes, are generally resolution-dependent \citep[e.g.][]{van18a}, we need to choose our limiting spatial and mass resolution accordingly. In their models, \citet[][]{die17a} adopt particle masses of $4 \times 10^4$ \Msun\ and $10^6$ \Msun, and minimum softening length of roughly 40 and 215 pc, for the stellar and DM components, respectively. We set the DM and stellar particle masses to $\sim 1.4 \times 10^4$ \Msun\ and $\sim 3.6 \times 10^3$ \Msun, respectively, and adopt a nominal limiting spatial resolution of about 250 pc for these components, and 60 pc for the gas components. In AMR simulations, the effective mass resolution is dictated by the adopted refinement strategy, for collisionless components usually defined in terms of the number of particles within a cell: in our experiment, if this number exceeds 40, the cell is refined. Thus, we have an effective mass resolution of order $10^5 - 10^6$ \Msun\ for the collisionless components. The maximum gas mass per resolution element is on the order of $10^6$ \Msun. Both our limiting spatial and mass resolution are therefore consistent with \citet[][]{die17a}. We shall refer to these as our `standard' resolution settings.

We assess the effect of our choice of limiting spatial resolution by running our reference model at higher resolution, corresponding to a minimum linear cell size of roughly 60 pc and 15 pc for the collisionless and gas components, respectively, with effective particle and gas masses both on the order of $10^5$ \Msun.

\section{Results} \label{sec:res}

Whenever possible, we compare our model results to the corresponding data.
In order to map the simulation results to observed space, we adopt a right-handed, Cartesian coordinate frame with the origin at the initial geometric centre of the Galaxy, with its positive $x$-axis pointing towards the infall point of the Sgr precursor ($\vec{r}_0$; see Sec.~\ref{sec:mod}), and its positive $z$-axis pointing towards the North Galactic Pole. In this frame, the Sun's position is identified {\em at any time}\footnote{This is equivalent to assuming that the Sun is fixed in space, rather than rotating around the Galaxy and oscillating across the Galactic plane.} by the vector $\vec{r}_{\rm sun} = \left( -8.3,\, 0,\, 0.025 \right)$ \kpc\ \citep[][]{bla16a}.
Note that we do not adopt the optimal orientation of the Galactic disc with respect to $\vec{r}_0$ determined by \citet[][]{die17a}, which  roughly corresponds to a rotation of 180 degrees around the $z$-axis, and a small tilt of the Galactic plane. We adopt a solar motion relative to the Local Standard of Rest (LSR) given by $\left( U, \, V, \, W \right) \approx \left( -11.1, \, 12.2, \, 7.3 \right)$ \kms\ \citep[][]{sch10b}, and the rotation velocity of the Galaxy at the Sun's location is taken to be 220 \kms\ \citep[][]{bov12b}. It is important to mention that our adopted structural and kinematic parameter values may in general differ slightly from those adopted by observational studies we use to compare the outcome of our experiments. A reasonable amount of disagreement between the model and data is thus expected from the onset.

\subsection{DM core, stellar remnant, and tidal debris} \label{sec:sdm}

We follow the infall of the Sgr precursor until the Galactocentric distance and velocity of its core roughly match their corresponding observed values, with its location along its orbit currently just past pericentre \citep[][]{iba97a}. The position and velocity of the DM core and of the stellar remnant in our model are taken to be those of the densest Sgr (DM / stellar) element at any time. This approach breaks down when tidal disruption becomes important and compression of the tidal streams along their orbit takes place, which in our model takes place at $t \gtrsim 5 $ Gyr since infall (not shown).\footnote{We denote simulation time with a lower case $t$, and lookback time with a Greek letter $\tau$; the epoch of the infall event is identified by $\tau_0$, and thus $\tau = \tau_0 - t$.} The kinematics of the DM core and stellar remnant in our model are shown in the left panel of Fig.~\ref{fig:kinC}. Apart from some slight differences in the resulting orbits and thus the inferred infall epoch, the results discussed here are comparable in the case of our reference run at the standard and high resolution, and between these and the gas-rich precursor. We will therefore limit our presentation to the results of the reference model at our standard resolution.\footnote{ Animations of the evolution of Sgr and additional material can be found at \url{http://www.physics.usyd.edu.au/\textasciitilde tepper/proj\_sgr\_paper.html}. }

The Galactocentric distance of Sgr's bound core has been estimated thus far to lie in the range $\sim 16 - 20$ kpc (see Sec.~\ref{sec:intro}). We adopt as reference the most common value of $d \approx 17$ kpc. In our model, Sgr reaches this distance at around $t \approx 3.6 ~\Gyr \equiv \tau_0$ since infall, moving at a total space velocity (with respect to the Galactic Standard of Rest or GSR) of $v \approx 370$ \kms. For reference, \citet[][]{die17a} estimate the observed total space velocity magnitude at $v = 333 \pm 30$ \kms. Thus, our model implies that Sgr entered the virial radius of the Galaxy at $z \approx 0.35$. Such a delayed infall is perfectly consistent with constraints on the timescales of stripping based on the trends of abundance ratios with metallicity observed in Sgr's nucleus \citep[][]{muc17a} and its star formation history \citep[SFH;][]{de-15c}, and is broadly consistent with the accretion history of subhaloes onto the Galaxy host halo \citep[][]{wet15b}.

\begin{figure*}
\centering
\includegraphics[width=0.33\textwidth]{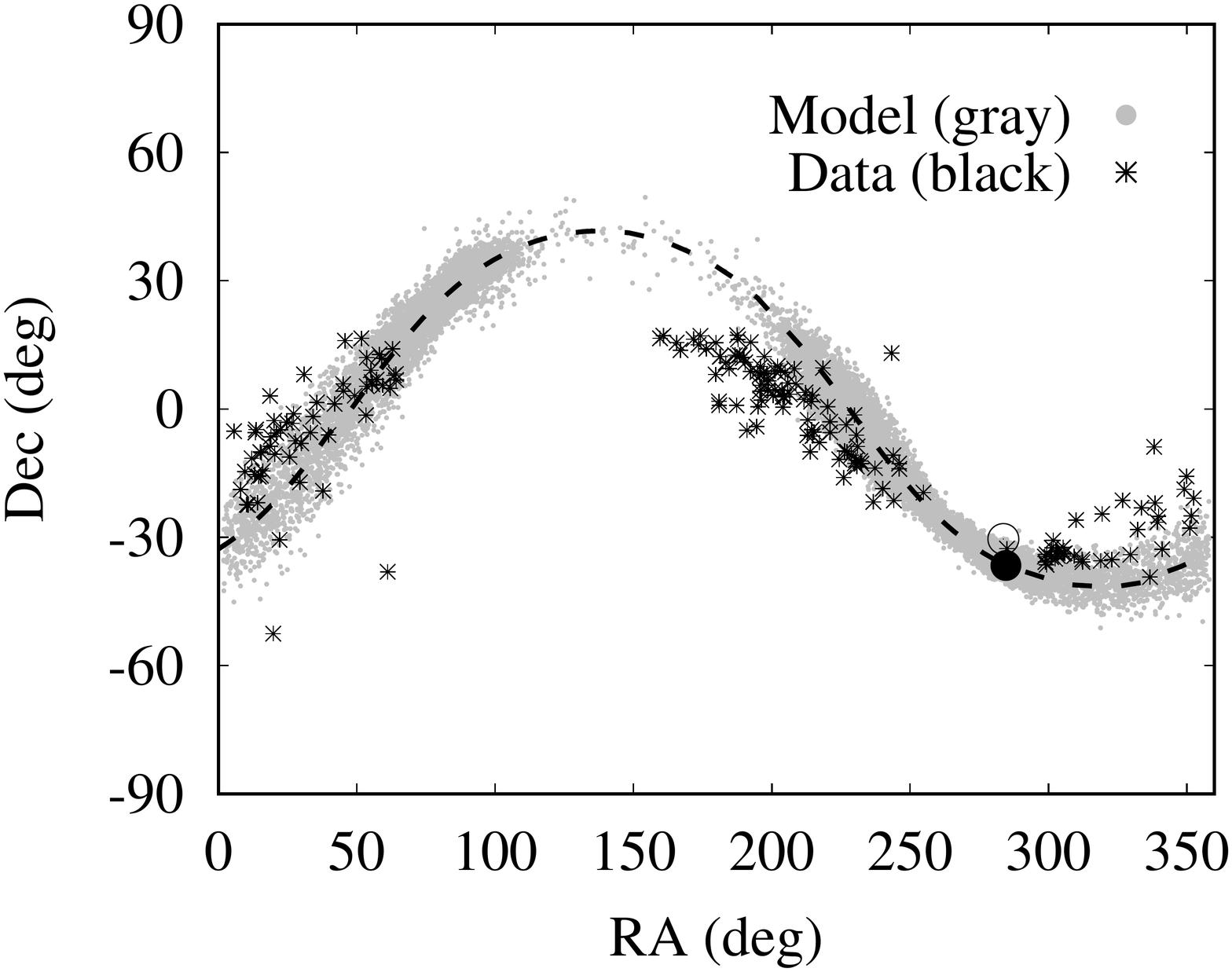}
\hfill
\includegraphics[width=0.33\textwidth]{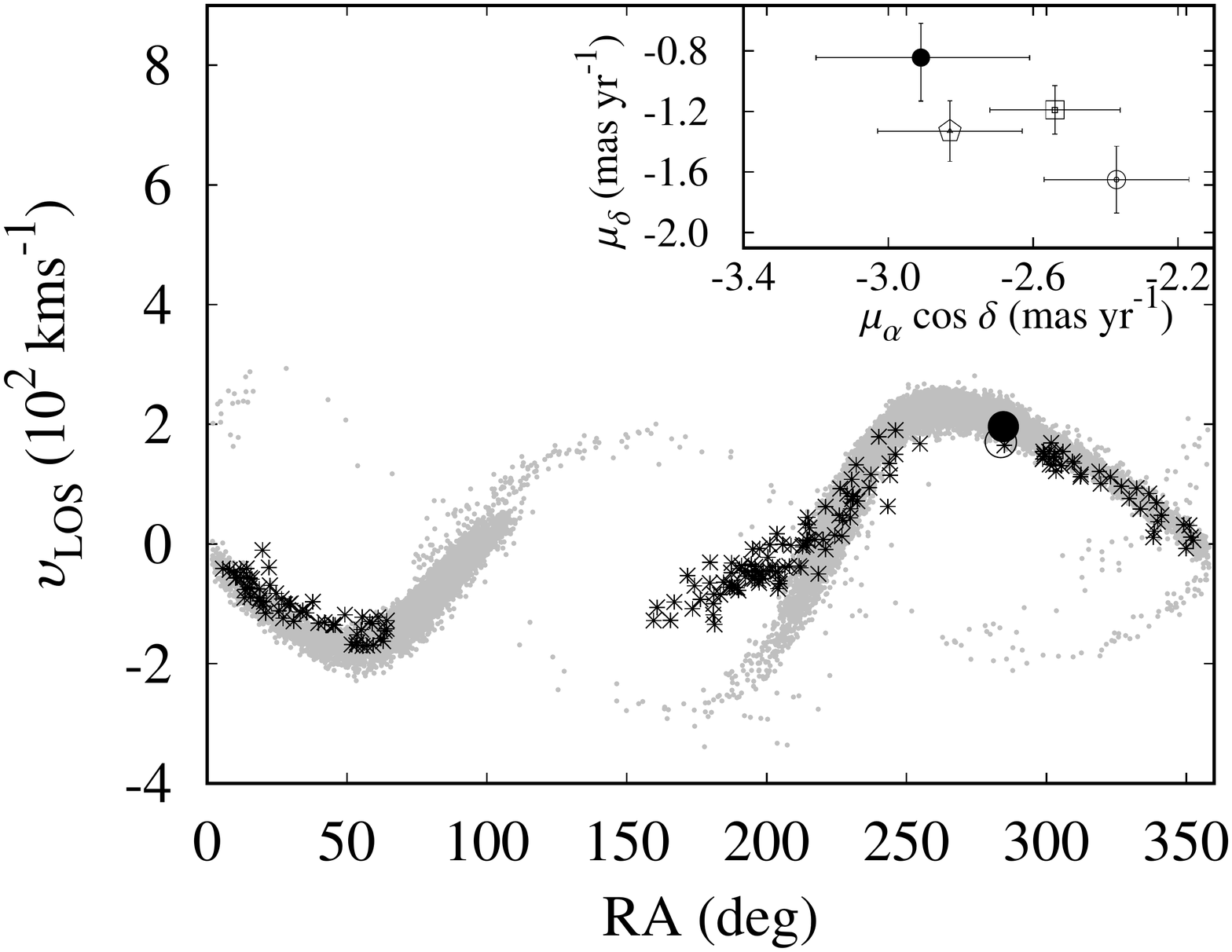}
\hfill
\includegraphics[width=0.33\textwidth]{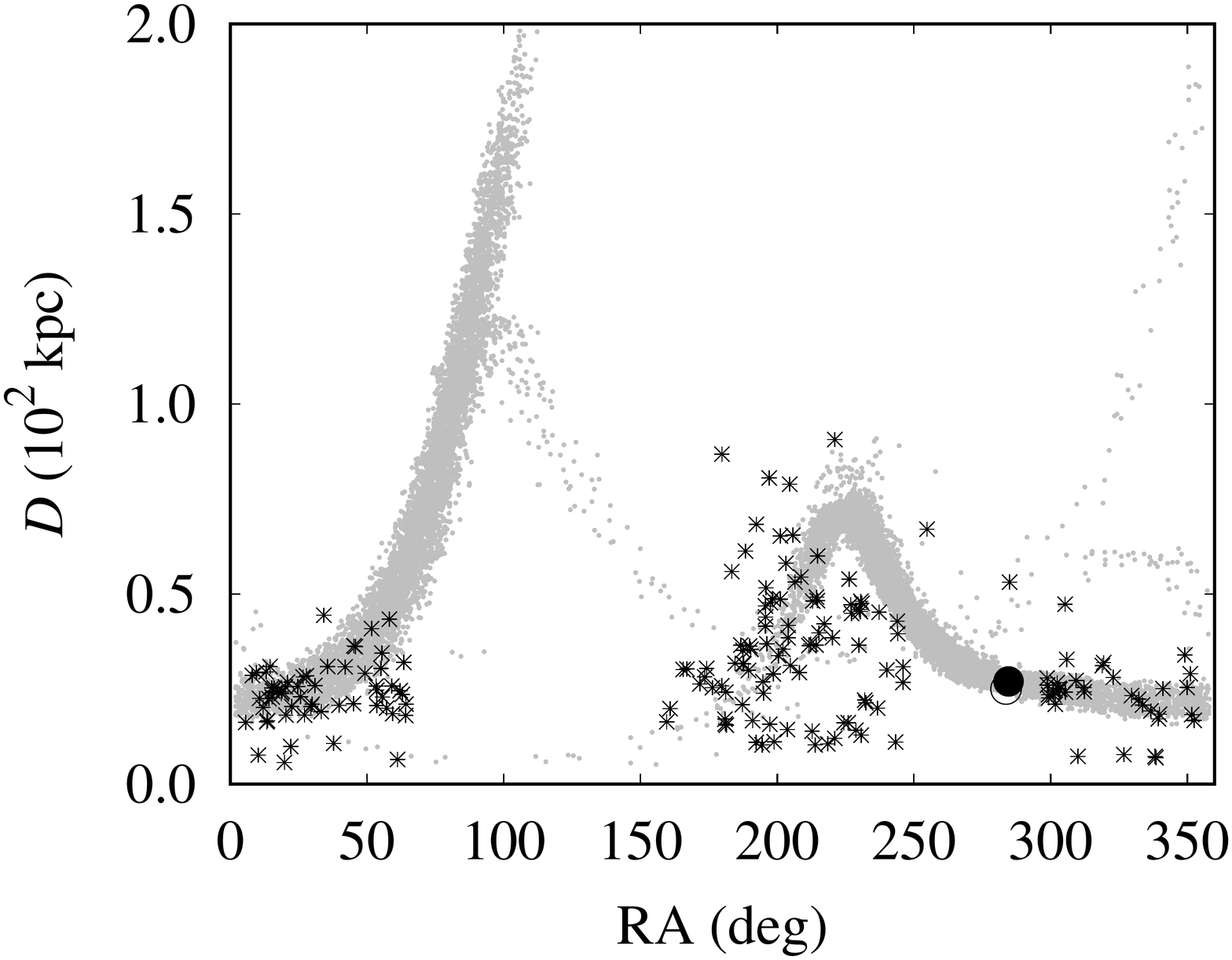}
\caption[ Model C Sgr sky distribution and data ]{ Present-day distribution of the stellar component (streams and core) of the Sgr model (grey dots), and data (black asterisks) corresponding to the stars in the tidal and leading streams of Sgr. Left: All-sky distribution (equatorial coordinates). A black filled (open) circle indicates the location of the Sgr core in the model (data). The dashed line indicates the orbit of the model Sgr core. Centre: line of sight velocity in the GSR vs. Right Ascension. Right: Heliocentric distance vs. Right Ascension. The inset in the central panel displays the proper motion components of the Sgr core. The open symbols and associated error bars correspond to different measurements of Sgr core's proper motion and its uncertainty: \citet[][polygon]{din05a}; \citet[][circle]{pry10a}; \citet[][square]{mas13a}. The uncertainty in our model is estimated using the corresponding values of adjacent snapshots that are $\sim 10$ Myr apart. Stellar stream data from \citet[][]{maj04a};\protect\footnotemark\ Sgr core coordinates from \citet[][]{maj03a}; line of sight velocity of the core in the GSR calculated from the heliocentric radial velocity measured by \citet[][]{bel08a}; core's heliocentric distance from \citet[][]{kun09a}. }
\label{fig:sgrCdata}
\end{figure*}

As anticipated above, despite the similarities in the physical and numerical properties of the Galaxy model as well as of Sgr's precursor and its orbital initial conditions between \citet[][]{die17a}'s model and ours, the orbital history of the dwarf is rather different between these models. Indeed, in ours the best match to the observed distance and velocity of Sgr is found after only three (rather than five) pericentric passages. In other words, our gas-dynamical model shows a faster orbital decay compared to a pure $N$-body model. Apparently, as Sgr moves through a gaseous medium -- and as long as the dwarf holds onto its gas --, its orbit is affected by hydrodynamical drag,\footnote{Gas drag scales with the background density $\rho$, the dwarf's speed $v$, its bound gas mass $m$ and cross section $A$ as $\sim\,A\,\rho\,v^2 / m$ \citep[][]{ben97a}.} in addition to dynamical friction. The braking effect of gas is present everywhere along the dwarf's orbit through the Galactic corona and the Galactic disc. In either case, its influence on the dwarf's dynamics is important. The coronal is rather diffuse much along Sgr's orbit, but its interaction with the dwarf is long and persistent. The interaction of Sgr with the gas disc is rather short, but strong nevertheless, because of the significantly higher density of the disc compared to the corona. The interaction with the disc is in fact so strong that, in addition to affecting Sgr's orbit, it is responsible for most of the gas stripping experienced by the dwarf along its orbit within the virial radius of the Galaxy (see Sec.~\ref{sec:gloss}).

The central and right panels Fig.~\ref{fig:kinC} show the present-day configuration of the DM component (centre) and stellar streams (right) of the model Sgr in physical space on an edge-on view of the system. For reference, the orbit of the precursor from infall to its present position is included as a dashed curve. In this projection, the Sgr precursor has fallen in from the right, and the (DM / stellar) core is instantly moving towards the top-right. Note that the stellar and DM components track each other well, both spatially and kinematically. The tidal debris remain on the orbital plane, but {\em not} necessarily along the orbit, as their Galactocentric distance varies with time (see also Fig.~\ref{fig:sgrCdata}, left and right panels). Because Sgr has not had but a few pericentric passages, the tidal disruption is not as dramatic and the tidal streams are not as developed as in previous models \citep[e.g.][]{gib14a}. This could also be in part consequence of the lack of a disc in the precursor, or the choice of profile for the stellar bulge.

\begin{figure}
\centering
\includegraphics[width=0.45\textwidth]{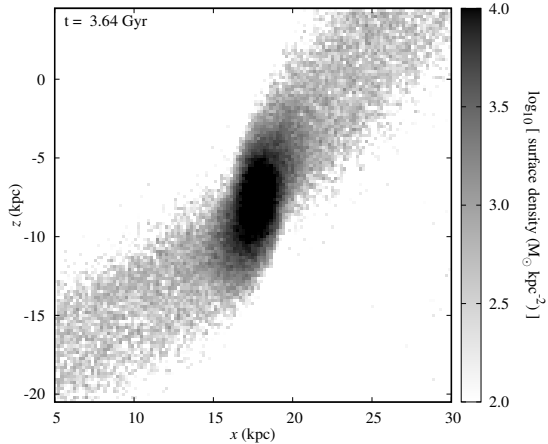}
\caption[ Model C Sgr core ]{ Present-day projected stellar density of the model Sgr remnant. }
\label{fig:core}
\end{figure}

Fig.~\ref{fig:sgrCdata} compares the distribution in observed space of the stellar component (streams and core) of the Sgr model to the data\footnotetext{As provided in \url{http://faculty.virginia.edu/srm4n/Sgr/}.}. \citet[][]{maj03a} estimate the centre of the Sgr remnant at $\left( \alpha, \, \delta \right) \approx ( 284, \, -30.5)$ deg. In our model, the Sgr core is found at $\left( \alpha, \, \delta \right) \approx ( 285, \, -36.6)$ deg.
The radial (heliocentric) velocity of the Sgr core is $v_r \approx 140$ \kms\ \citep[][]{bel08a}. Using the equatorial coordinates for the Sgr core, the radial velocity translates into a line of sight velocity in the GSR of $v_{\textnormal{\sc los}} \approx 170$ \kms, consistent with \citet[][]{iba97a}. In our model, we find $v_r \approx 187$ \kms and $v_{\textnormal{\sc los}} \approx 196$ \kms. It is worth noting that some degree of mismatch between our model and the data in terms of the position (declination) or the GSR velocity of the tidal debris at $\alpha \approx 150 - 250$ deg is a common feature of other models \citep[e.g.][]{pur11a}, even in those that contemplate a discy Sgr progenitor \citep[e.g.][]{lok10a}; that consider the inclination of the orbit with respect to a perfect polar orbit \citep[][]{die17a}; or that take the perturbing effect of the LMC to the overall gravitational potential of the system into account \citep[][]{gom15a}.

The inset in the central panel of Fig.~\ref{fig:sgrCdata} shows the proper motion components of the model Sgr core on the plane of the sky. We find $\left( \mu_\alpha \cos \delta, \, \mu_\delta \right) = \left( -2.90,\, -0.85 \right)$ \maspyr. For comparison, \citet[][]{die17a} obtain\footnote{Values extracted from their Figure 7, and assuming $\delta = -30.5$ deg.} $\left( \mu_\alpha \cos \delta, \, \mu_\delta \right) = \left( -1.72,\, -0.90 \right)$ \maspyr. As noted by \citet[][]{mas13a}, different measurements of Sgr's proper motion may not be directly comparable to one another, as the may not correspond to the same region of the stellar remnant (in addition to different studies generally adopting different structural and kinematic parameters values, as mentioned above). Since it would be difficult to match the model Sgr core with the region used in any of these measurements, here we present the measurements as given in their respective reference, together with the result from our model, to show that the scatter in the measured values is comparable to the difference between them and our model result.

The model Sgr stellar remnant at the present epoch is displayed in Fig.~\ref{fig:core}. Adopting a \citet[][]{kin62a} profile, \citet[][]{maj03a} measured the extension of the Sgr core to be $\sim 30$ degrees (or roughly 14 kpc at $D = 25$ kpc), and its ellipticity, $e = 0.65$. Assuming a prolate body shape, these measurements imply an axes ratio of roughly 3:1:1, consistent with \citet[][]{iba97a}. Hence, the core width can be estimated at $\sim 5$ kpc. In our model, the core has approximately the appropriate dimensions, and is comparable to other model results \citep[cf.][]{lok10a}. Thus, and contrary to the conjecture by these authors, the observed elongated shape of Sgr's stellar remnant can in fact be recovered from an initially spherical, disc-less precursor. The same behaviour is seen in the disc-crossing halo of \citet[][]{nic14b}.

\subsection{The Sagittarius gas stream} \label{sec:gas}

As Sgr spirals down the potential well of the Galaxy, the interaction with the Galactic corona removes the gas within Sgr that is less bound. Because the corona is predominantly diffuse along Sgr's orbit ($\lesssim 10^{-4}$ \pcc) and most of the gas is held within the dwarf's deep potential well, the fraction of gas that is removed is low. Substantial gas stripping occurs in fact only during disc transits -- which roughly coincide with its pericentric passages --, where the gas within Sgr is subject to strong shocks as it collides with the denser Galactic gas disc \citep[the `kinetic argument';][]{bla09a}.

\begin{figure}
\centering
\includegraphics[width=0.45\textwidth]{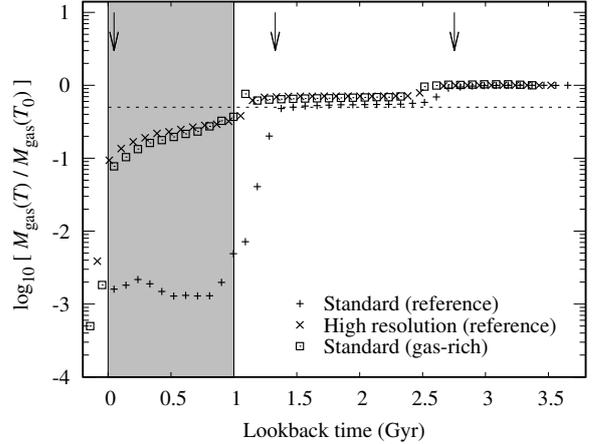}
\caption[ Model Sgr gas content ]{ Evolution of the mass of gas associated to Sgr with lookback time. Plus signs (+) indicate the mass in gas bound to Sgr within a radius of 5 kpc around its stellar core, relative to its initial value ($M_{\rm gas}(\tau_0) \sim 6  \times 10^8$ \Msun); crosses ($\times$) indicate the results of the higher resolution run. The result for a gas-rich precursor ($M_{\rm gas}(\tau_0) \sim 9  \times 10^8$ \Msun) at our standard resolution are indicated by the open squares. The shaded area indicates roughly the extension of the most recent star formation episode in Sgr suggested by the age range of its youngest stellar population. The dashed horizontal line indicates half the fraction of the initial gas mass. The systematic, stepwise decrements in gas are in line with Sgr's disc transits (which follow its pericentric passages, approximately flagged by the vertical arrows in the case of the standard model, for illustration; cf. left panel of Fig.~\ref{fig:kinC}). }
\label{fig:gas}
\end{figure}

\subsubsection{Gas stripping} \label{sec:gloss}

In order to quantify the fate of all the gas associate to Sgr, we consider separately, at any given time: i) the gas bound to its stellar core; and ii) all gas, either bound or stripped, initially associated to its precursor, that is `detectable'. The mass in gas bound to Sgr is quantified by calculating at any given time the amount of gas within 5 kpc around its stellar core. The detectability of gas that has been stripped is estimated as follows. We attach a passive scalar (tracer) to the gas initially bound to the infalling galaxy, i.e. to its gas halo, with an initial large (albeit arbitrary) value of 10. The Galactic gas disc and the Galactic corona have been initially assigned a tracer value of 0. As a result of the interaction with the corona and the gas disc, and the consequent gas mixing \citep[][]{tep18a}, the value of the gas tracer attached to Sgr's gas will generally decrease. At any given time, we consider any stripped gas initially associated to Sgr to be detectable if its tracer value is larger than a (arbitrary) threshold of 2. It should be mentioned that our results are qualitatively insensitive to the exact adopted tracer value. A higher (lower) threshold simply translates into a slightly lower (higher) fraction of detectable gas. Thus, our choice of threshold value does not affect our analysis, but merely improves the clarity of the graphical representation of our results.

Fig.~\ref{fig:gas} shows the evolution of the gas bound to Sgr, relative to its initial total gas mass: Plus signs (+) correspond to our reference model at the standard resolution with an initial gas mass $M_{\rm gas}(\tau_0) \sim 6  \times 10^8$ \Msun; crosses ($\times$) indicate the corresponding result for the reference model at higher resolution. The results for a gas-rich precursor ($M_{\rm gas}(\tau_0) \sim 9  \times 10^8$ \Msun) at the standard resolution are indicated by open squares. In either case, the mass of gas bound to Sgr decreases substantially after each disc transit, which follow its pericentric passages (approximately flagged by the vertical arrows), but it remains practically unchanged in between. This clearly demonstrates the importance of the interaction of Sgr with the Galactic gas disc in removing gas from the dwarf. It also shows, as we found before \citep[][]{tep18a}, that a subhalo transiting the Galactic gas disc does not accrete any significant amount of gas. This is consistent with the conclusion reached by \citet[][]{sie07a} that the observed age-metallicity relation of Sgr can be explained assuming a closed-box model for the dwarf.

In the standard resolution case, Sgr virtually loses all of its gas after its second disc transit. The higher resolution run shows, however, that gas loss is in fact highly overestimated at our standard resolution. Indeed, in this case, about 10 percent of the original gas mass remains within the core up to the present-day, inconsistent with observations. The gas content of an initially gas-rich precursor experiences a similar evolution. The gas-rich precursor completely loses its gas after its third disc transit (as does the reference model at higher resolution), which explains the sudden drop in either case at $\tau < 0$ Gyr. In general, gas stripping is more dramatic after the second (and eventually the third) transit because the average gas density within the precursor is progressively lower as a result of previous episodes of significant mass stripping, while the gas column density of the Galactic disc at each crossing point -- progressively closer to the GC -- is higher as a result of its radially exponential surface density profile.

\begin{figure*}
\centering
\includegraphics[width=0.33\textwidth]{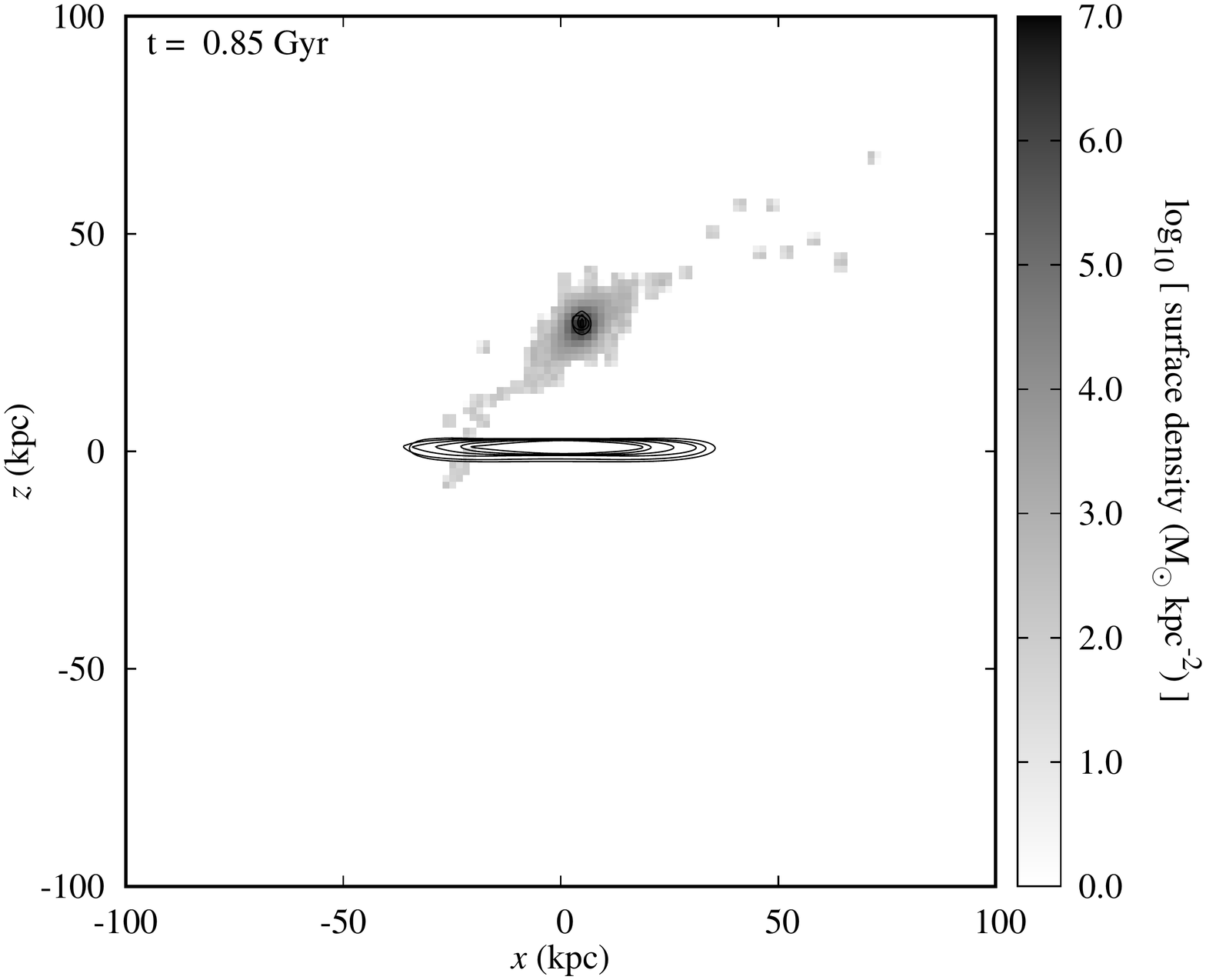}
\includegraphics[width=0.33\textwidth]{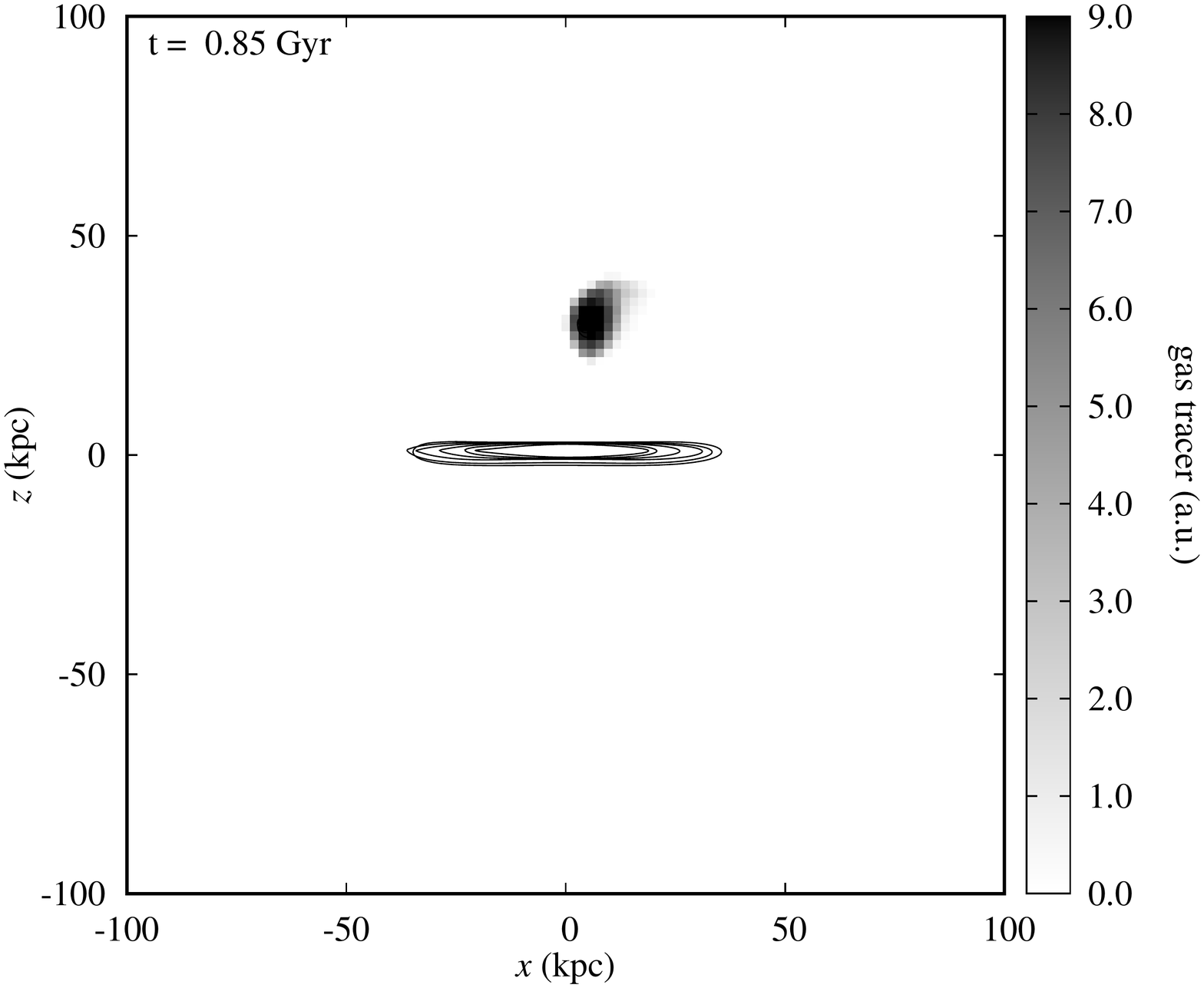}
\hfill
\includegraphics[width=0.33\textwidth]{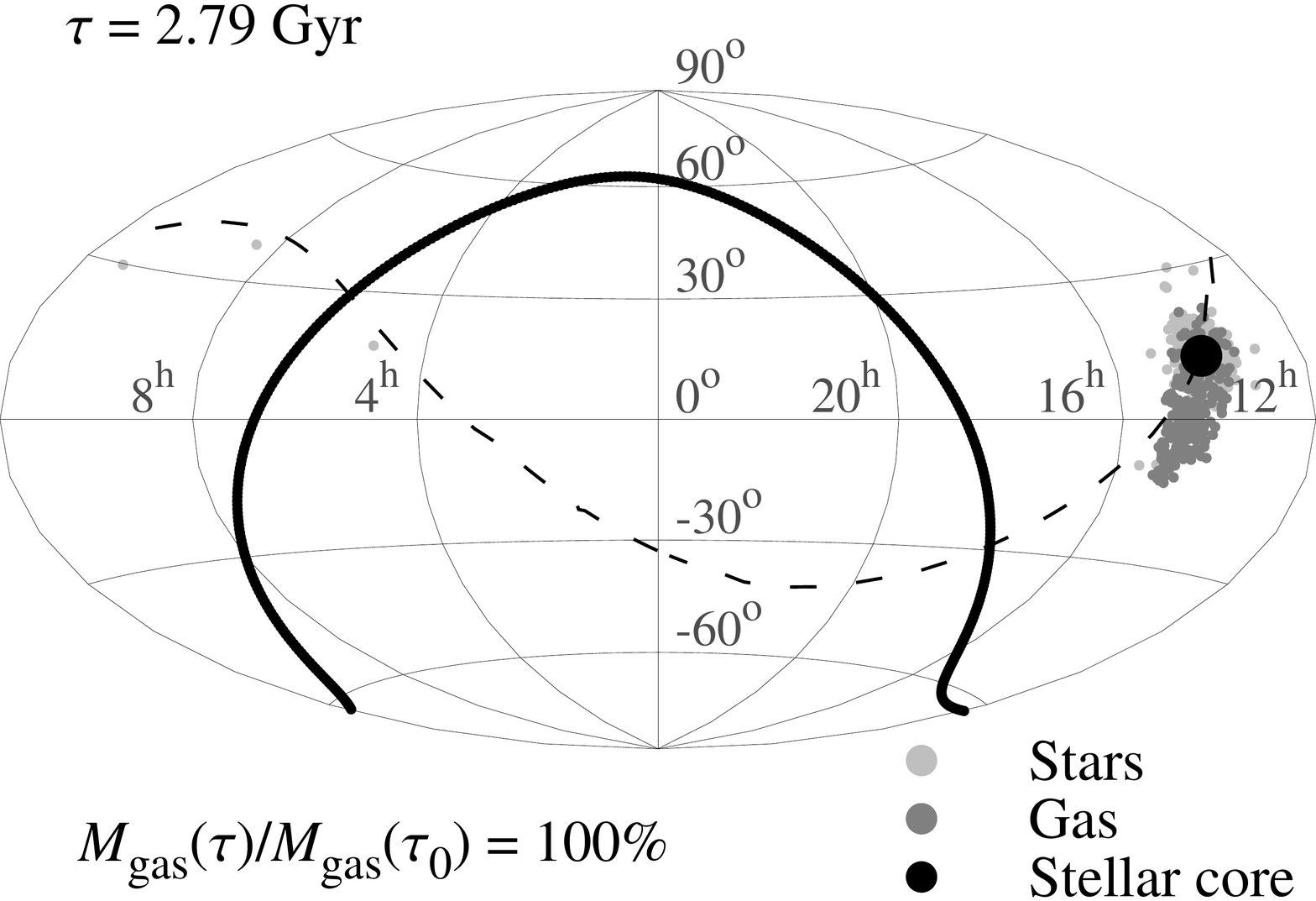}\\
\vspace{-5pt}
\includegraphics[width=0.33\textwidth]{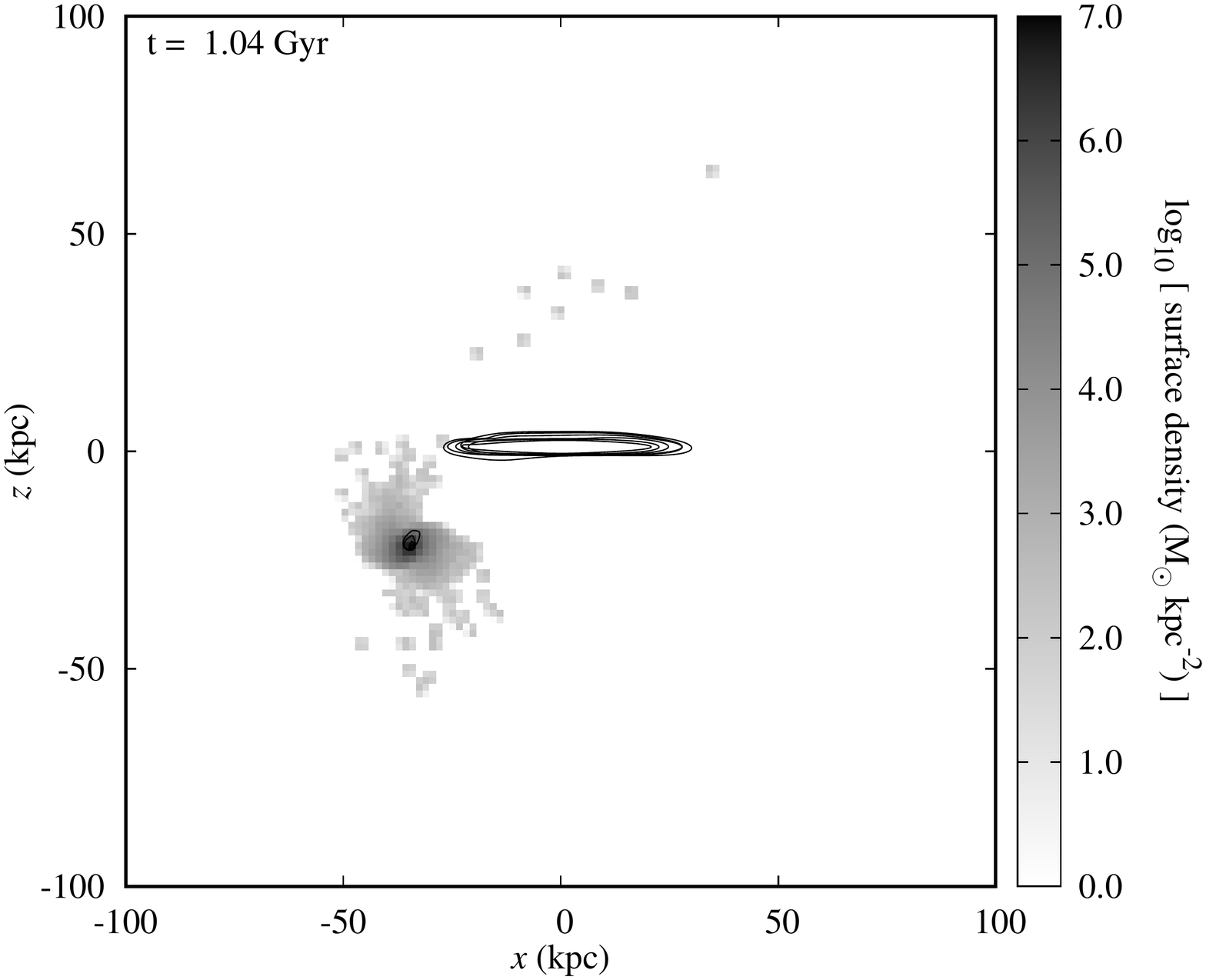}
\includegraphics[width=0.33\textwidth]{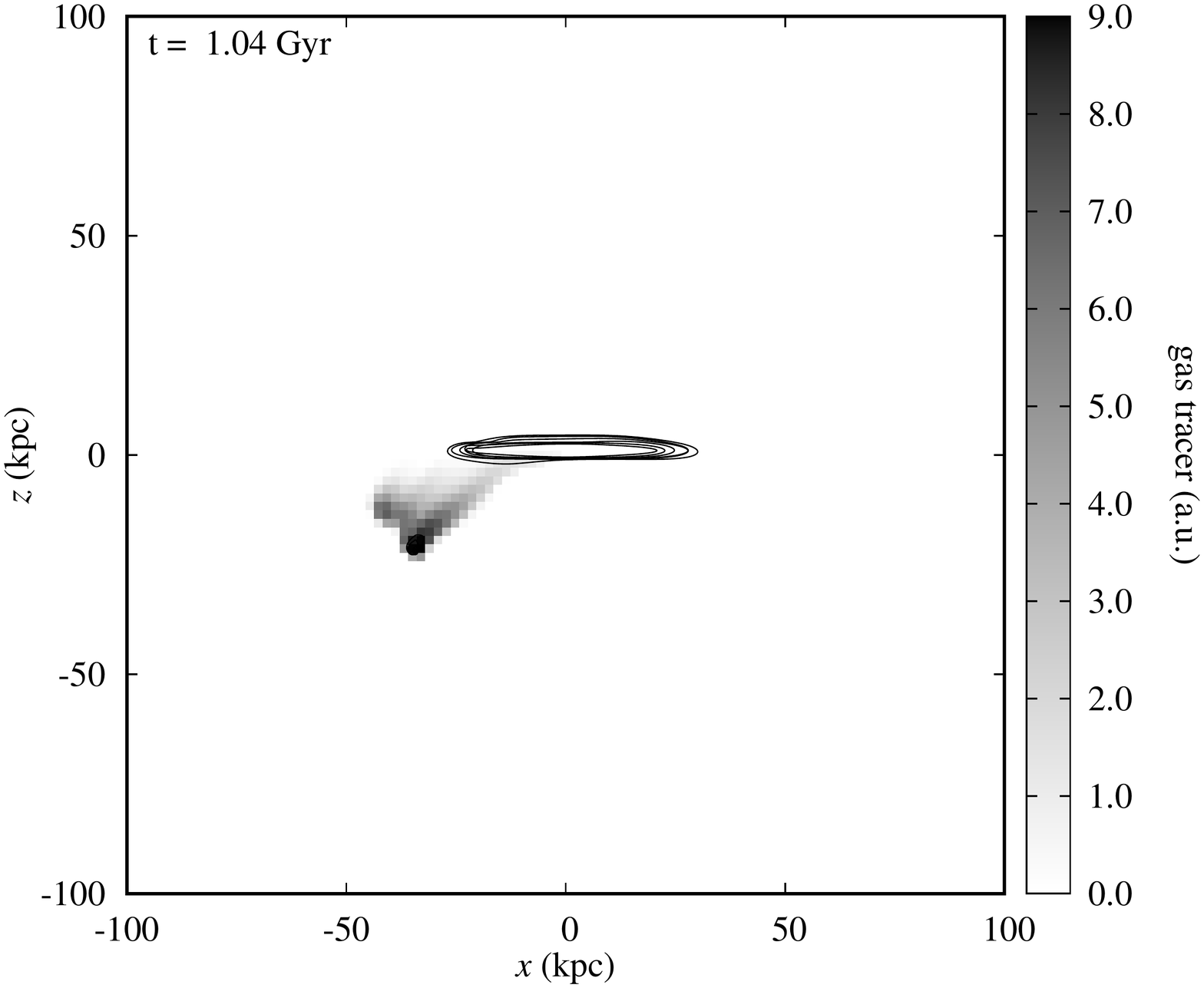}
\hfill
\includegraphics[width=0.33\textwidth]{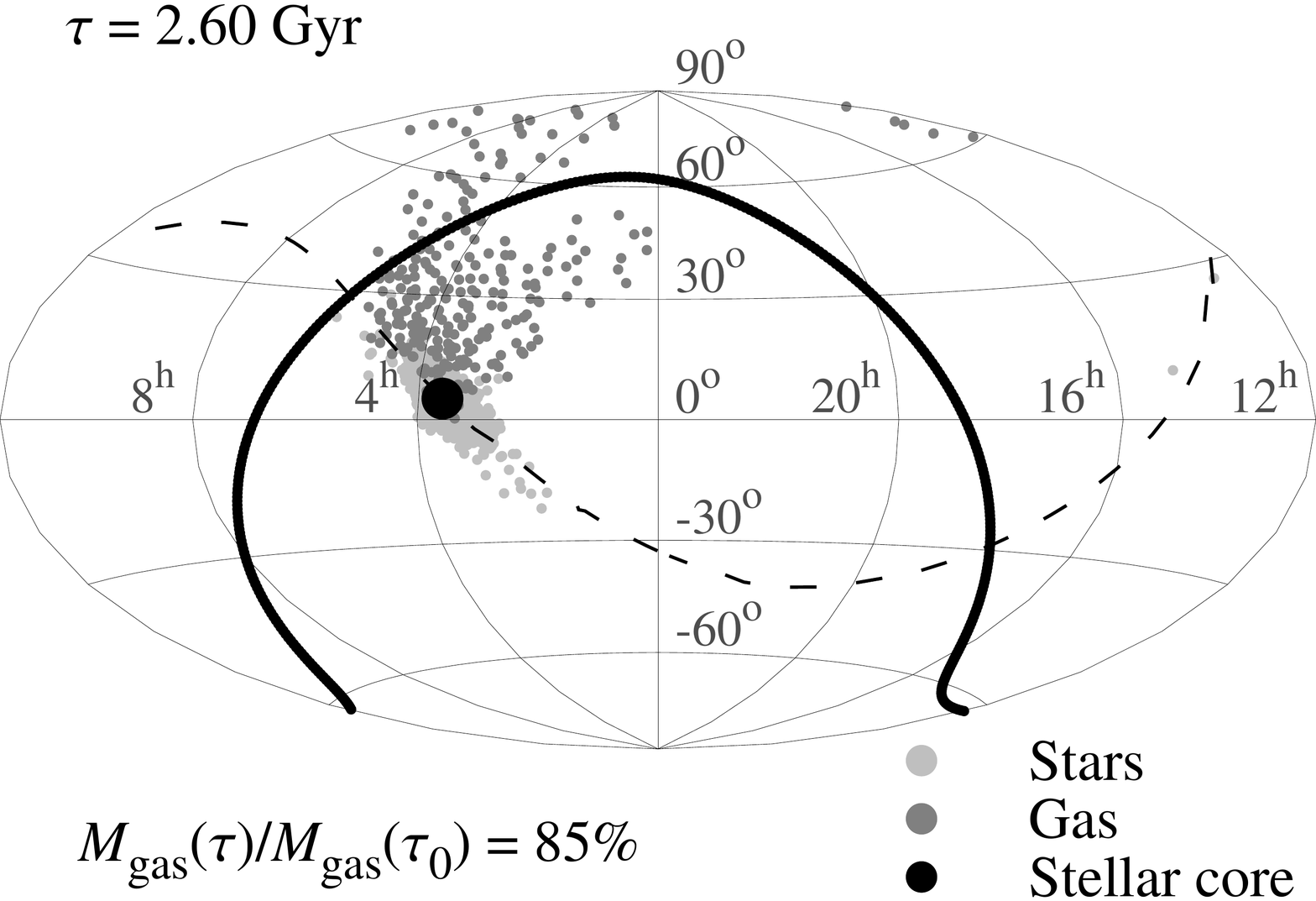}\\
\vspace{-5pt}
\includegraphics[width=0.33\textwidth]{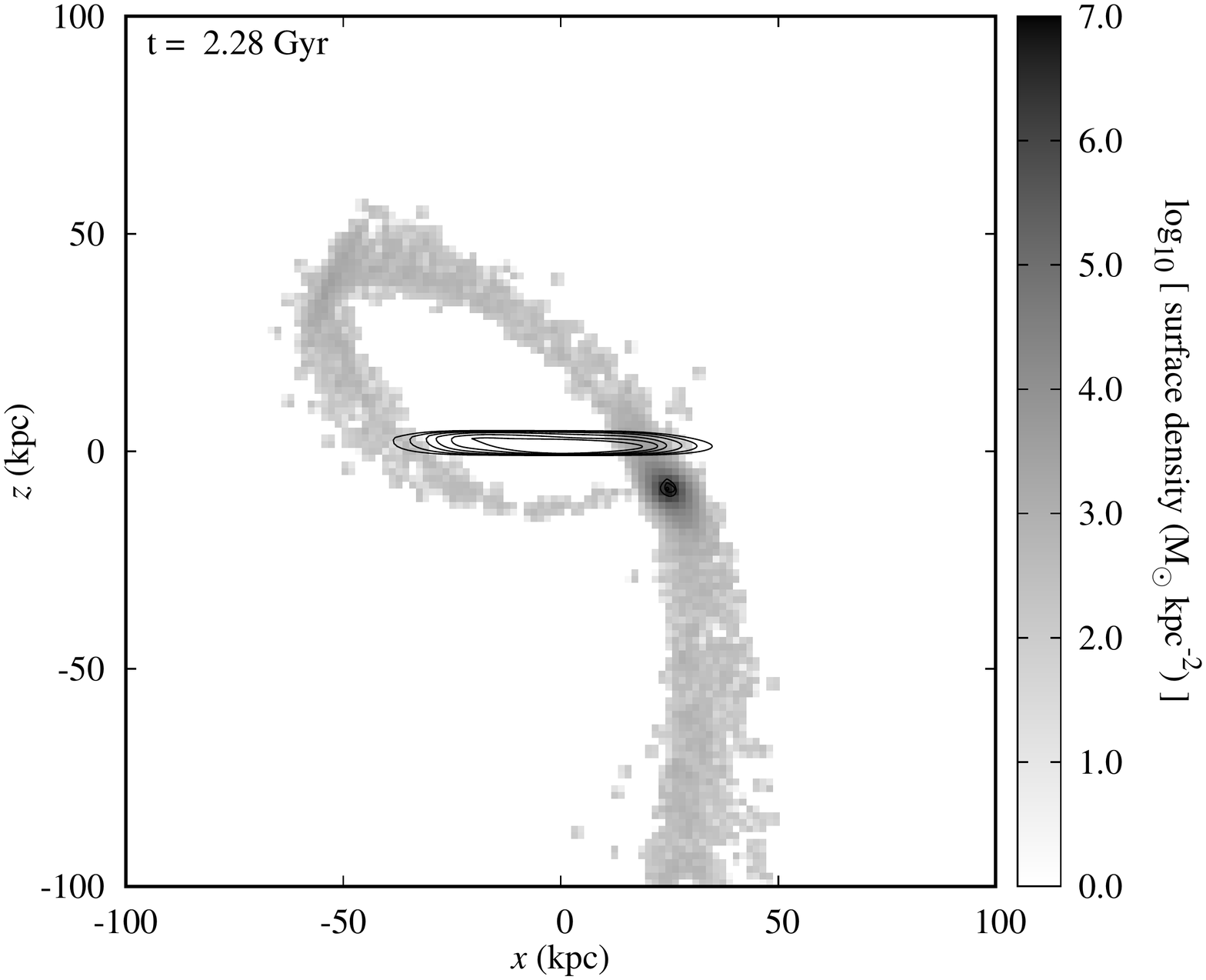}
\includegraphics[width=0.33\textwidth]{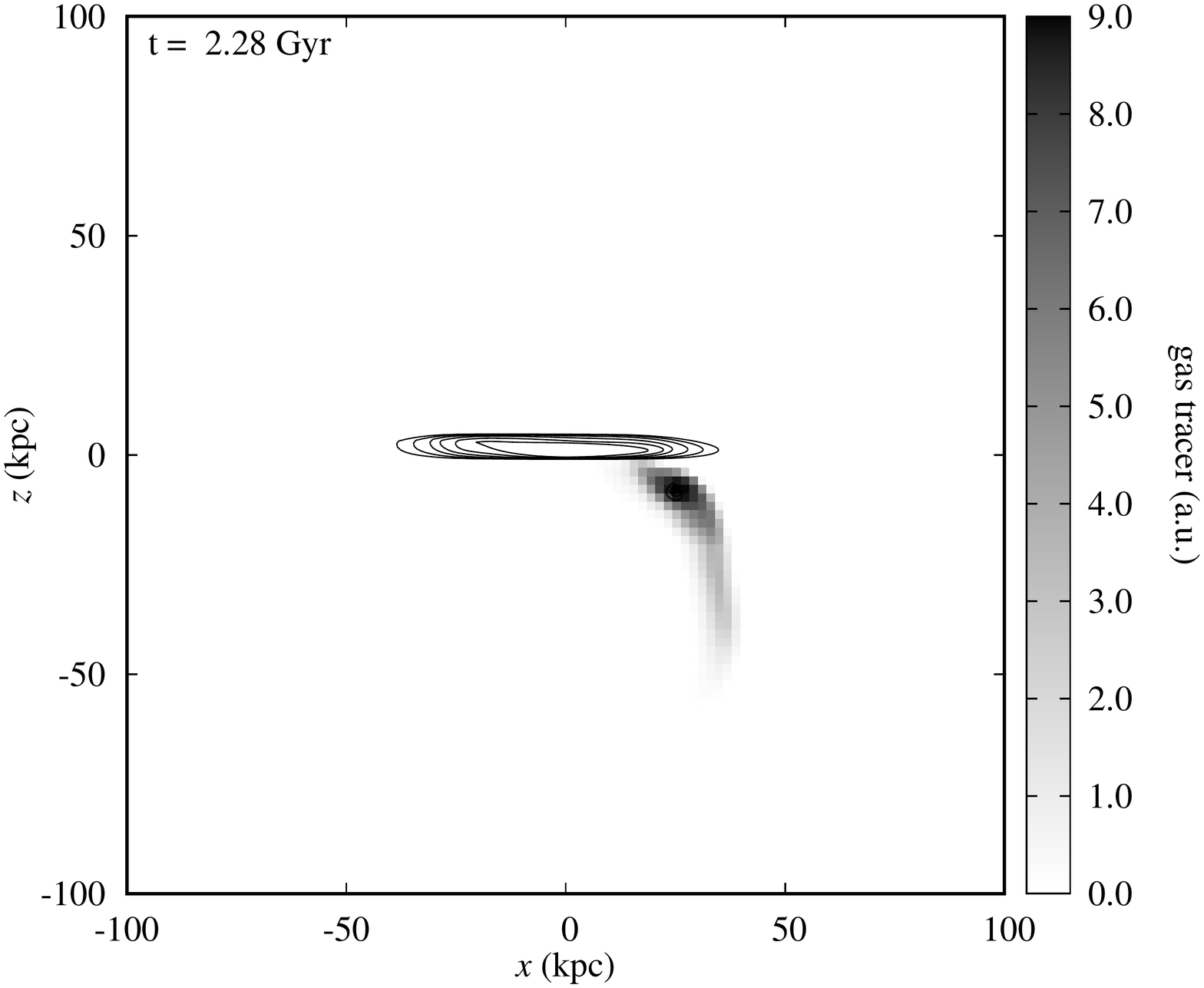}
\hfill
\includegraphics[width=0.33\textwidth]{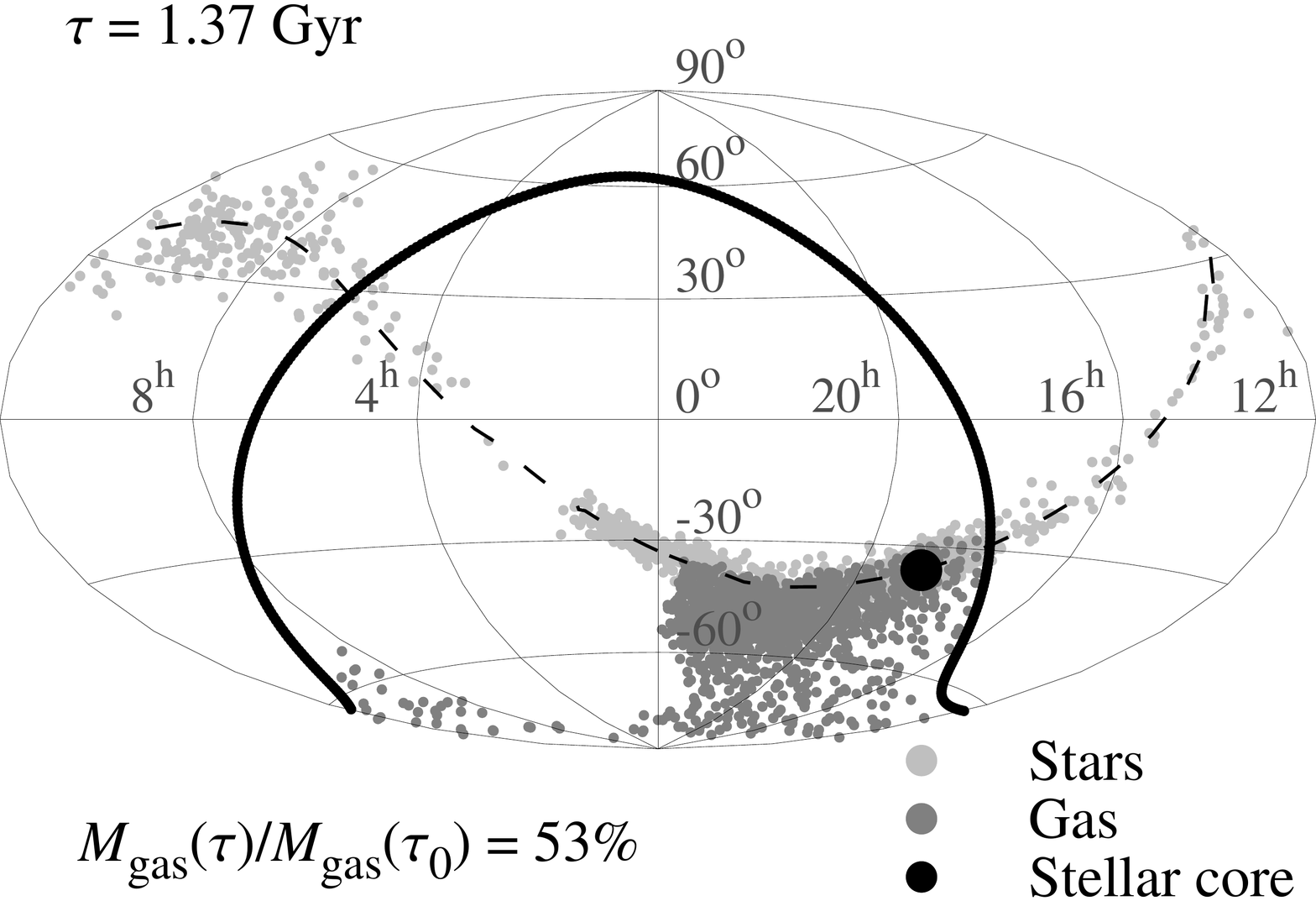}\\
\vspace{-5pt}
\includegraphics[width=0.33\textwidth]{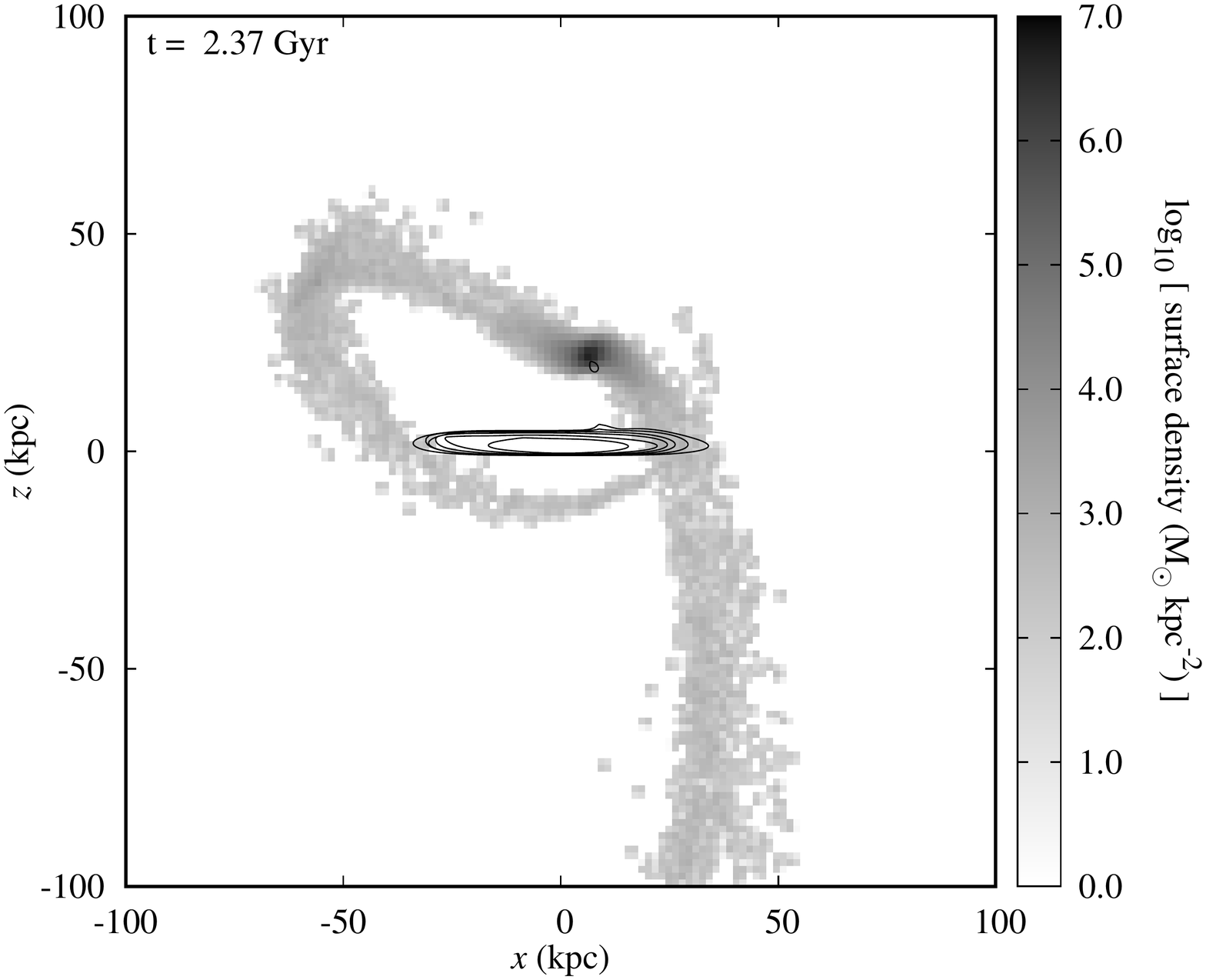}
\includegraphics[width=0.33\textwidth]{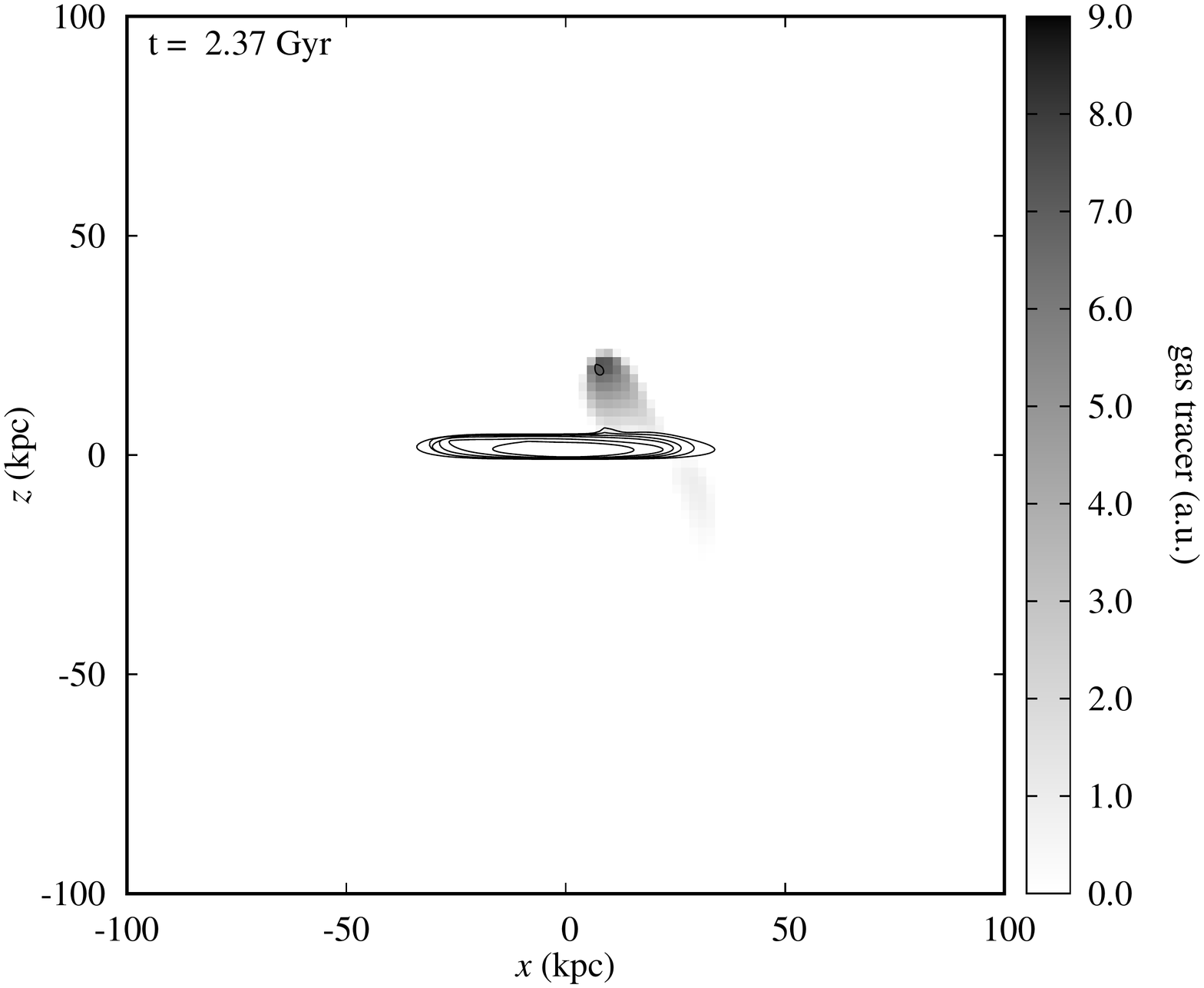}
\hfill
\includegraphics[width=0.33\textwidth]{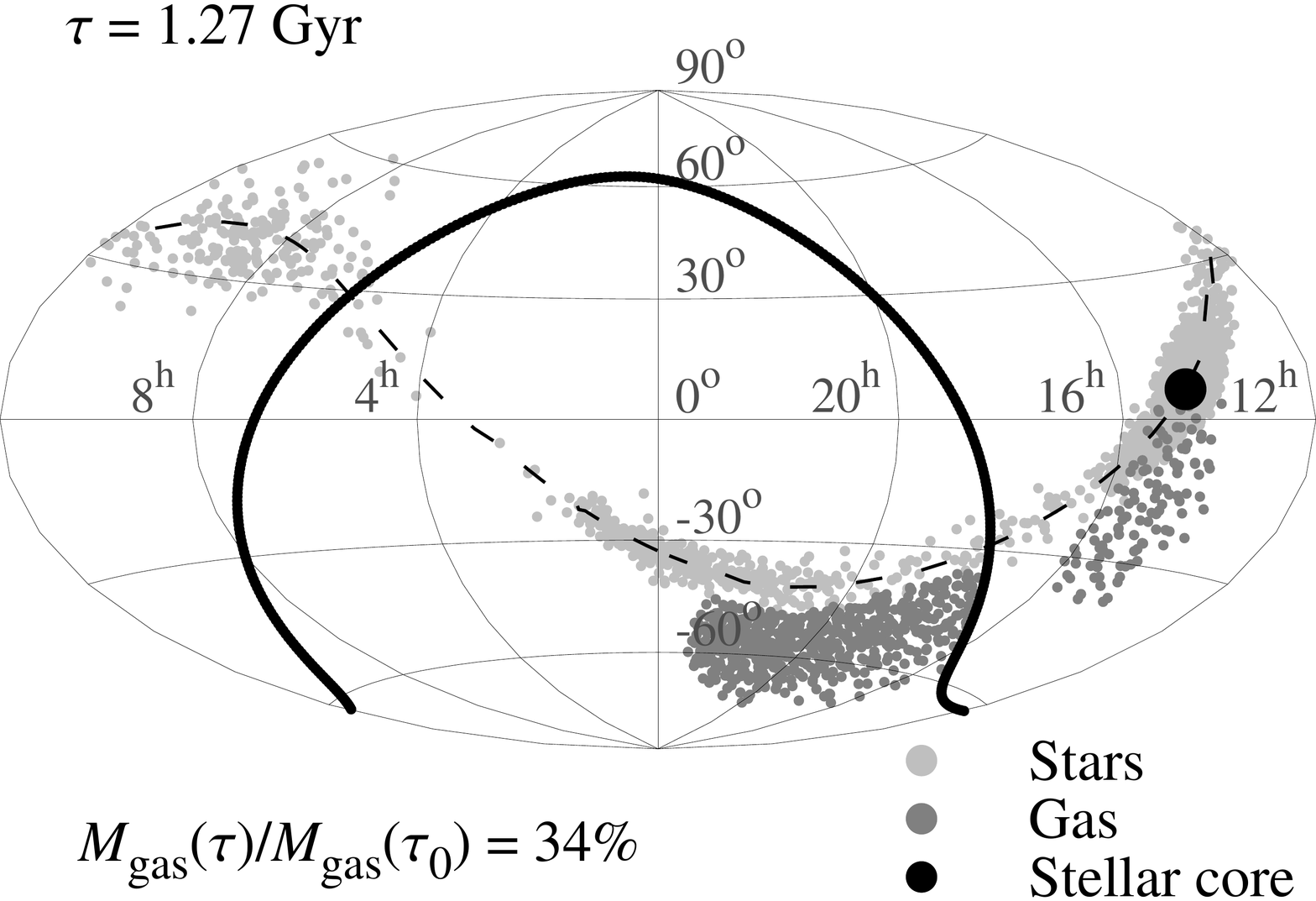}\\
\vspace{-5pt}
\includegraphics[width=0.33\textwidth]{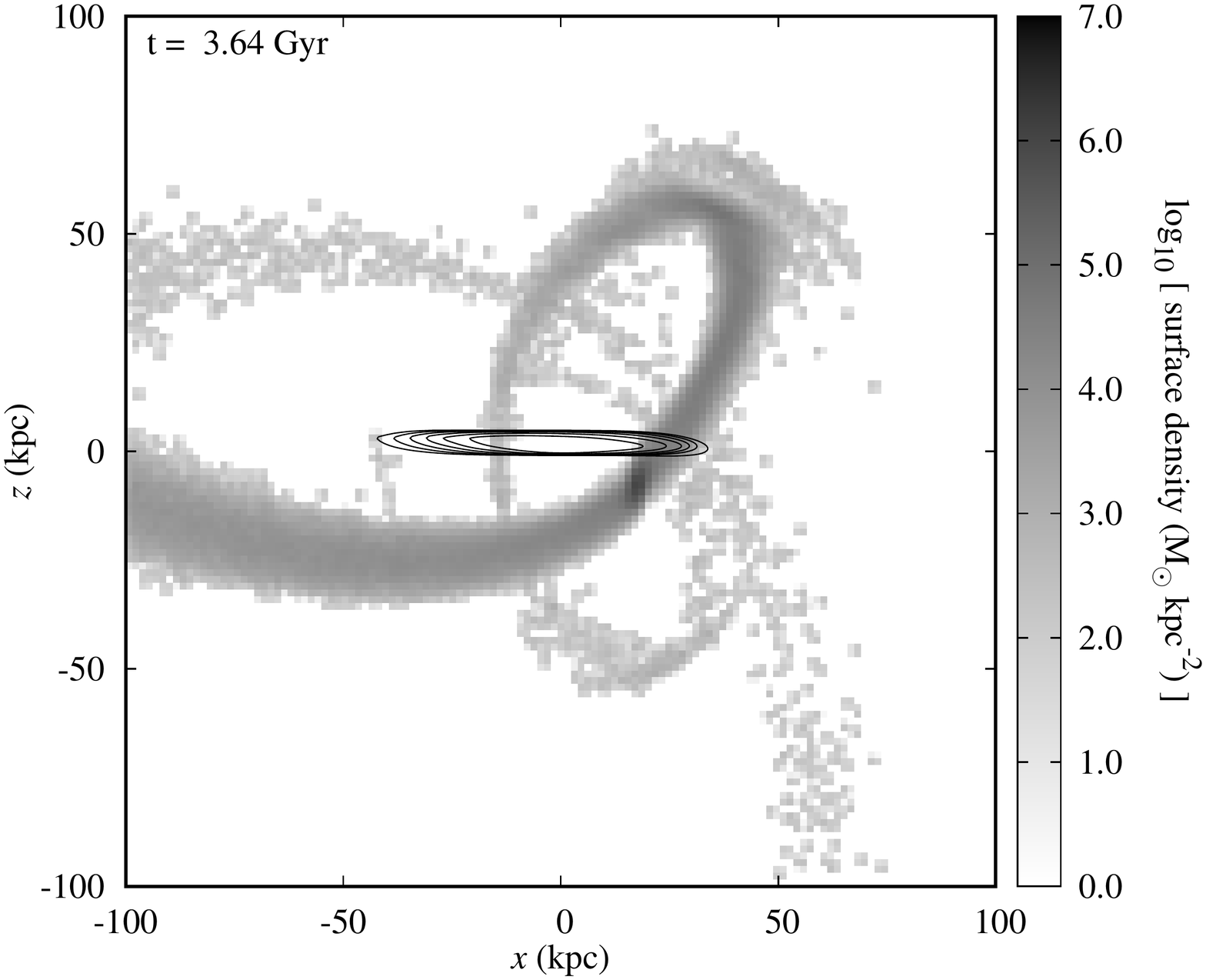}
\includegraphics[width=0.33\textwidth]{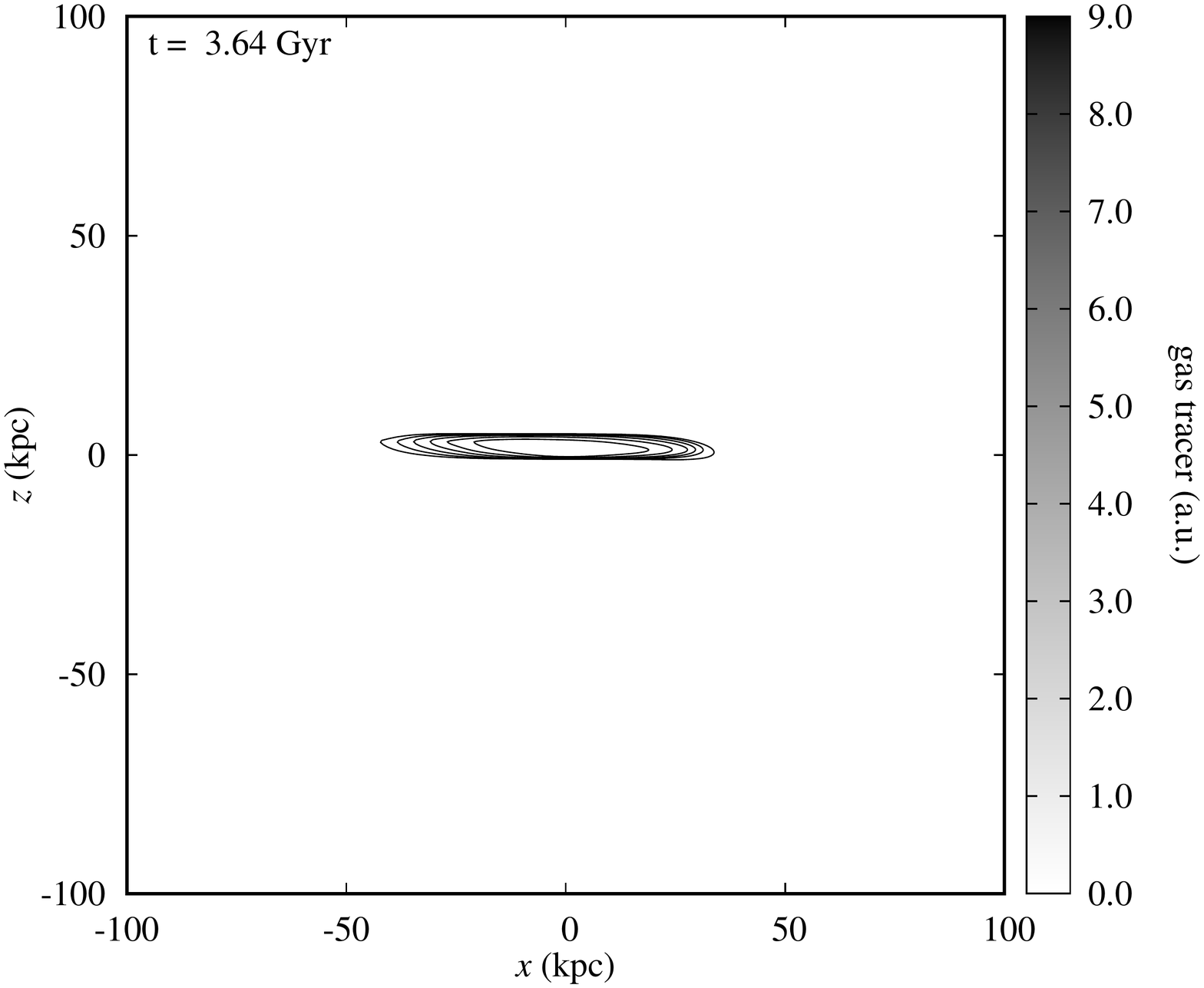}
\hfill
\includegraphics[width=0.33\textwidth]{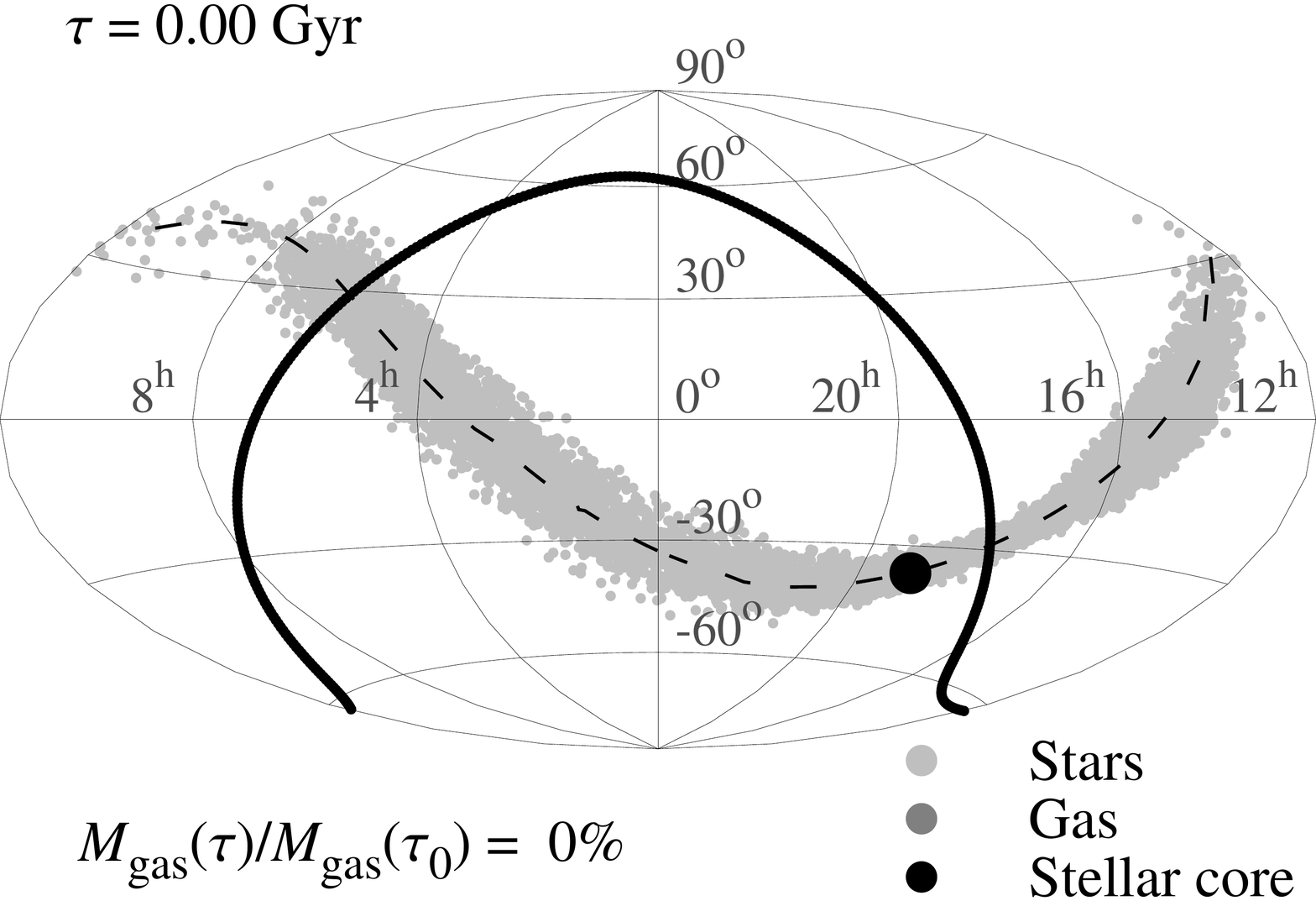}\\
\vspace{-5pt}
\caption[ Model C All-sky distribution ]{ Stellar (left) and gas (centre) components of Sgr in physical space on an edge-on view of the Galaxy. Each pair of rows from the top correspond to snapshots just before and after a disc transit. The last row corresponds to the present epoch. Contours indicate the total gas column density in the range $10^{21} - 10^{22}$ \psc. The Galactic gas disc can be seen as a thick horizontal slab. The panels on the right show the location of Sgr's stellar core, its associated tidal debris and its gas streams on an all-sky Hammer-Aitoff projection in equatorial coordinates. Mind the potential overlap between both components. The thick solid and dashed curves indicate the locus of the Galactic plane and the full orbit of the stellar core, respectively. The quantity $M_{\rm gas}(\tau) / M_{\rm gas}(\tau_0)$ corresponds to the mass of {\em detectable} gas relative to the total initial gas mass in Sgr. Time tags indicate the time $t$ since infall (left / centre) or the lookback time ($\tau$; right). Sgr moves in the direction of decreasing $\alpha$. }
\label{fig:allskyC}
\end{figure*}

\begin{figure*}
\centering
\includegraphics[width=0.42\textwidth]{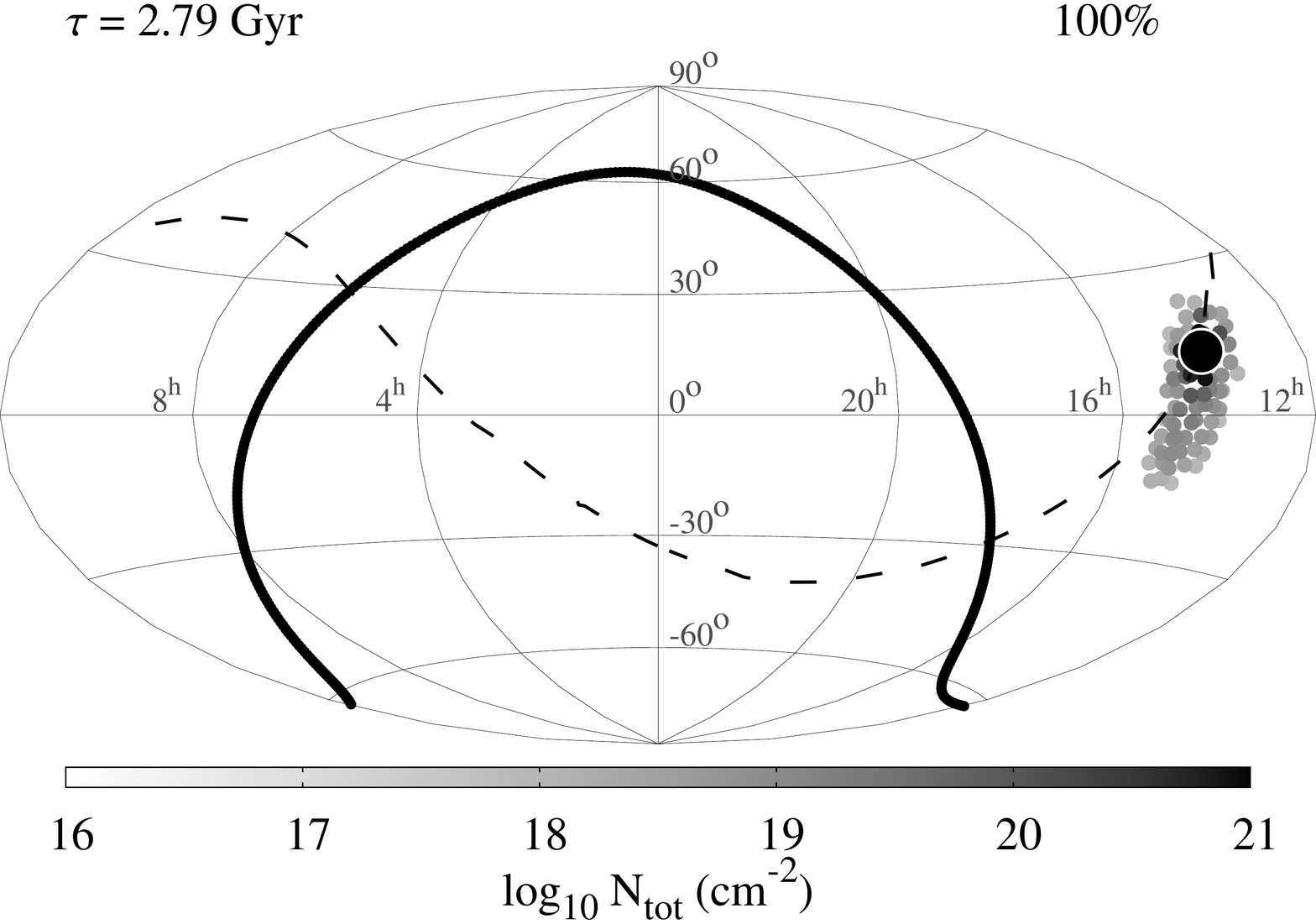}
\hspace{20pt}
\includegraphics[width=0.42\textwidth]{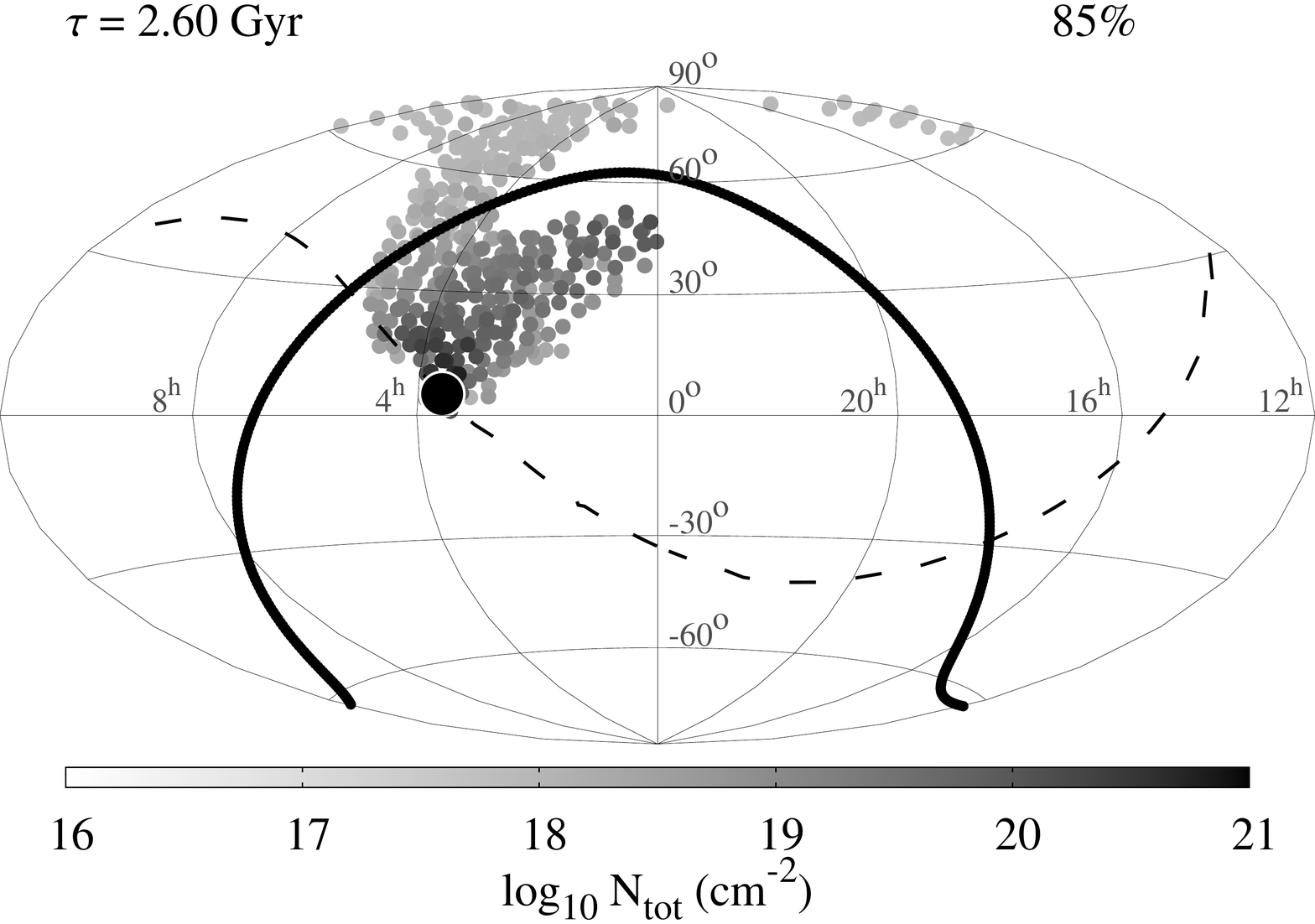}\\
\includegraphics[width=0.42\textwidth]{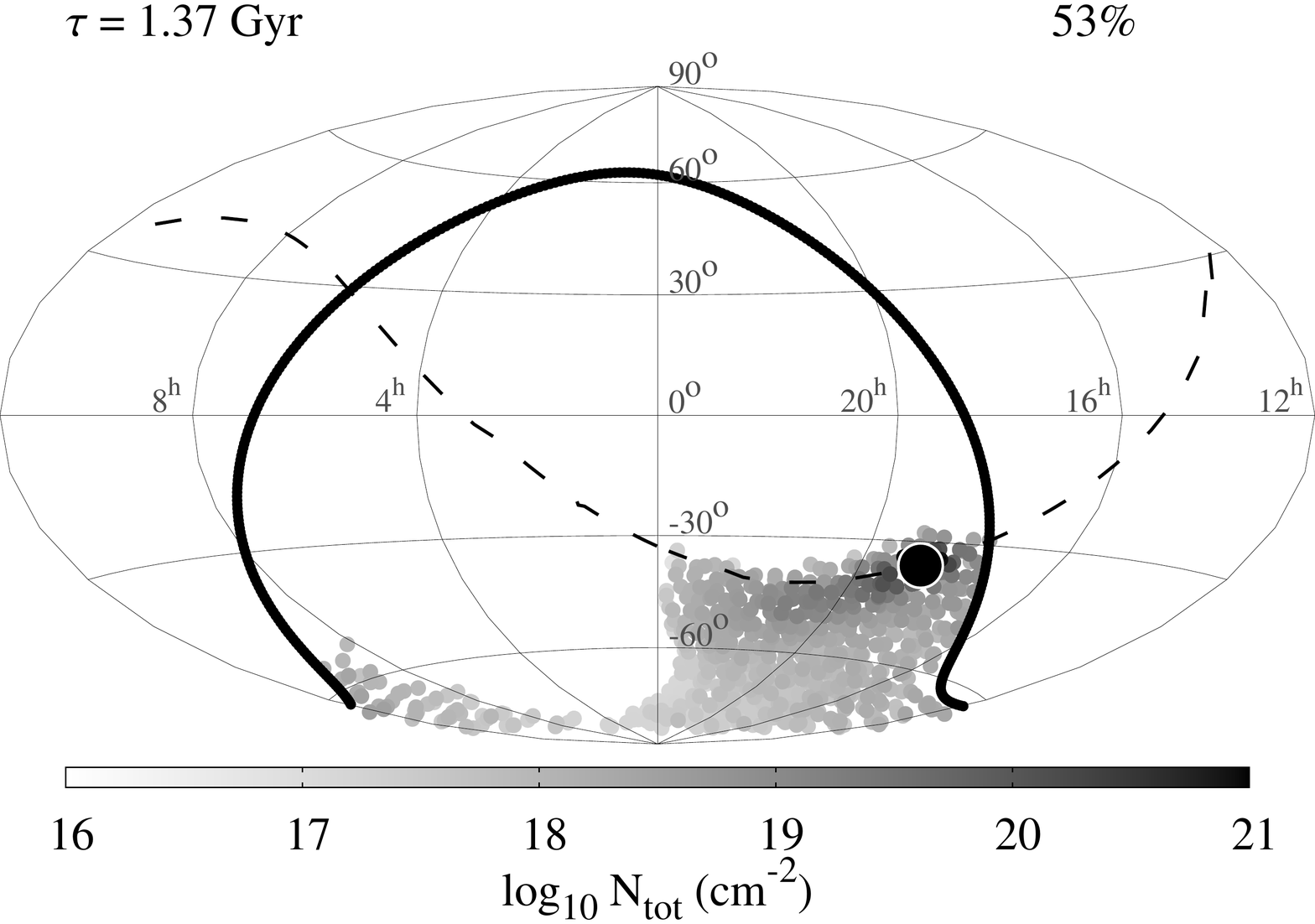}
\hspace{20pt}
\includegraphics[width=0.42\textwidth]{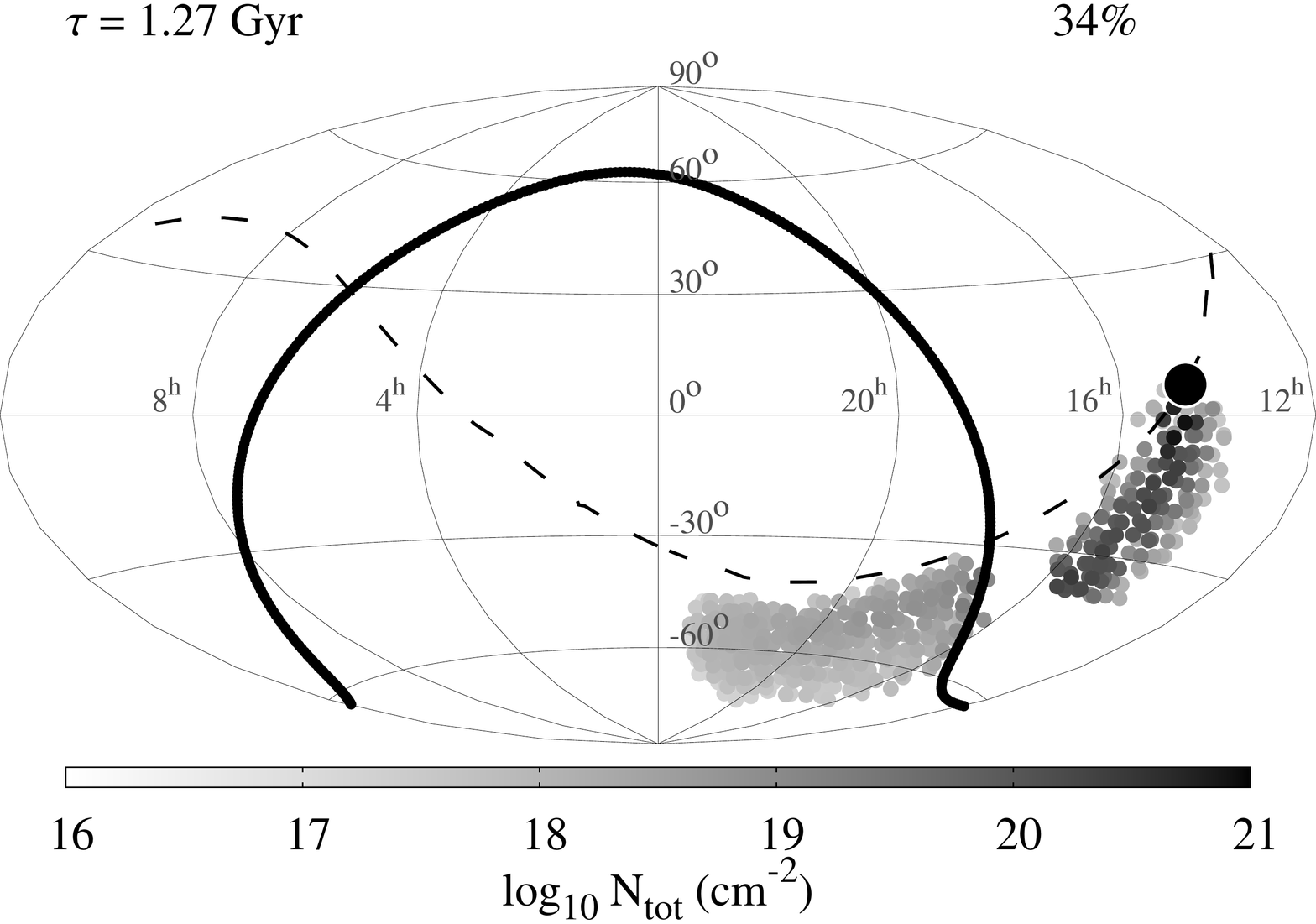}
\caption[ Model C All-sky distribution with gas total col.dens.]{ Total column density of the Sagittarius gas stream just before (left) and after (right) the first (top) and second (bottom) disc transit (i.e. corresponding to the first four rows of Fig.~\ref{fig:allskyC}). The percentage on the top right corner of each panel indicates the mass fraction of the detectable gas relative to the initial gas content of the precursor. Note that the fraction is in each case slightly lower than that given in the corresponding panel of the right column in Fig.~\ref{fig:allskyC}, as a result of the binning of the data projected onto observed space (see text for details). Line and symbol meanings as in the right panels of Fig.~\ref{fig:allskyC}. } 
\label{fig:allskyC2}
\end{figure*}

Based on the previous results, it is reasonable to assume that running either of our models at an even higher resolution would increase the fraction of gas retained within Sgr. This implies that if our model is to agree with the observed lack of gas in Sgr at the present epoch, then its gas mass fraction at infall cannot have been higher than what we have assumed here.\footnote{This argument relies on the inefficiency of stellar feedback -- as we assume here -- to assist gas removal from the dwarf (see Sec.~\ref{sec:mod}).} Thus our reference model sets an upper bound on the gas {\em within} Sgr at any time after infall -- if run at an arbitrarily high resolution. In the limit of low resolution, our model puts a stringent upper limit on the fraction of gas {\em stripped} away from Sgr. In other words, our reference model at our standard resolution {\em overestimates} the presence of gas beyond Sgr's core, and is thus adequate to estimate the maximum possible mass in Sgr's gas streams, everywhere around the Galaxy and at any epoch.

Taken at face value, our models indicate that within the last $\sim 1$ Gyr of evolution, Sgr retained between 0.1 and 10 percent of its initial gas mass, which presumably allowed for a last burst of stars to be formed. Conversely, this implies that at least 90 per cent of the initial gas mass within the precursor was stripped 
from Sgr over a timeline from infall to the present epoch.

\begin{figure*}
\centering
\includegraphics[width=0.33\textwidth]{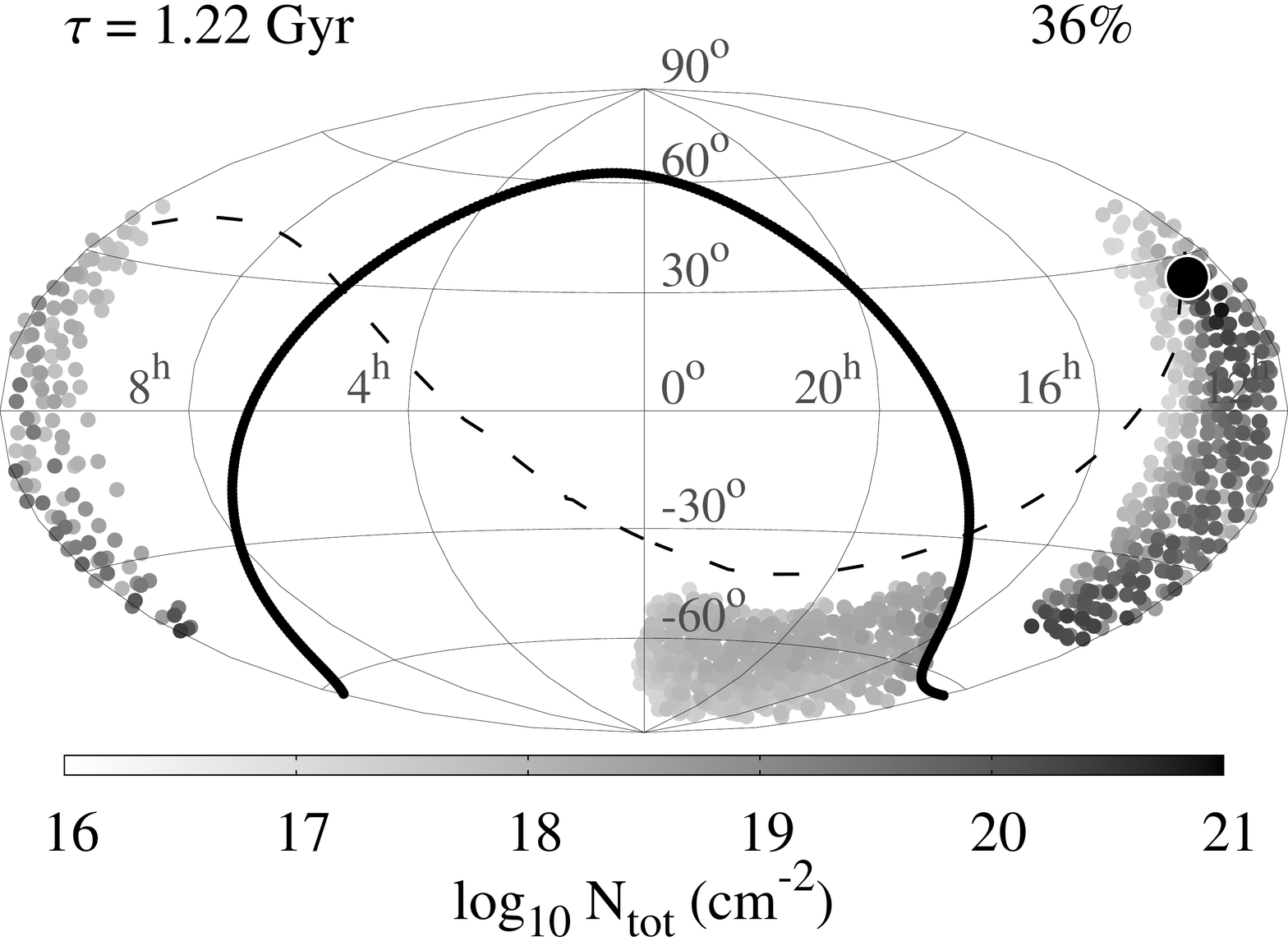}
\hfill
\includegraphics[width=0.33\textwidth]{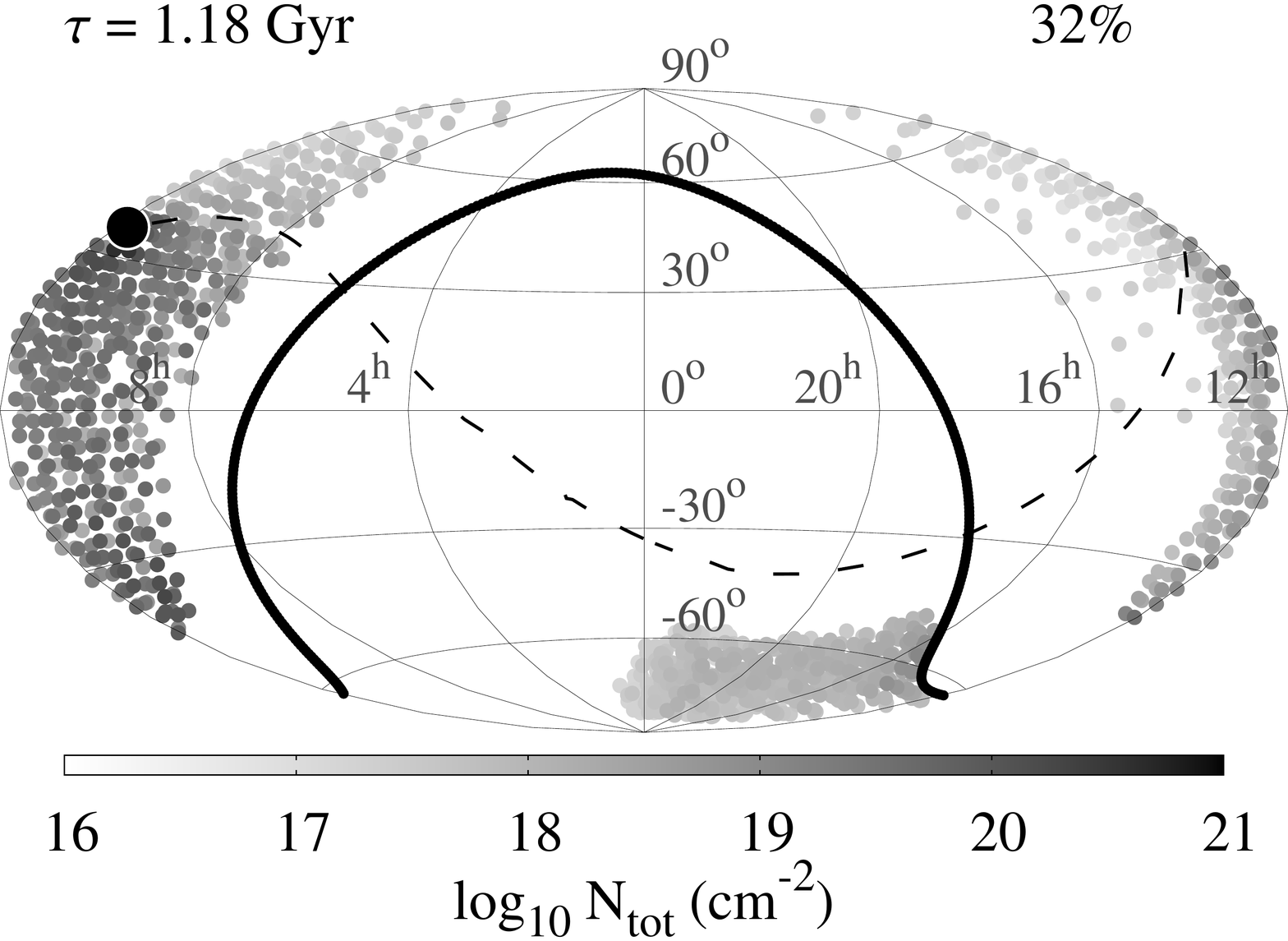}
\hfill
\includegraphics[width=0.33\textwidth]{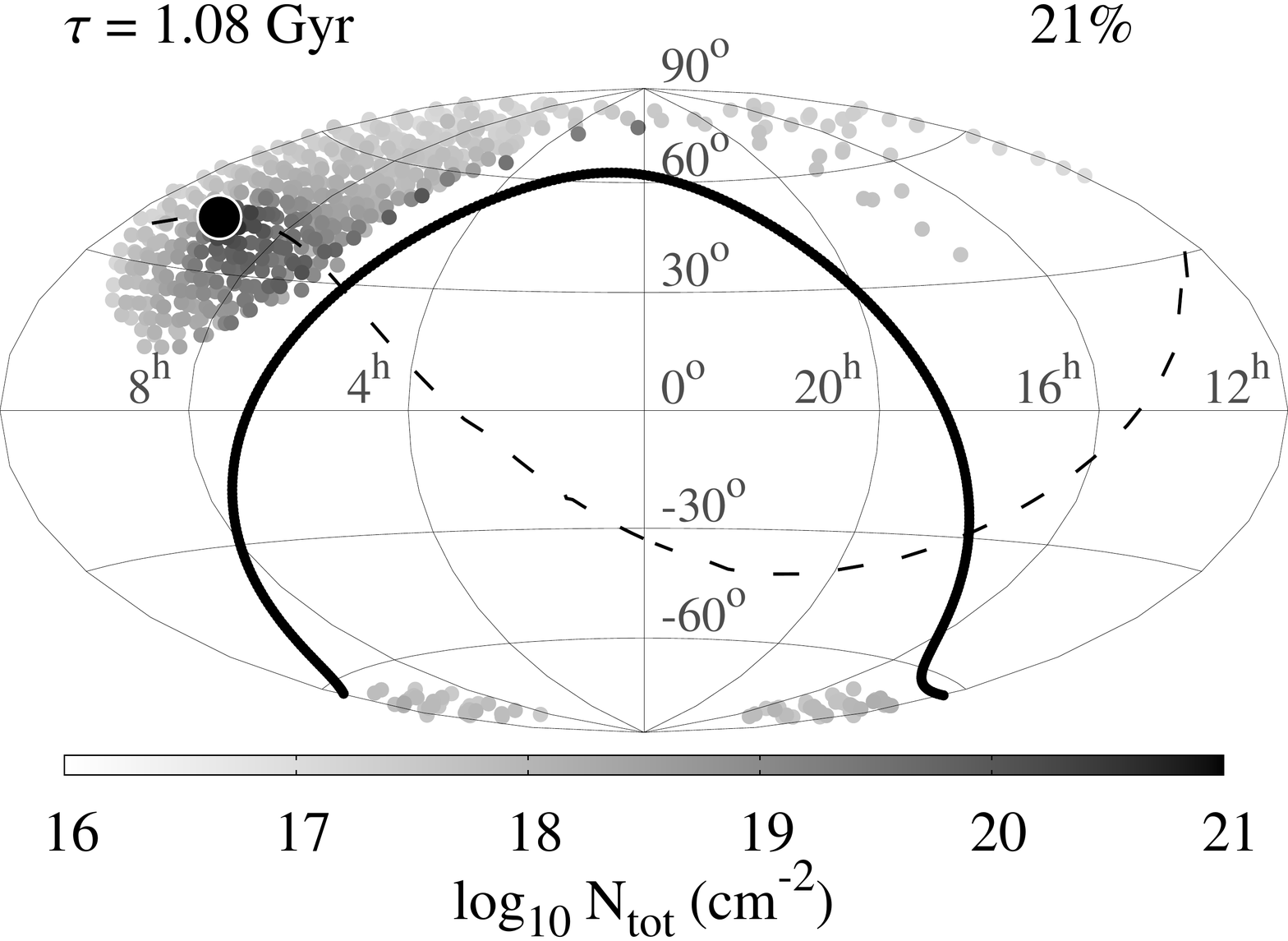}
\hfill
\includegraphics[width=0.33\textwidth]{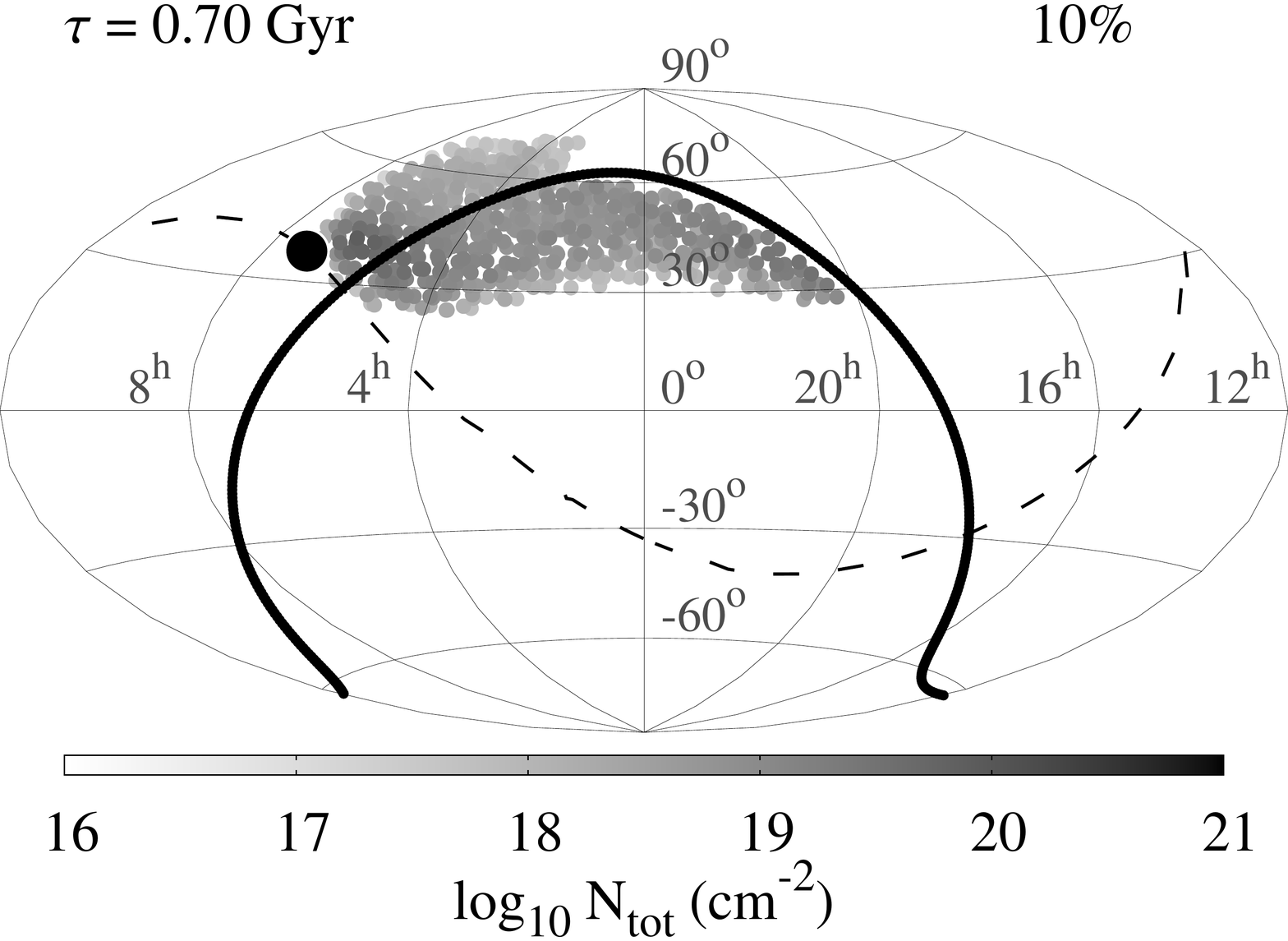}
\hfill
\includegraphics[width=0.33\textwidth]{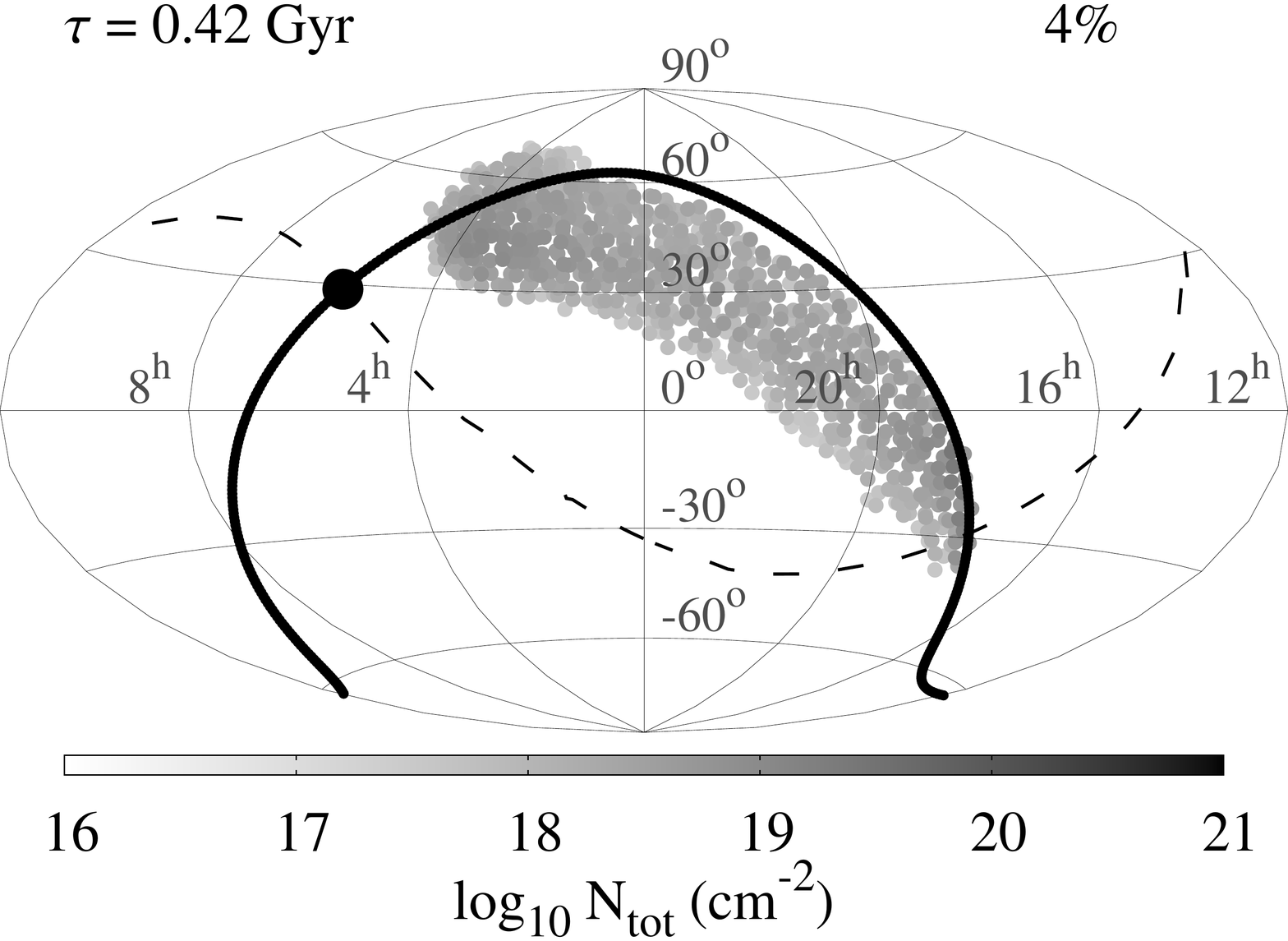}
\hfill
\includegraphics[width=0.33\textwidth]{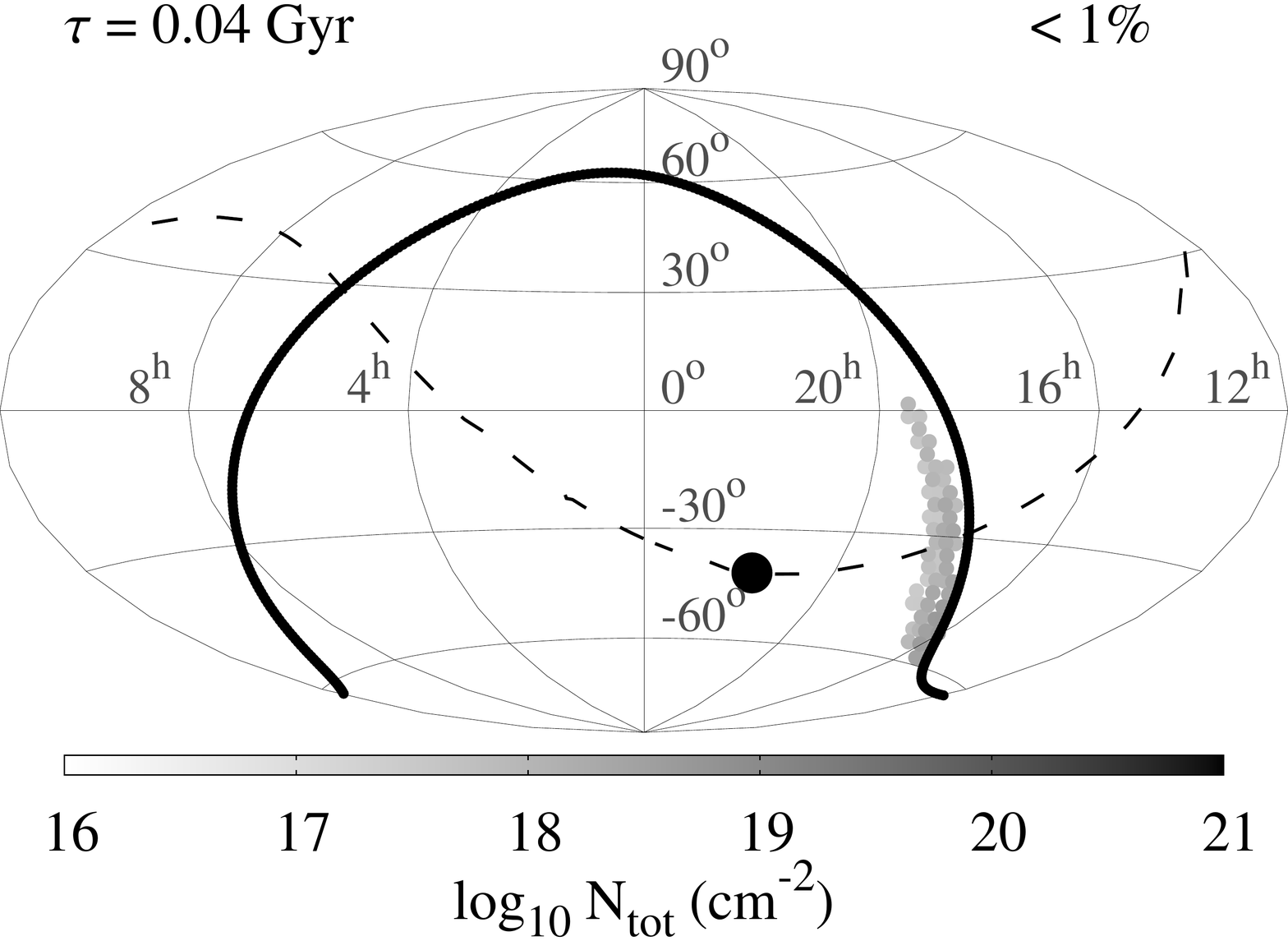}
\caption[ Model C All-sky distribution with gas total col.dens.]{ The Sagittarius gas stream during the last billion years of the dwarf's evolution. The detectability of gas in this case is defined by a threshold value of 1 (rather than 2) on the gas tracer. The stellar component has been omitted here for clarity; only the orbit and the instantaneous position of the stellar core are indicated (dashed curve, filled black circle). Sgr moves in the direction of decreasing $\alpha$. Note that the snapshots are {\em not} equally spaced in time. Clearly, the stripped gas {\em does not} follow Sgr's orbit. After it detaches from the dwarf's main body, the gas slowly settles onto the Galactic plane (thick solid curve), and disappears from view over a time scale of about 0.5 Gyr as it mixes with the gas in the disc (and its gas tracer value consequently drops below 1). See also Sec.~\ref{sec:gacc2}, and Figs.~\ref{fig:allskyC3_vel} - \ref{fig:allskyC3_dist}.}
\label{fig:allskyC3_snpart}
\end{figure*}

\subsubsection{Accretion onto the Galaxy} \label{sec:gacc}

But what about the fate of stripped gas? Has it been accreted by the Galaxy? Is there some left in the form of HVCs or IVCs? If so, are the clouds spread throughout the halo? In order to answer these questions we now turn our attention to the distribution of gas associated to Sgr in relation to its stellar debris. As argued above, the reference model at standard resolution favours the stripping of gas away from Sgr.  Therefore, in what follows, we will focus on the discussion of the results of this model only.

The instantaneous configuration of the composite system just before and after each of the first disc transits is shown in Fig.~\ref{fig:allskyC}. The last row shows its present-day state. These particular epochs are chosen based on our previous discussion on the importance of disc transits on the overall evolution of Sgr's gas content. The left (central) panels in the figure show the stellar (gas) distribution on an edge-on view projection of the Galaxy. The Galactic gas disc at total gas column densities in the range $10^{21} - 10^{22}$ \psc\ can be seen as a thick horizontal slab. To allow for a direct comparison to observations, we display the distribution of stars (grey dots) and gas (dark-grey dots) in an all-sky Hammer-Aitoff projection \citep[][]{ait89a,ham92a} in equatorial (i.e. in the International Celestial Reference System or ICRS) coordinates in the right panels. For reference, we include there the locus corresponding to the Galactic plane (thick solid curve), the instantaneous position of Sgr's stellar core (black filled circle), and its full orbit (dashed curve). The total mass in the gas stream\footnote{ {\em Above} our adopted detection limit; see Sec.~\ref{sec:gloss}. } relative to the total initial gas mass of Sgr is indicated in the bottom left corner of each panel on the right column. The discrete nature of the gas stream is due to the (arbitrary) pixelation of the simulation volume, and the consequent discreteness of the spatial and sky coordinates.

The total gas column density at the four first epochs (i.e. before and after each of the two disc transits) is displayed in Fig.~\ref{fig:allskyC2}. It is estimated as follows. The simulation data is projected onto the plane of the sky, and binned adopting a beam with (rather coarse) angular resolution\footnote{This value roughly corresponds to the angular size of a cell of size 500 kpc / 512 (the resolution we adopt to map the AMR output onto a regular Cartesian grid for the analysis) at an average distance of 20 kpc.} of 3\deg. The column density in each beam pointing is then calculated by integrating the column density of individual volume elements (or voxels) along the line of sight. It is worth noting that the result is not dramatically changed if we adopt a different (e.g. finer) angular resolution. The gas tracer at each beam pointing is taken to be the median of the overlapping voxels. This naturally leads to a sightly smaller gas mass,\footnote{Compare for instance mass fractions of gas indicated in the top right corner of each panel in Fig.~\ref{fig:allskyC2} and the corresponding panel in Fig.~\ref{fig:allskyC}. } as some beam pointings will drop below the adopted threshold of 2.

\begin{figure*}
\centering
\includegraphics[width=0.33\textwidth]{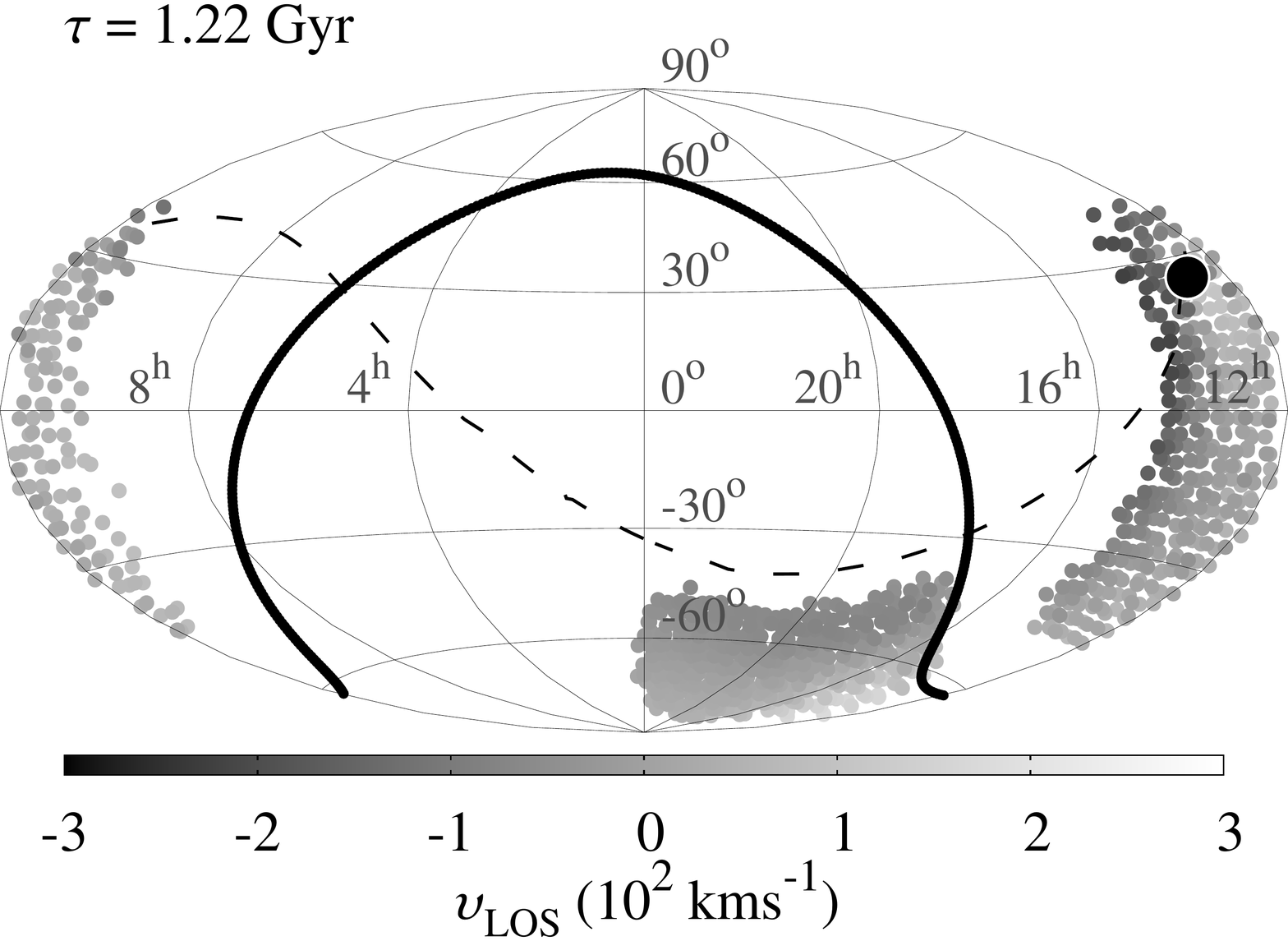}
\hfill
\includegraphics[width=0.33\textwidth]{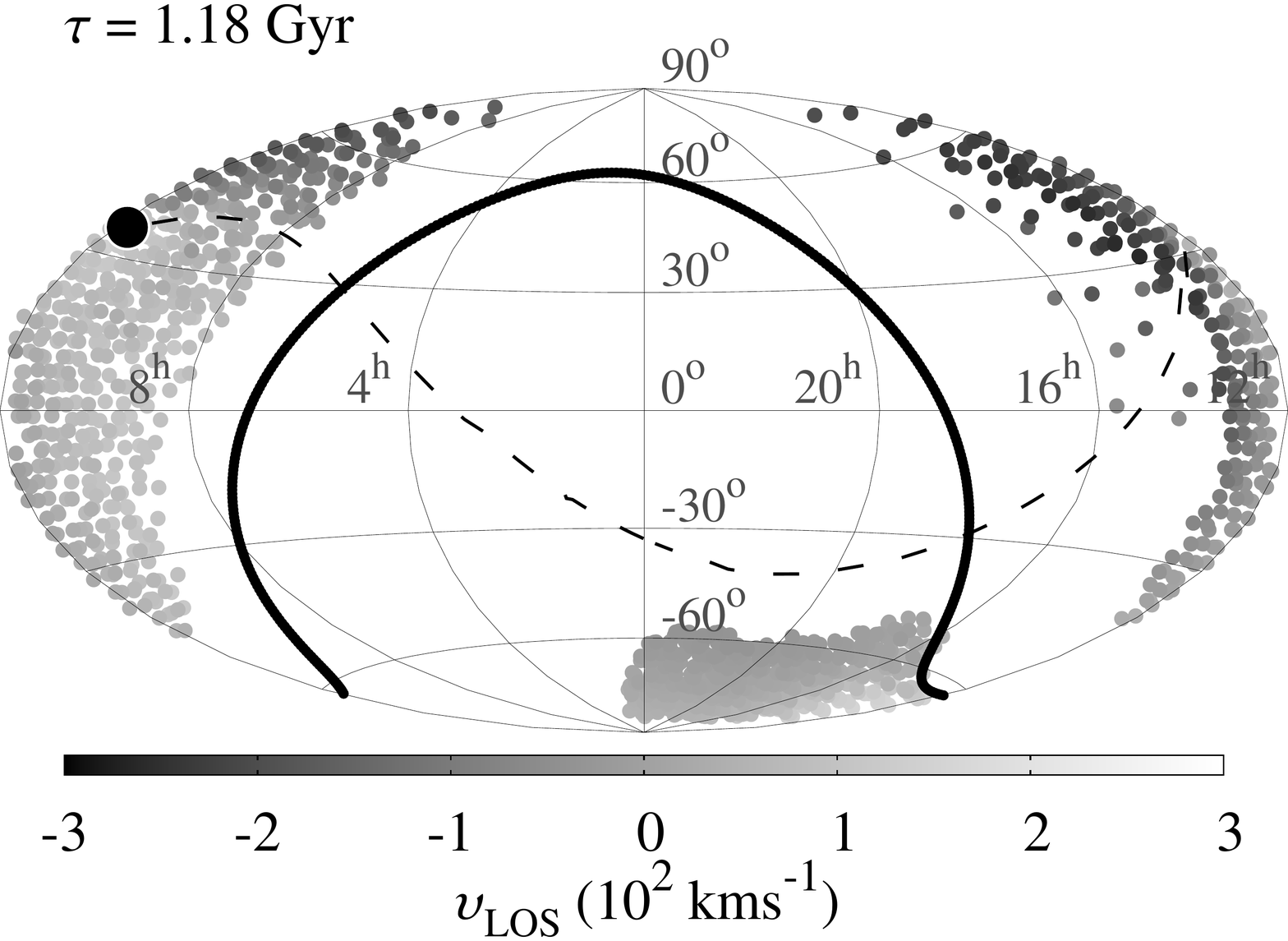}
\hfill
\includegraphics[width=0.33\textwidth]{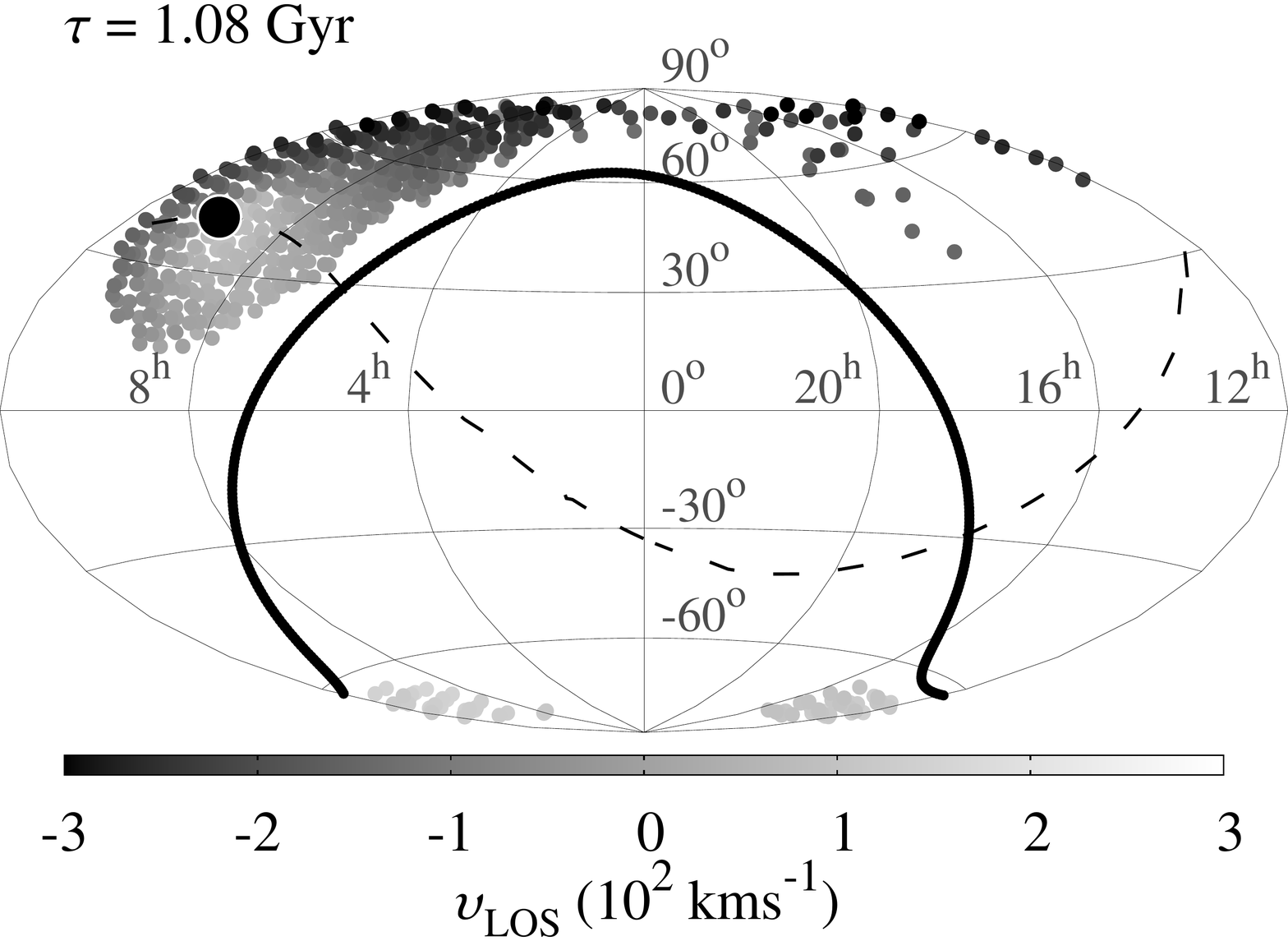}
\hfill
\includegraphics[width=0.33\textwidth]{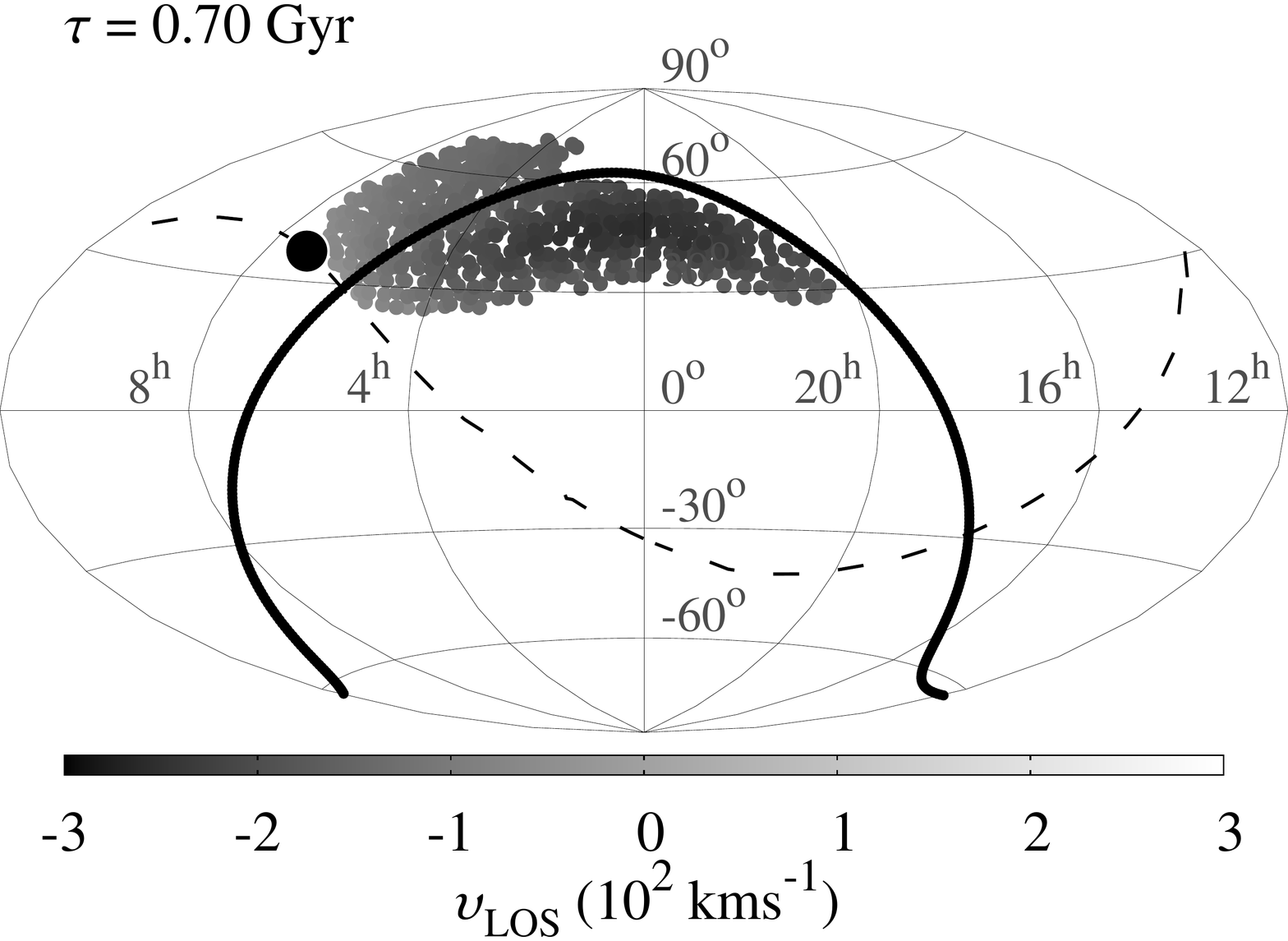}
\hfill
\includegraphics[width=0.33\textwidth]{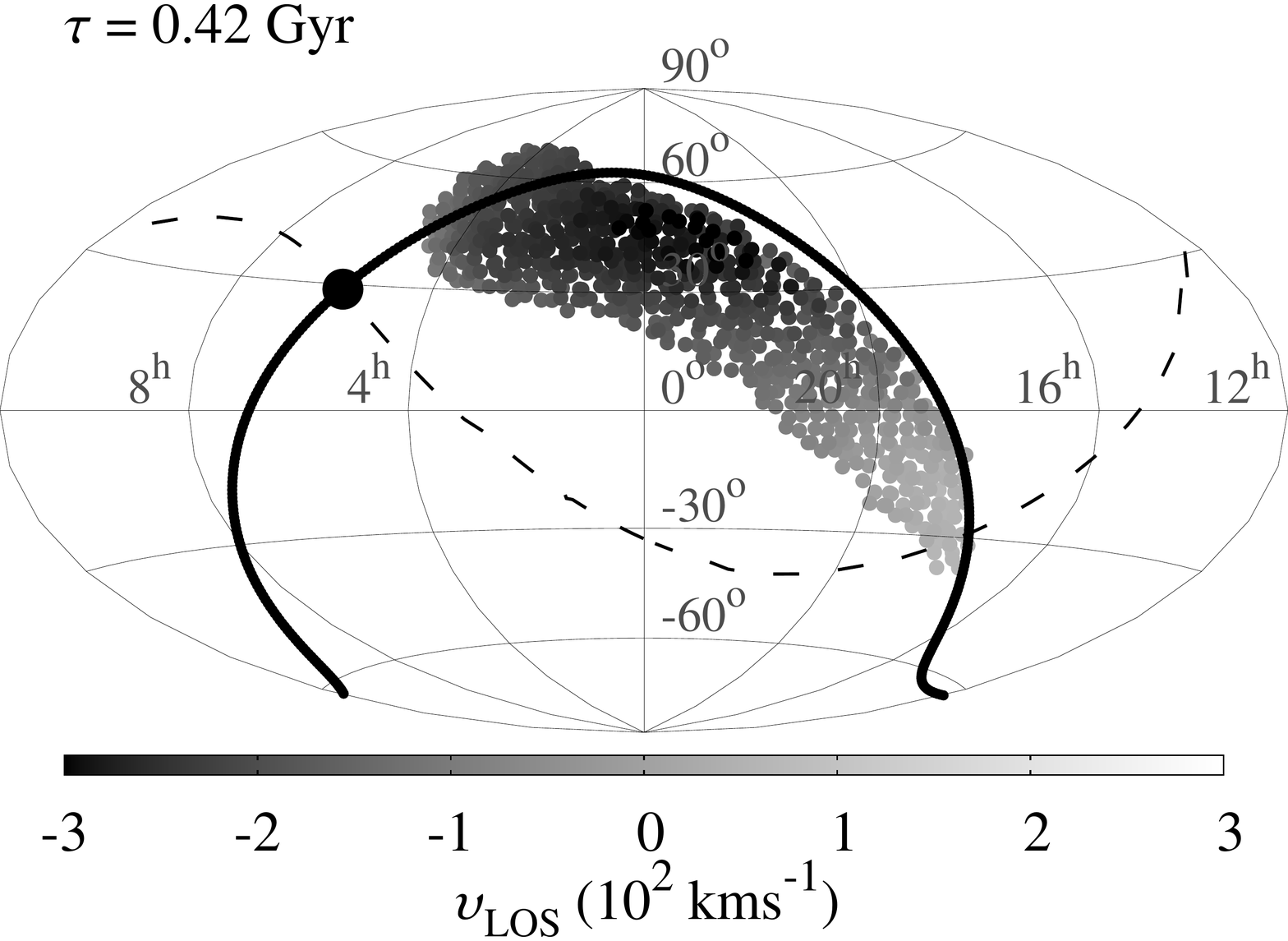}
\hfill
\includegraphics[width=0.33\textwidth]{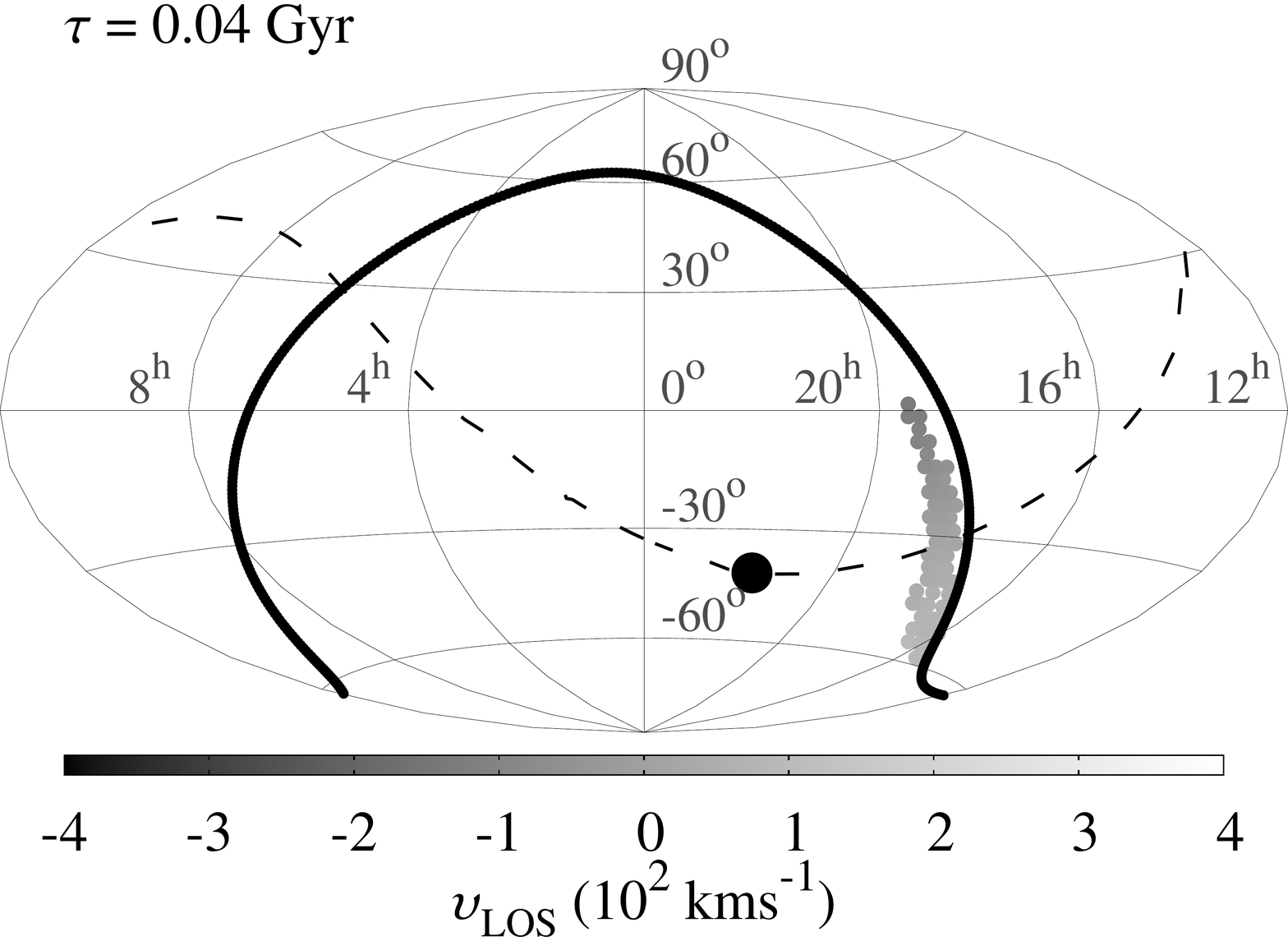}
\caption[ Model C All-sky distribution with gas velocity ]{ Line of sight velocity of the Sagittarius gas stream (measured in the LSR) during the last billion years of the dwarf's evolution. Most of the gas has negative velocities in this projection. See also Figs.~\ref{fig:allskyC3_snpart},\ref{fig:allskyC3_temp}, and \ref{fig:allskyC3_dist}. }
\label{fig:allskyC3_vel}
\end{figure*}

All previous results collectively suggest the following picture about the evolution of Sgr: As it approached the Galaxy, its stellar component remained tightly bound, at first. In contrast, its gas began to be stripped right away due to ram pressure exerted by the tenuous corona, creating a distinctive trailing stream. After the first transit at the outskirts of the Galactic disc some 2.5 Gyr ago, Sgr's body was stretched along its orbit due to tidal forces, and it lost roughly half of its gas content. Then the gas stream extended over several tens of degrees on the sky, nearly perpendicular to Sgr's orbit at that point. As the dwarf approached the disc for a second time, falling deeper into the potential of the Galaxy, more stars got tidally stripped, creating two elongated streams, one trailing and another leading the system. Similarly, a more pronounced trailing stream of gas formed. The second disc transit removed nearly all of its gas (Fig.~\ref{fig:gas}), leaving behind a compact stellar core followed by a dense gaseous stream (Fig.~\ref{fig:allskyC2}). The core orbited once more around the Galaxy, being further disrupted by the strong tidal interaction, and its associated gas slowly disappearing from view. Today, the core is on its way to colliding anew with the Galactic disc, and will perhaps be fully destroyed after the event.

\begin{figure*}
\centering
\includegraphics[width=0.33\textwidth]{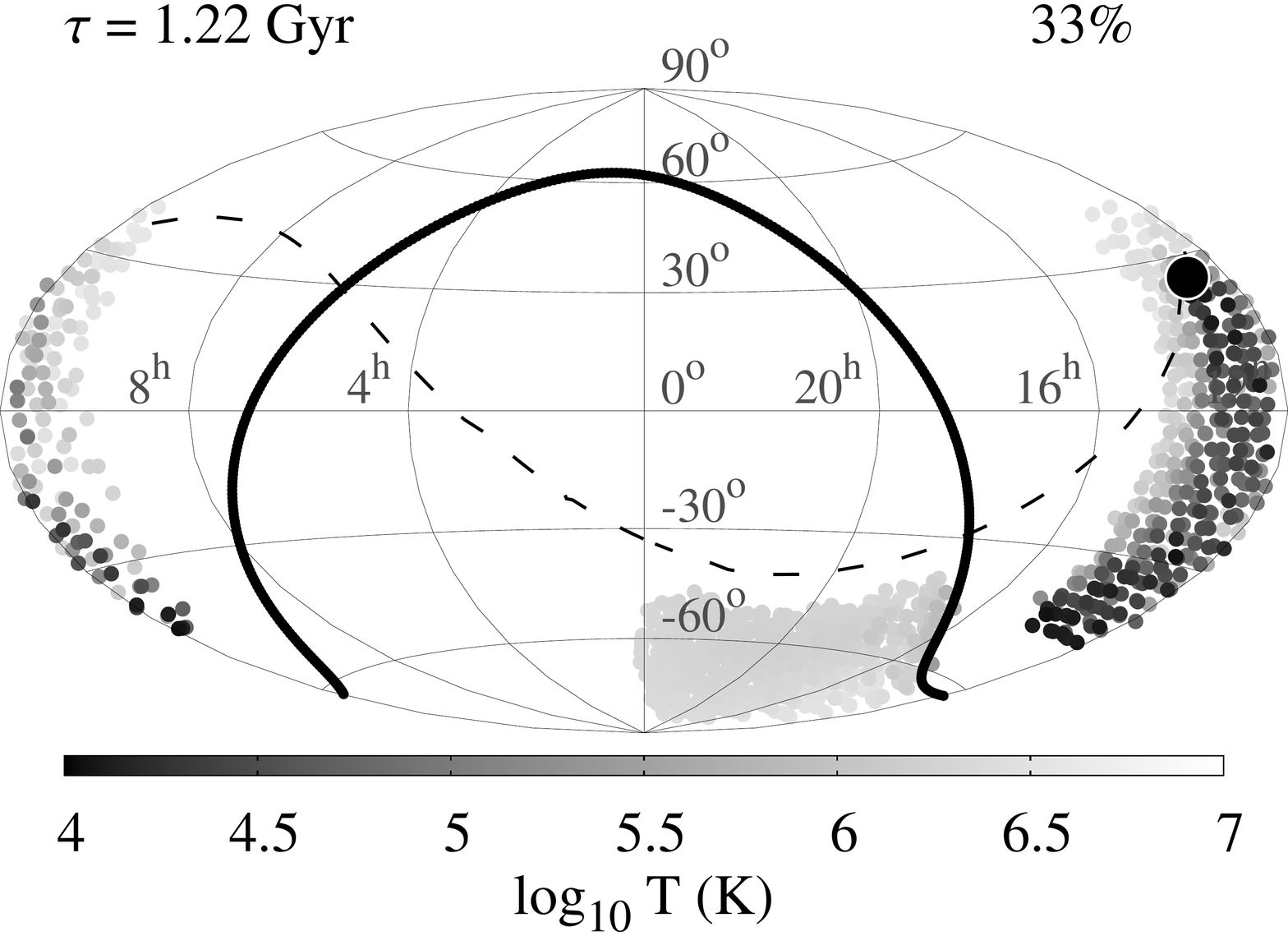}
\hfill
\includegraphics[width=0.33\textwidth]{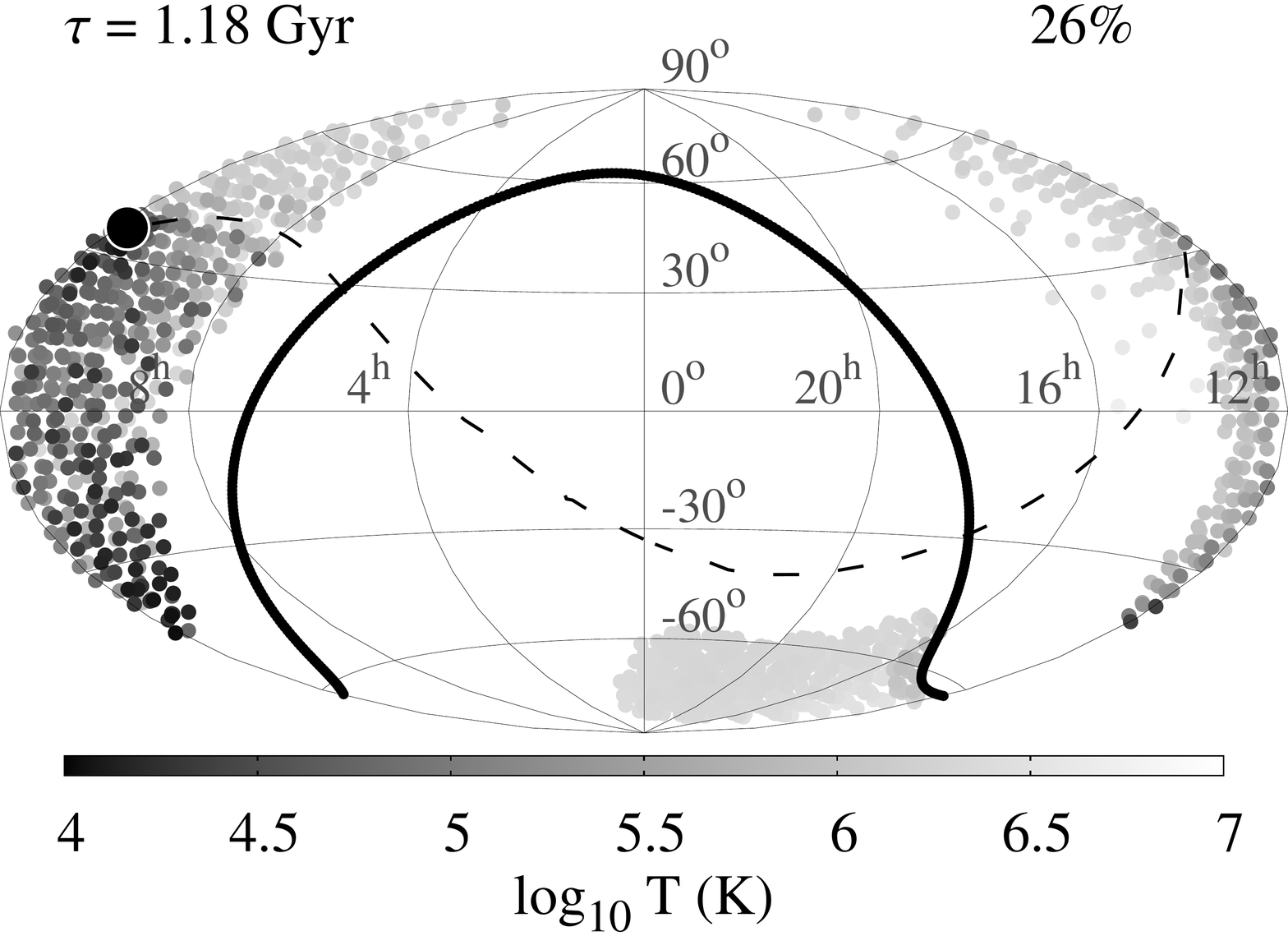}
\hfill
\includegraphics[width=0.33\textwidth]{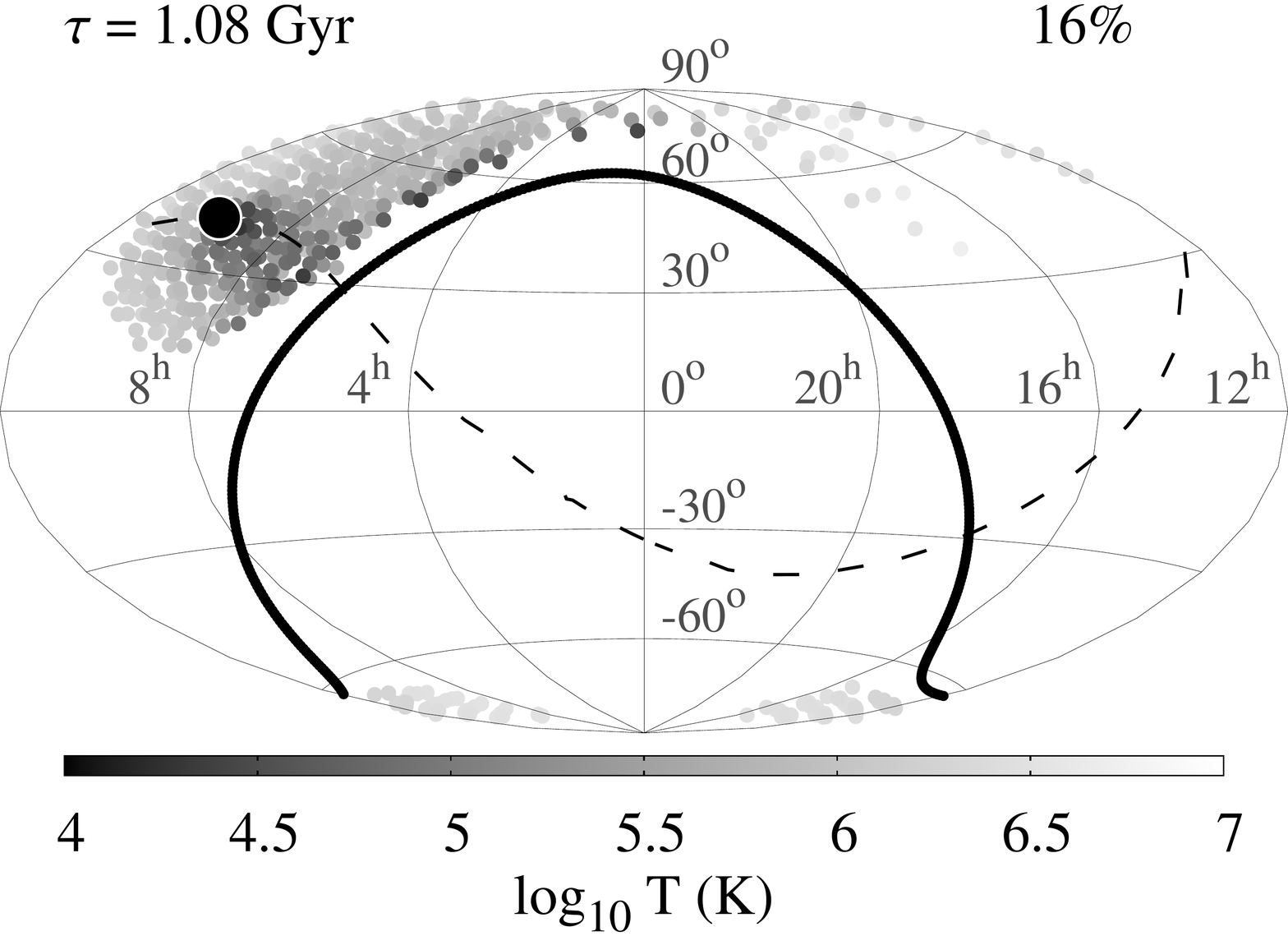}
\hfill
\includegraphics[width=0.33\textwidth]{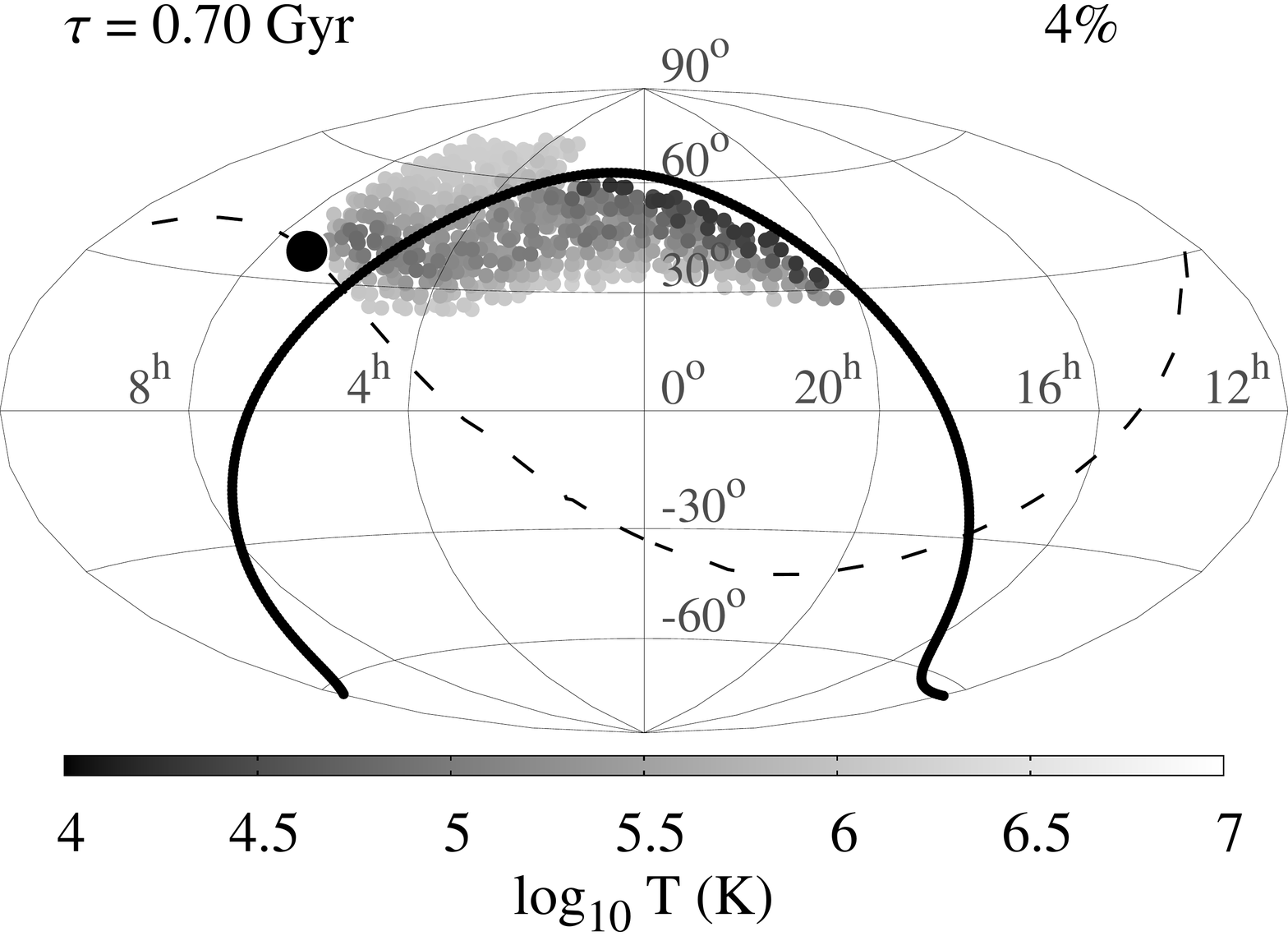}
\hfill
\includegraphics[width=0.33\textwidth]{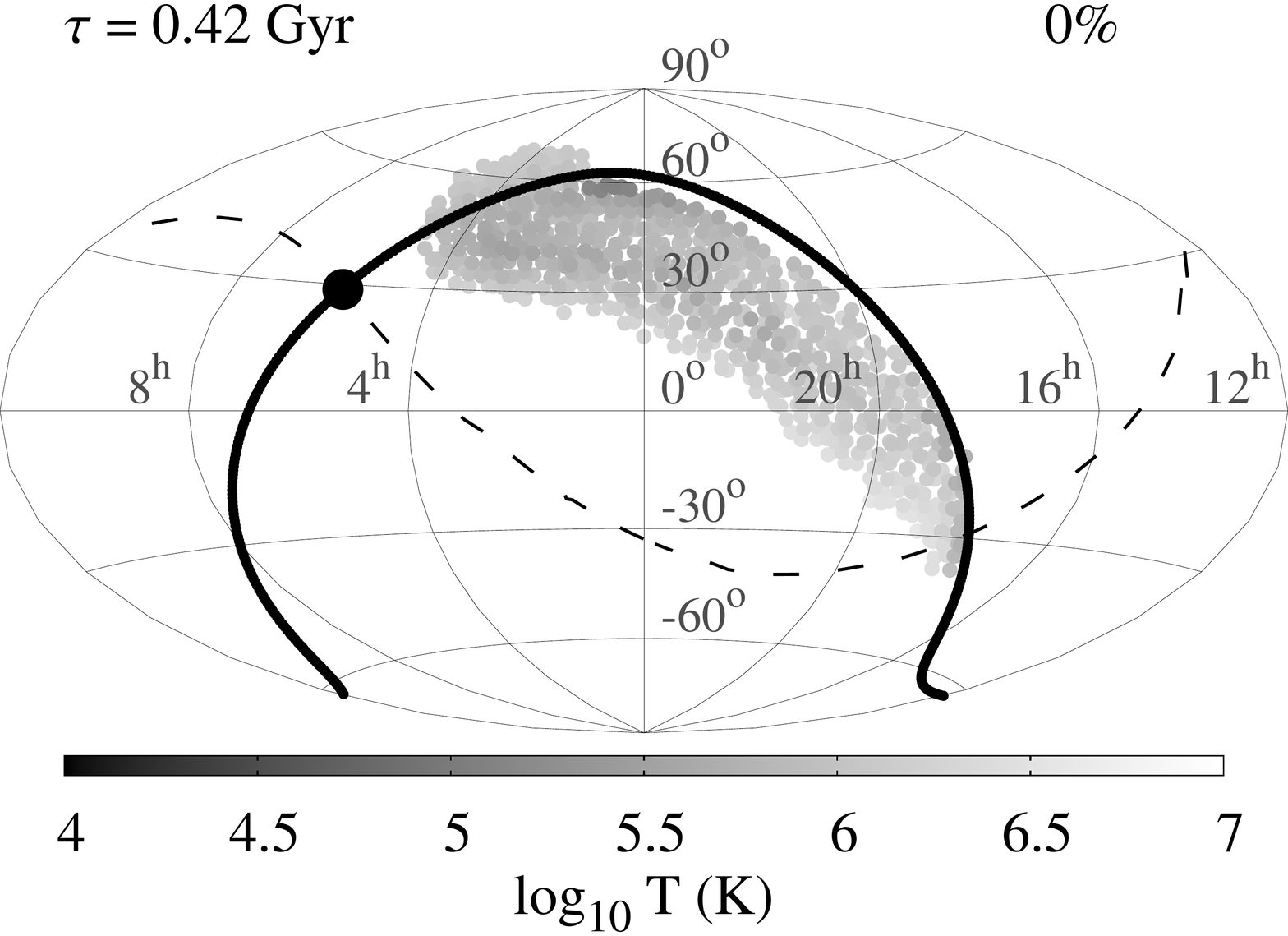}
\hfill
\includegraphics[width=0.33\textwidth]{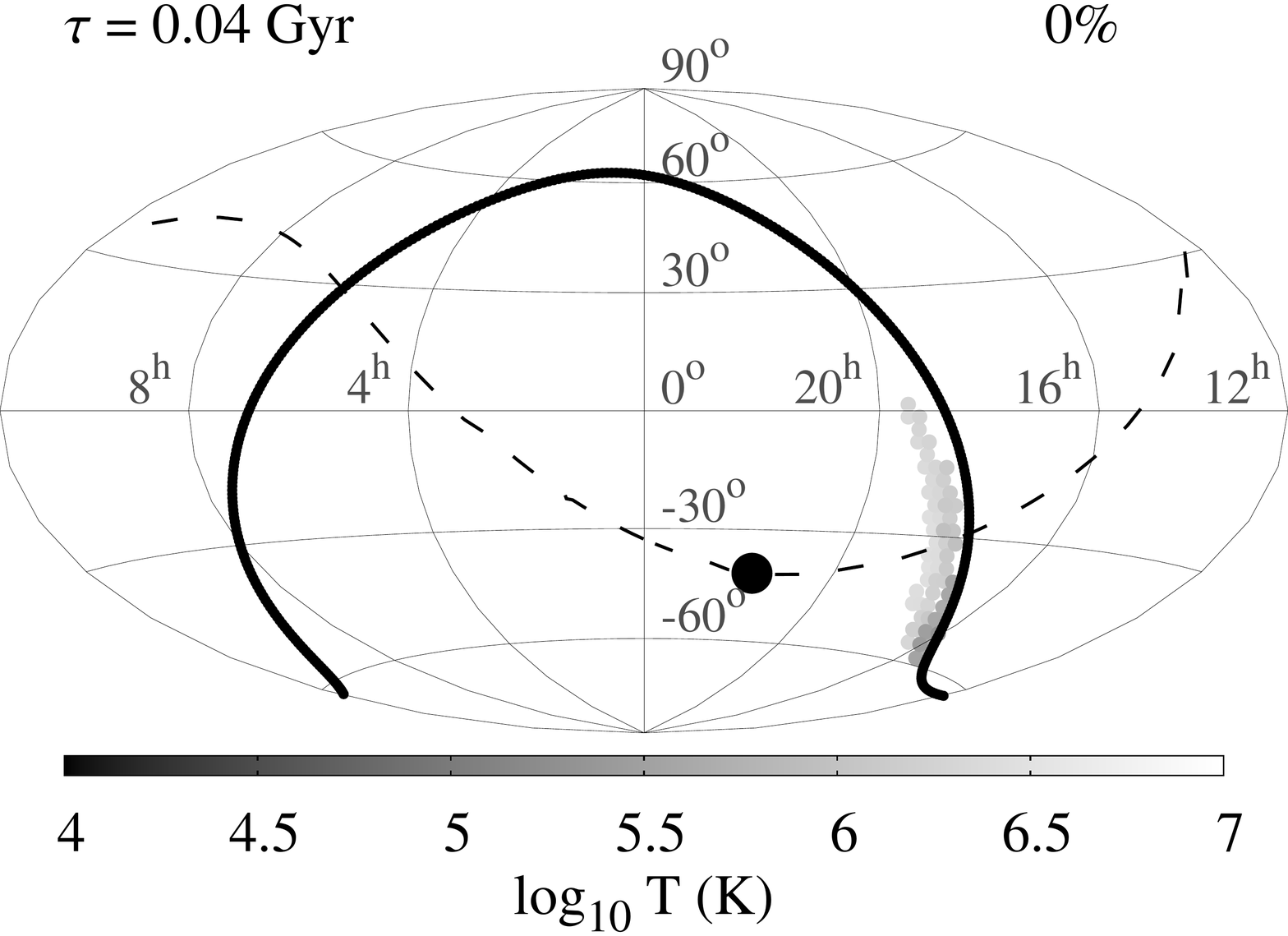}
\caption[ Model C All-sky distribution with gas temperature ]{ Temperature of the Sagittarius stream gas during the last billion years of
the dwarf's evolution. The percentage given in the top right corner of each panel indicates the mass fraction of gas with $T < 10^5$ K. Clearly, it decreases systematically with time, indicating that the stripped gas beyond the Galactic disc that is {\em detectable} mixes and heats up as a result of the interaction with the Galactic corona. See also Figs.~\ref{fig:allskyC3_snpart},\ref{fig:allskyC3_vel}, and \ref{fig:allskyC3_dist}.  }
\label{fig:allskyC3_temp}
\end{figure*}

\begin{figure*}
\centering
\includegraphics[width=0.33\textwidth]{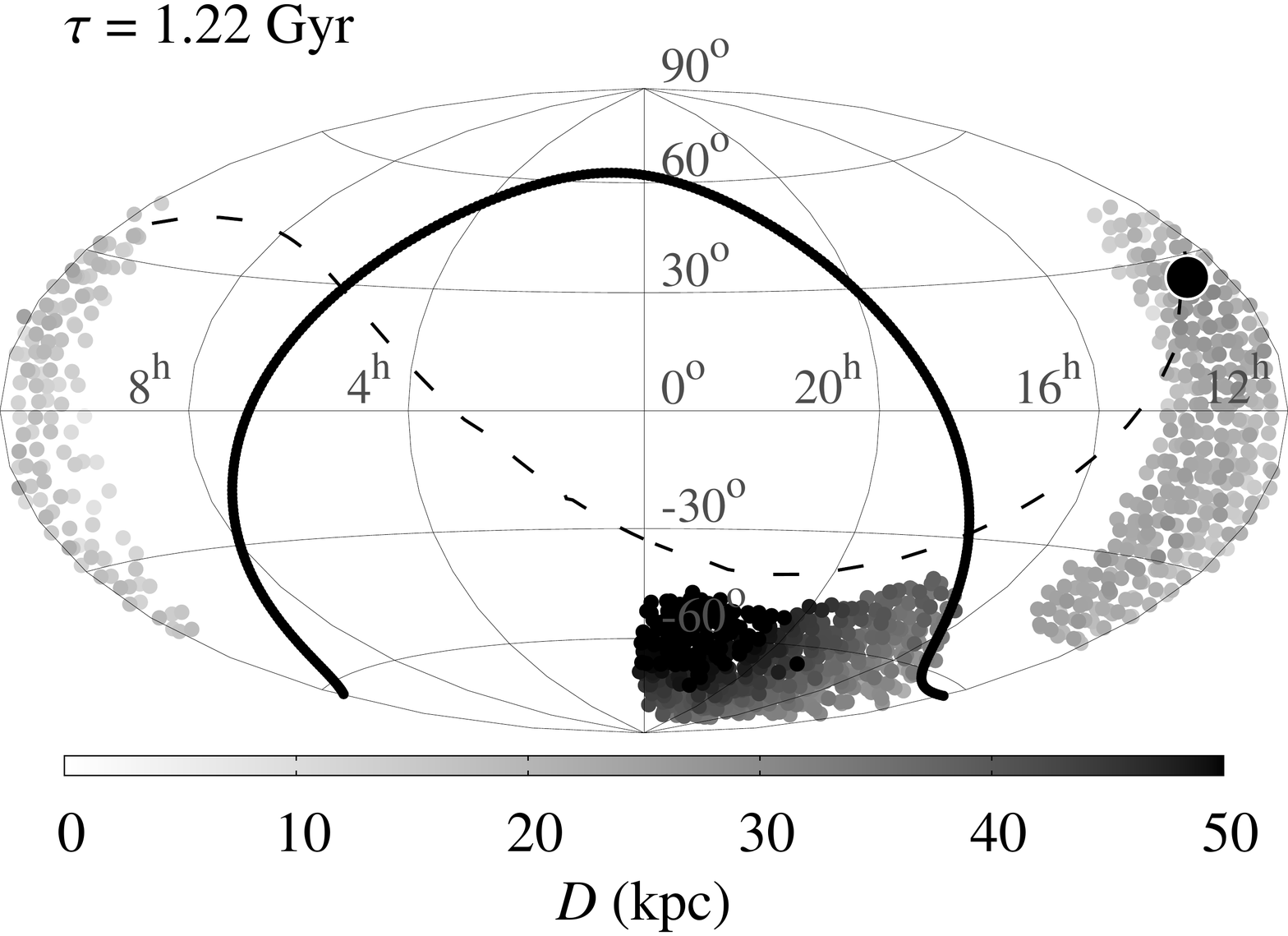}
\hfill
\includegraphics[width=0.33\textwidth]{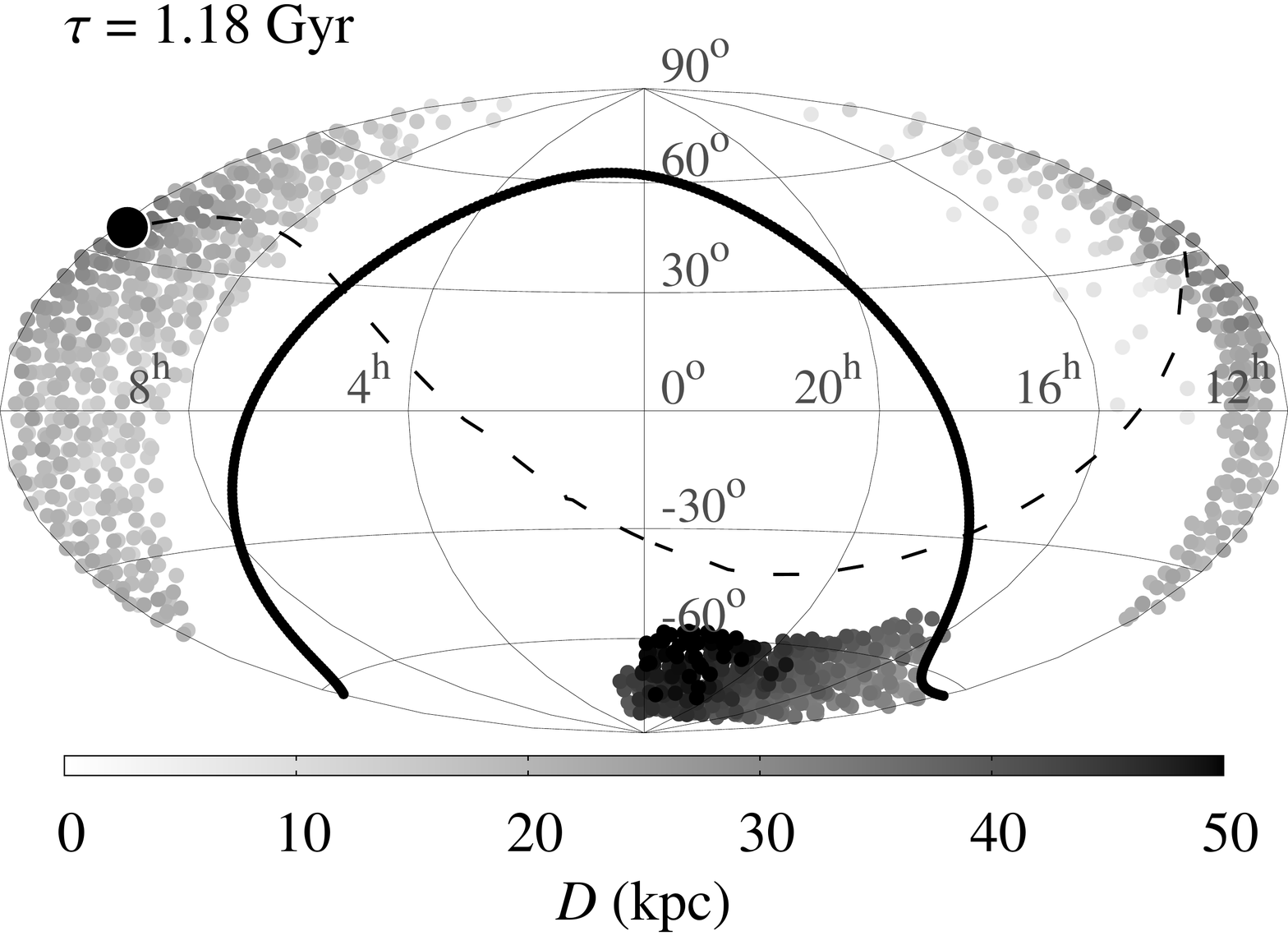}
\hfill
\includegraphics[width=0.33\textwidth]{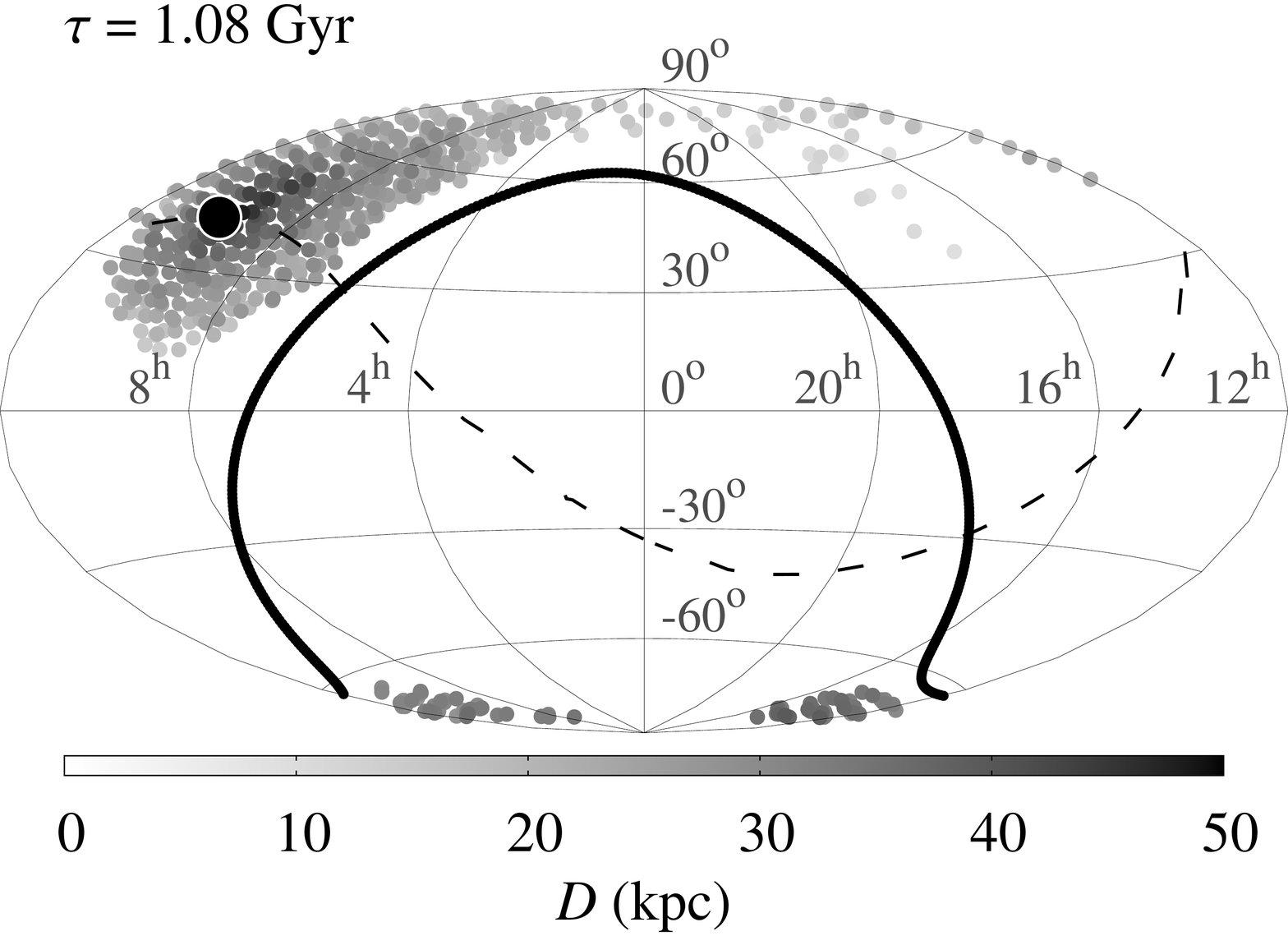}
\hfill
\includegraphics[width=0.33\textwidth]{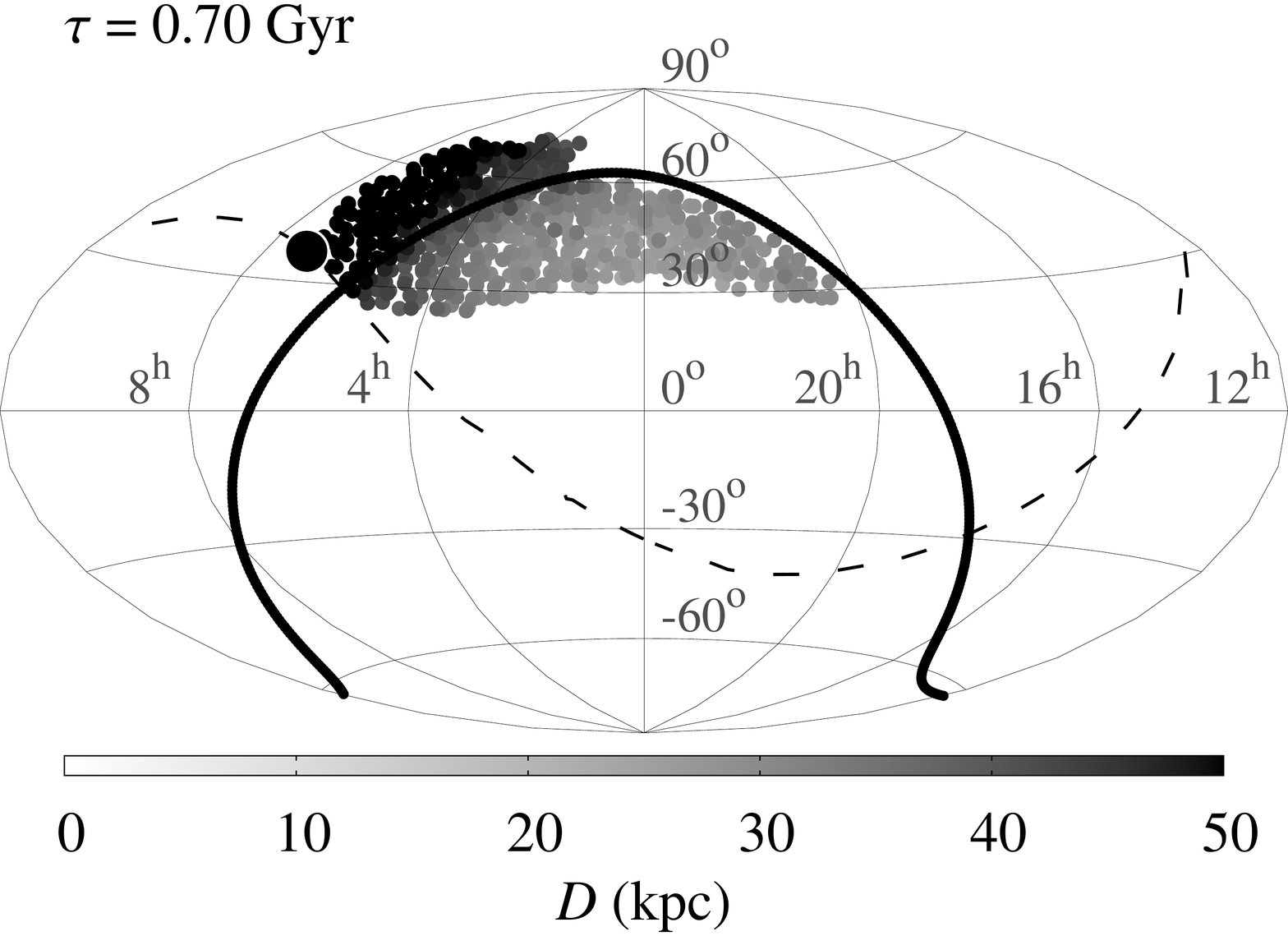}
\hfill
\includegraphics[width=0.33\textwidth]{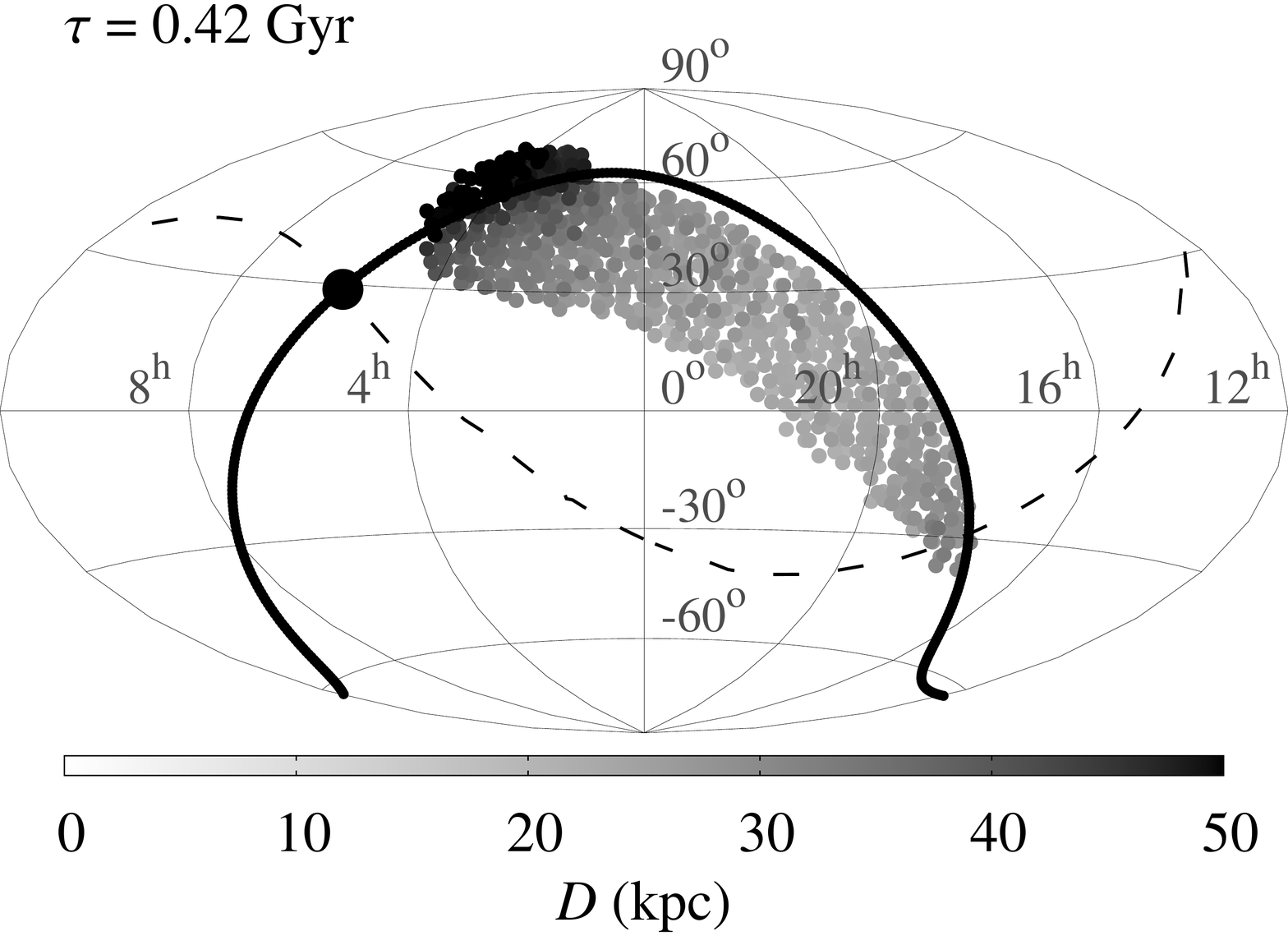}
\hfill
\includegraphics[width=0.33\textwidth]{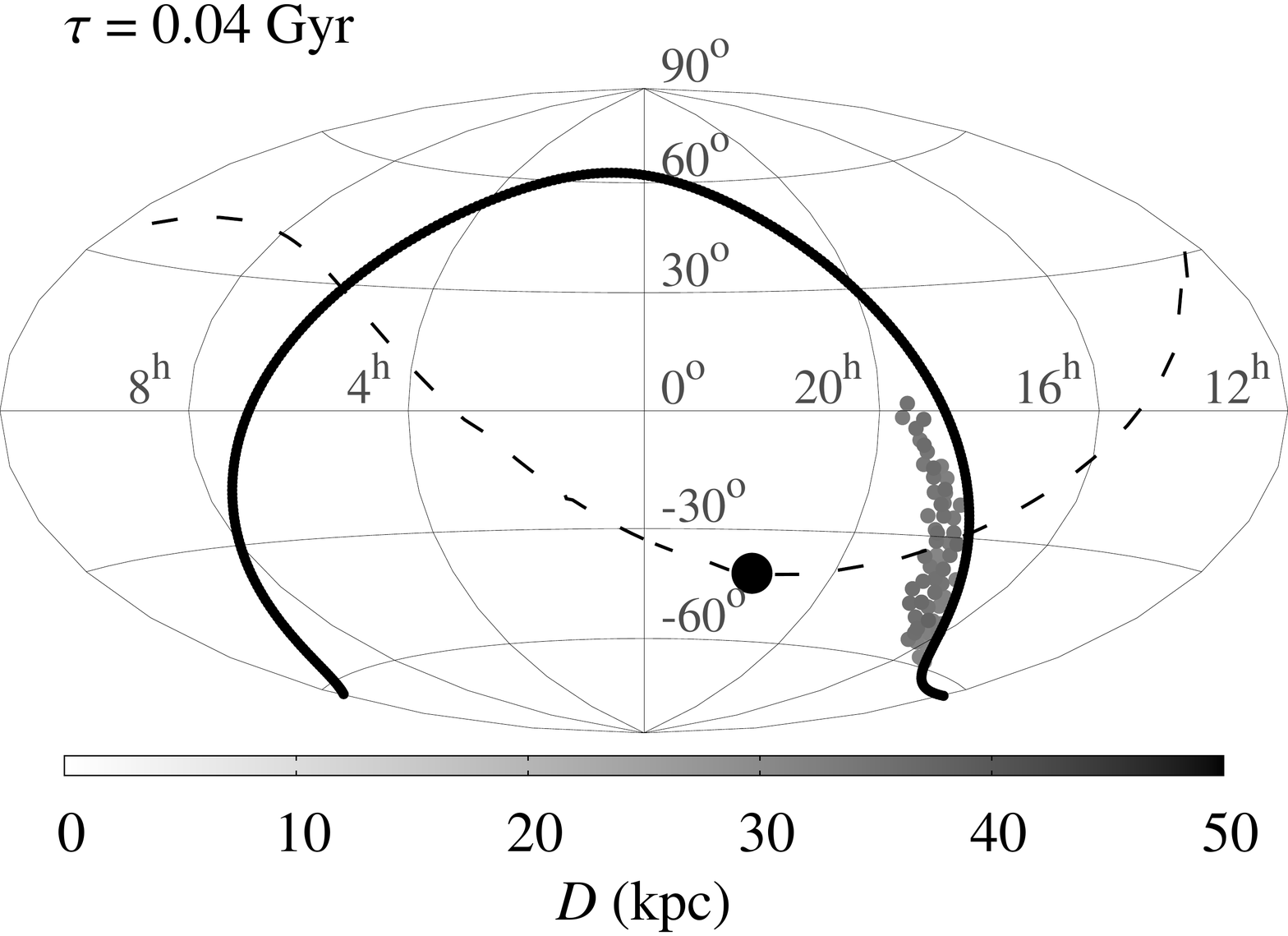}
\caption[ Model C All-sky distribution with gas distance ]{ Heliocentric distance of the Sagittarius gas stream during the last billion years of
the dwarf's evolution. See also Figs.~\ref{fig:allskyC3_snpart} - \ref{fig:allskyC3_temp}. }
\label{fig:allskyC3_dist}
\end{figure*}

Between the second disc transit and the present epoch (i.e. between the snapshots shown in the fourth and the last rows of Fig.~\ref{fig:allskyC}),  corresponding to roughly the last Gyr of Sgr's orbit about the Galaxy, the stripped gas experienced a dramatic evolution. Figs.~\ref{fig:allskyC3_snpart} - \ref{fig:allskyC3_dist} display its all-sky distribution at a series of selected epochs within this time frame, with additional information about two direct observables: the total gas column density (Fig.~\ref{fig:allskyC3_snpart}), and the line of sight velocity (Fig.~\ref{fig:allskyC3_vel}); and two quantities that in principle can be inferred from observation: the temperature (Fig.~\ref{fig:allskyC3_temp}), and the distance (Fig.~\ref{fig:allskyC3_dist}) of the gas. All of the latter quantities correspond to density-weighted averages of the voxels overlapping along the line of sight. For clarity, its associated stellar component has been omitted in these panels. Also, here we have decreased the gas tracer threshold value from 2 to 1 since at this point the value of the tracer of most of the gas has dropped below 2 as a result of gas mixing.

Take for instance the gas that detached from Sgr (located at $\alpha \approx 0 - 16^{\rm h}$, $\delta < -30^\deg$ and $D \gtrsim 50$ kpc at $\tau \approx 1.2$ Gyr) and was left behind after the last transit. Being diffuse (${\rm N_{tot}} < 10^{19}$ \psc), it was subject to strong heating due to mixing with the coronal gas, its temperature rising well above $10^5$ K, and disappeared quickly, in a matter of $\sim 200$ Myr.

The gas stream, still attached to the main body of the dwarf, eventually disappeared as well, but more slowly. Being denser, some of the gas overcame the strong heating via radiative cooling, and thus remained warm and denser for a longer time. As it kept falling towards the Galactic plane, some 400 Myr ago, the Sagittarius gas stream finally detached from its stellar core, and gradually settled onto the Galactic disc, with a small fraction likely dissolving into the Galactic corona. Here, the gas disappears from view because it falls below the detection limit in terms of a gas tracer threshold as it mixes with the gas in the disc. A more detailed discussion of the accretion process onto the Galaxy is deferred to the Appendix \ref{sec:gacc2}.

From our vantage point, most of the gas would have appeared in the form of gigantic, high-velocity gas complexes with negative velocities in the range $\sim -50$ to -300 \kms, at distances on the order of 20 - 40 kpc, and up to $\sim 50$ kpc while still attached to the stellar core. However, our model suggests that this gas has long settled onto the Galactic disc. In other words, we find that virtually no gas associated with Sgr survived in the Galactic halo to the present.

\section{Concluding remarks} \label{sec:conc}

We now discuss briefly some of our results and their consequences within a broader context.

{\em Did gas from Sgr trigger star formation in the Galaxy?} Determinations of the star formation history (SFH) of the Galactic disc indicate clear episodes of enhanced star formation at roughly $0 - 1$ Gyr and $2 - 5$ Gyr ago \citep[][]{roc00b}. This is consistent with the reconstruction the SFH of the stars in the Solar neighbourhood, indicating that the Galaxy was forming stars more actively in the past, with a significant increase around 3 Gyr ago \citep[][]{cig06a} These authors attribute the origin of the last episode to the likely interaction with a satellite system. According to our model, the first pericentric passages of Sgr (at $d \approx 20$ kpc) took place between $2.5 - 3$ ago (left panel of Fig.~\ref{fig:kinC}), consistent with the time of the last epoch of enhanced SF in the Galaxy. However, we speculate that this SF episode would not result from elevated gas accretion but rather from the gravitational disturbance inflicted on the Galactic gas disc \citep[e.g.][]{pet17a}. We stress that Sgr is a more powerful trigger than the either of the Magellanic Clouds. The strong influence of Sgr on the Galaxy, in particular on its disc, was first recognised by \citet[][]{jia00a}, and was demonstrated with the help of $N$-body simulations by \citet[][]{pur11a}. We see this effect in our model in the form of a slight warp of the Galactic gas disc (see central panel in the last row of Fig.~\ref{fig:allskyC}). Whether the accretion of the gas stripped away from Sgr contributed to the star formation episodes is unclear at this point. Still, we can estimate its contribution to the gas budget of the Galaxy. In our model, on the order of $10^8$ \Msun\ are removed from Sgr after each disc transit, and the gas settles on a time scale of a few $10^8$ yr. This implies an accretion rate of $\sim 1$ \Msun~\pyr{}, comparable to the {\em present-day} star formation rate of the Galaxy \citep[e.g.][]{rob10a}.

{\em Did the interaction with the Galaxy induce star forming episodes in Sgr?} It is believed that pericentric passages are associated with bursts of star formation \citep[e.g.][]{may01a}. \citet[][]{sie07a} identify at least two distinct stellar populations in Sgr, with ages $\sim 2.3$ Gyr and $\sim 0.1 - 0.8$. In our model, Sgr experiences pericentric passages at $\sim 2.8$ Gyr and $\sim 1.3$ Gyr, which would be consistent with delayed star formation episodes at the epochs determined from observations. Moreover, pericentric passages are closely related to disc transits that, as our results show, remove a significant fraction of the gas within a dwarf system. Thus, our model suggests that the last pericentric passage of Sgr about the Galaxy triggered the last burst of star formation within the dwarf prior to its disc crossing, just before the dwarf was fully stripped of its remaining gas.

{\em Was the Sgr precursor a disc galaxy?} \citet[][]{lok10a} speculate that the observed geometric and kinematic structure of Sgr's stellar remnant cannot be recovered unless the precursor had a disc. But is has been argued that a residual kinematic signature should be observable in Sgr stellar remnant if it had been a disc galaxy \citep[][]{pen10b}, which does not appear to be the case \citep[][]{pen11b}. Our results show that Sgr's present-day morphology can be recovered from a gas-bearing, spheroidal progenitor. But we cannot make any statement about its kinematics. Yet, we argue that Sgr need not have had a significant disc component prior to infall, and could even have been a gas-rich dwarf spheroidal, which have been observed \citep[][]{bli00a}.

{\em Is the gas stripped away from Sgr currently observable as high- or intermediate-velocity gas?} Under the assumption that stellar feedback within the dwarf does not play in important role in removing gas, we argued that the mass of the stripped gas around the Galaxy as given by our model may be regarded as an upper limit at any given time. And even more so since we ignore the onset of starburst- or nuclear-driven winds within the Galaxy which impact negatively the lifetime of gas clouds \citep[e.g.][]{coo09a}. In other words, our model favours the survivability within the Galactic halo of the gas stripped away from Sgr. Given the resulting lack of gas associated to Sgr in the halo in our model, we conclude that is its unlikely that any of the gas stripped away from Sgr survived to be observable as HVCs or IVCs, as been claimed in the past \citep[][]{bla98a,put04a}.

In this respect, it is worth emphasising our finding that the Sagittarius gas stream is never quite aligned with its associated stellar streams, or with the orbital plane of Sgr for that matter. Significant offsets between gas and stars along the extended tidal features of interacting systems are not at all uncommon \citep[e.g.][]{smi97a,hib99a,hib00a}, and have been explained for the most extreme cases as the consequence of the differing radial distribution between gas and stars in the galaxy discs coupled to the dissipative nature of gas \citep[][]{mih01a}. In our Sgr model, the misalignment is primarily a direct consequence of the {\em transverse} drag exerted by the spinning Galactic corona upon the stripped gas as Sgr moves along its {\em polar} orbit. But it is also partly due to the instantaneous torque the gas experiences when transiting the Galactic gas disc. We speculate that a faster spinning corona, as observed \citep[][]{hod16a}, would lead to an even stronger misalignment. We shall explore this in more detail in future work.

We thus reason that, regardless of whether the gas stripped from Sgr is not longer present around the Galaxy -- as our model suggests--, the generic spatial offset between gas and stars casts doubt upon a putative association of the Smith Cloud \citep[][]{smi63a} with Sgr \citep[][]{bla98a}, and upon the interpretation of \citet[][]{put04a} about the nature of the HVC anti-centre complexes ACHV and ACVHV \citep[][]{wak91b}. In particular, \citeauthor[][]{put04a}'s interpretation relies on the observed alignment (in projection) of these HVCs with of Sgr's orbit. As mentioned above, however, an alignment between the stripped stars and gas is not expected, at least not at considerable angular separations from the stellar remnant. Therefore, if the gas stripped from Sgr has survived to date in the form of HVCs, these have yet to be identified in existing catalogues \citep[e.g.][]{wak91b,mos13a}.

We would like to remind our reader that our conclusions are based on a model partly constrained by assuming a lack of {\em total} gas within Sgr's stellar remnant.  However, observations so far constrain only the {\em neutral} (\HI) gas content of Sgr. It is unknown whether any amount of ionised gas is associated to the remnant. Conversely, as it stands, our model cannot say anything about the specific \HI\ content of Sgr or its associated gas stream, as calculating the ionisation state of the gas requires a careful account of its chemical composition, of the UV radiation both within the dwarf and from the Galaxy, and their evolution with time, ideally using radiative transfer and non-equilibrium calculations, all of which is beyond the scope of this paper. Also, as we show in our related magneto-hydrodynamic study \citep[][see also \citealt{kon02a}]{gro17a}, magnetic fields can increase the destruction timescales of HVCs, or affect the overall accretion rate of gas onto the disc \citep[][]{bir09a}, and is worth considering in future simulations.

An implication of our study independent of the above considerations and relevant to earlier and future models of Sgr is the following. We showed that there is a significant difference in the orbital history resulting from pure $N$-body models and our model which incorporates gas components and its associated drag, even though both start from virtually identical initial conditions. This clearly demonstrates that a full treatment of all relevant components, both in the Galaxy and the precursor, is required if we are to arrive at a consistent dynamical model of the disruption of Sgr, and of other comparable dynamical systems. In addition, in future models it may be appropriate to take into account processes that we have neglected here such as star formation and its associates feedback, both within the dwarf and the Galaxy.

We close by linking this work to our recent study on the Smith Cloud \citep[][]{tep18a}. First, this new work confirms the short timescale (of order a few $100$ Myr) for the Smith Cloud gas in the earlier work. But also importantly, these studies together are suggestive of a slowly emerging, unifying picture of gas accretion at Galactic scales. On the one hand, we have a massive \HI\ gas structure on the brink of being accreted by the Galaxy, with no clear stellar (or DM) counterpart. On the other hand, we have Sgr, a stellar system with no associated gas component just about to collide anew with the Galaxy. Yet, this and our previous study demonstrate that these apparently very disparate systems -- at least in terms of their basic dynamic, kinematic, and structural properties -- can be accommodated within a common framework: DM-assisted gas accretion within a realistic multi-phase
model of the Galaxy. In future papers, we extend this to all Galactic dwarfs and look at the totality of gas accretion via satellites over cosmic time.

\section*{Acknowledgments}
We thank the referee for carefully reading our manuscript and providing insightful comments.
TTG acknowledges financial support from the Australian Research Council (ARC) through an Australian Laureate Fellowship awarded to JBH. JBH acknowledges a Miller Professorship from the Miller Institute, University of California Berkeley.
We acknowledge the facilities, and the scientific and technical assistance of the Sydney Informatics Hub (SIH) at the University of Sydney and, in particular, access to the high-performance computing facility Artemis and additional resources on the National Computational Infrastructure (NCI),  which is supported by the Australian Government, through the University of Sydney's Grand Challenge Program the {\em Astrophysics Grand Challenge: From Large to Small} (CIs: Geraint F.~Lewis and JBH).
All figures and movie frames created with {\sc Gnuplot}, originally written by Thomas Williams and Colin Kelley.\footnote{ \url{http://www.gnuplot.info} }
All animations assembled with {\sc Splicer}, written by R.~S. Sutherland.\footnote{ \url{https://miocene.anu.edu.au/splicer} }
This research has made use of NASA's Astrophysics Data System (ADS) Bibliographic Services\footnote{ \url{http://adsabs.harvard.edu} }, and of {\sc Astropy},\footnote{ \url{http://www.astropy.org} } a community-developed core {\sc Python} package for Astronomy \citepalias{ast13a}.


\begin{thebibliography}{}
\makeatletter
\relax
\def\mn@urlcharsother{\let\do\@makeother \do\$\do\&\do\#\do\^\do\_\do\%\do\~}
\def\mn@doi{\begingroup\mn@urlcharsother \@ifnextchar [ {\mn@doi@}
  {\mn@doi@[]}}
\def\mn@doi@[#1]#2{\def\@tempa{#1}\ifx\@tempa\@empty \href
  {http://dx.doi.org/#2} {doi:#2}\else \href {http://dx.doi.org/#2} {#1}\fi
  \endgroup}
\def\mn@eprint#1#2{\mn@eprint@#1:#2::\@nil}
\def\mn@eprint@arXiv#1{\href {http://arxiv.org/abs/#1} {{\tt arXiv:#1}}}
\def\mn@eprint@dblp#1{\href {http://dblp.uni-trier.de/rec/bibtex/#1.xml}
  {dblp:#1}}
\def\mn@eprint@#1:#2:#3:#4\@nil{\def\@tempa {#1}\def\@tempb {#2}\def\@tempc
  {#3}\ifx \@tempc \@empty \let \@tempc \@tempb \let \@tempb \@tempa \fi \ifx
  \@tempb \@empty \def\@tempb {arXiv}\fi \@ifundefined
  {mn@eprint@\@tempb}{\@tempb:\@tempc}{\expandafter \expandafter \csname
  mn@eprint@\@tempb\endcsname \expandafter{\@tempc}}}

\bibitem[\protect\citeauthoryear{{Aitoff}}{{Aitoff}}{1889}]{ait89a}
{Aitoff} D.,  1889, {Projections des cartes g{\'e}ographiques}.
Hachette Paris

\bibitem[\protect\citeauthoryear{{Astropy Collaboration} et~al.,}{{Astropy
  Collaboration} et~al.}{2013}]{ast13a}
{Astropy Collaboration} et~al., 2013, \mn@doi [\aap]
  {10.1051/0004-6361/201322068}, \href
  {http://adsabs.harvard.edu/abs/2013A%26A...558A..33A} {558, A33}

\bibitem[\protect\citeauthoryear{{Bellazzini} et~al.,}{{Bellazzini}
  et~al.}{2008}]{bel08a}
{Bellazzini} M.,  et~al., 2008, \mn@doi [\aj] {10.1088/0004-6256/136/3/1147},
  \href {http://adsabs.harvard.edu/abs/2008AJ....136.1147B} {136, 1147}

\bibitem[\protect\citeauthoryear{{Belokurov} et~al.,}{{Belokurov}
  et~al.}{2014}]{bel14a}
{Belokurov} V.,  et~al., 2014, \mn@doi [\mnras] {10.1093/mnras/stt1862}, \href
  {http://adsabs.harvard.edu/abs/2014MNRAS.437..116B} {437, 116}

\bibitem[\protect\citeauthoryear{{Benjamin} \& {Danly}}{{Benjamin} \&
  {Danly}}{1997}]{ben97a}
{Benjamin} R.~A.,  {Danly} L.,  1997, \mn@doi [\apj] {10.1086/304078}, \href
  {http://adsabs.harvard.edu/abs/1997ApJ...481..764B} {481, 764}

\bibitem[\protect\citeauthoryear{{Bernstein-Cooper} et~al.,}{{Bernstein-Cooper}
  et~al.}{2014}]{ber14b}
{Bernstein-Cooper} E.~Z.,  et~al., 2014, \mn@doi [\aj]
  {10.1088/0004-6256/148/2/35}, \href
  {http://adsabs.harvard.edu/abs/2014AJ....148...35B} {148, 35}

\bibitem[\protect\citeauthoryear{{Besla}, {Kallivayalil}, {Hernquist},
  {Robertson}, {Cox}, {van der Marel}  \& {Alcock}}{{Besla}
  et~al.}{2007}]{bes07a}
{Besla} G.,  {Kallivayalil} N.,  {Hernquist} L.,  {Robertson} B.,  {Cox} T.~J.,
   {van der Marel} R.~P.,   {Alcock} C.,  2007, \mn@doi [\apj]
  {10.1086/521385}, \href {http://adsabs.harvard.edu/abs/2007ApJ...668..949B}
  {668, 949}

\bibitem[\protect\citeauthoryear{{Birnboim}}{{Birnboim}}{2009}]{bir09a}
{Birnboim} Y.,  2009, \mn@doi [\apjl] {10.1088/0004-637X/702/2/L101}, \href
  {http://adsabs.harvard.edu/abs/2009ApJ...702L.101B} {702, L101}

\bibitem[\protect\citeauthoryear{{Bland-Hawthorn}}{{Bland-Hawthorn}}{2009}]{bla09a}
{Bland-Hawthorn} J.,  2009, in {Andersen} J.,  {Nordstr{\"o}ara} {m} B.,
  {Bland-Hawthorn} J.,  eds,  IAU Symposium Vol. 254, The Galaxy Disk in
  Cosmological Context Proceedings IAU Symposium No. 254. pp 241--254
  (\mn@eprint {arXiv} {0811.2467}), \mn@doi{10.1017/S174392130802766X}

\bibitem[\protect\citeauthoryear{Bland-Hawthorn \& Gerhard}{Bland-Hawthorn \&
  Gerhard}{2016}]{bla16a}
Bland-Hawthorn J.,  Gerhard O.,  2016, \mn@doi [Annual Review of Astronomy and
  Astrophysics] {10.1146/annurev-astro-081915-023441}, 54, 529

\bibitem[\protect\citeauthoryear{{Bland-Hawthorn}, {Veilleux}, {Cecil},
  {Putman}, {Gibson}  \& {Maloney}}{{Bland-Hawthorn} et~al.}{1998}]{bla98a}
{Bland-Hawthorn} J.,  {Veilleux} S.,  {Cecil} G.~N.,  {Putman} M.~E.,  {Gibson}
  B.~K.,   {Maloney} P.~R.,  1998, \mn@doi [\mnras]
  {10.1046/j.1365-8711.1998.01902.x}, \href
  {http://adsabs.harvard.edu/abs/1998MNRAS.299..611B} {299, 611}

\bibitem[\protect\citeauthoryear{{Blitz} \& {Robishaw}}{{Blitz} \&
  {Robishaw}}{2000}]{bli00a}
{Blitz} L.,  {Robishaw} T.,  2000, \mn@doi [\apj] {10.1086/309457}, \href
  {http://adsabs.harvard.edu/abs/2000ApJ...541..675B} {541, 675}

\bibitem[\protect\citeauthoryear{{Bovy} et~al.,}{{Bovy} et~al.}{2012}]{bov12b}
{Bovy} J.,  et~al., 2012, \mn@doi [\apj] {10.1088/0004-637X/759/2/131}, \href
  {http://adsabs.harvard.edu/abs/2012ApJ...759..131B} {759, 131}

\bibitem[\protect\citeauthoryear{{Bradford}, {Geha}  \& {Blanton}}{{Bradford}
  et~al.}{2015}]{bra15a}
{Bradford} J.~D.,  {Geha} M.~C.,   {Blanton} M.~R.,  2015, \mn@doi [\apj]
  {10.1088/0004-637X/809/2/146}, \href
  {http://adsabs.harvard.edu/abs/2015ApJ...809..146B} {809, 146}

\bibitem[\protect\citeauthoryear{{Burton} \& {Lockman}}{{Burton} \&
  {Lockman}}{1999}]{bur99b}
{Burton} W.~B.,  {Lockman} F.~J.,  1999, \aap, \href
  {http://adsabs.harvard.edu/abs/1999A%26A...349....7B} {349, 7}

\bibitem[\protect\citeauthoryear{{Byrd}, {Valtonen}, {McCall}  \&
  {Innanen}}{{Byrd} et~al.}{1994}]{byr94a}
{Byrd} G.,  {Valtonen} M.,  {McCall} M.,   {Innanen} K.,  1994, \mn@doi [\aj]
  {10.1086/117015}, \href {http://adsabs.harvard.edu/abs/1994AJ....107.2055B}
  {107, 2055}

\bibitem[\protect\citeauthoryear{{Cignoni}, {Degl'Innocenti}, {Prada Moroni}
  \& {Shore}}{{Cignoni} et~al.}{2006}]{cig06a}
{Cignoni} M.,  {Degl'Innocenti} S.,  {Prada Moroni} P.~G.,   {Shore} S.~N.,
  2006, \mn@doi [\aap] {10.1051/0004-6361:20065645}, \href
  {http://adsabs.harvard.edu/abs/2006A%26A...459..783C} {459, 783}

\bibitem[\protect\citeauthoryear{{Cooper}, {Bicknell}, {Sutherland}  \&
  {Bland-Hawthorn}}{{Cooper} et~al.}{2009}]{coo09a}
{Cooper} J.~L.,  {Bicknell} G.~V.,  {Sutherland} R.~S.,   {Bland-Hawthorn} J.,
  2009, \mn@doi [\apj] {10.1088/0004-637X/703/1/330}, \href
  {http://adsabs.harvard.edu/abs/2009ApJ...703..330C} {703, 330}

\bibitem[\protect\citeauthoryear{{Dierickx} \& {Loeb}}{{Dierickx} \&
  {Loeb}}{2017a}]{die17a}
{Dierickx} M.~I.~P.,  {Loeb} A.,  2017a, \mn@doi [\apj]
  {10.3847/1538-4357/836/1/92}, \href
  {http://adsabs.harvard.edu/abs/2017ApJ...836...92D} {836, 92}

\bibitem[\protect\citeauthoryear{{Dierickx} \& {Loeb}}{{Dierickx} \&
  {Loeb}}{2017b}]{die17b}
{Dierickx} M.~I.~P.,  {Loeb} A.,  2017b, \mn@doi [\apj]
  {10.3847/1538-4357/aa8767}, \href
  {http://adsabs.harvard.edu/abs/2017ApJ...847...42D} {847, 42}

\bibitem[\protect\citeauthoryear{{Dinescu}, {Girard}, {van Altena}  \&
  {L{\'o}pez}}{{Dinescu} et~al.}{2005}]{din05a}
{Dinescu} D.~I.,  {Girard} T.~M.,  {van Altena} W.~F.,   {L{\'o}pez} C.~E.,
  2005, \mn@doi [\apjl] {10.1086/427731}, \href
  {http://adsabs.harvard.edu/abs/2005ApJ...618L..25D} {618, L25}

\bibitem[\protect\citeauthoryear{{Fellhauer} et~al.,}{{Fellhauer}
  et~al.}{2006}]{fel06a}
{Fellhauer} M.,  et~al., 2006, \mn@doi [\apj] {10.1086/507128}, \href
  {http://adsabs.harvard.edu/abs/2006ApJ...651..167F} {651, 167}

\bibitem[\protect\citeauthoryear{{Gibbons}, {Belokurov}  \& {Evans}}{{Gibbons}
  et~al.}{2014}]{gib14a}
{Gibbons} S.~L.~J.,  {Belokurov} V.,   {Evans} N.~W.,  2014, \mn@doi [\mnras]
  {10.1093/mnras/stu1986}, \href
  {http://adsabs.harvard.edu/abs/2014MNRAS.445.3788G} {445, 3788}

\bibitem[\protect\citeauthoryear{{Gibbons}, {Belokurov}  \& {Evans}}{{Gibbons}
  et~al.}{2017}]{gib17a}
{Gibbons} S.~L.~J.,  {Belokurov} V.,   {Evans} N.~W.,  2017, \mn@doi [\mnras]
  {10.1093/mnras/stw2328}, \href
  {http://adsabs.harvard.edu/abs/2017MNRAS.464..794G} {464, 794}

\bibitem[\protect\citeauthoryear{{G{\'o}mez}, {Besla}, {Carpintero},
  {Villalobos}, {O'Shea}  \& {Bell}}{{G{\'o}mez} et~al.}{2015}]{gom15a}
{G{\'o}mez} F.~A.,  {Besla} G.,  {Carpintero} D.~D.,  {Villalobos} {\'A}.,
  {O'Shea} B.~W.,   {Bell} E.~F.,  2015, \mn@doi [\apj]
  {10.1088/0004-637X/802/2/128}, \href
  {http://adsabs.harvard.edu/abs/2015ApJ...802..128G} {802, 128}

\bibitem[\protect\citeauthoryear{{Gonz{\'a}lez-Samaniego}, {Col{\'{\i}}n},
  {Avila-Reese}, {Rodr{\'{\i}}guez-Puebla}  \&
  {Valenzuela}}{{Gonz{\'a}lez-Samaniego} et~al.}{2014}]{gon14a}
{Gonz{\'a}lez-Samaniego} A.,  {Col{\'{\i}}n} P.,  {Avila-Reese} V.,
  {Rodr{\'{\i}}guez-Puebla} A.,   {Valenzuela} O.,  2014, \mn@doi [\apj]
  {10.1088/0004-637X/785/1/58}, \href
  {http://adsabs.harvard.edu/abs/2014ApJ...785...58G} {785, 58}

\bibitem[\protect\citeauthoryear{{Grcevich} \& {Putman}}{{Grcevich} \&
  {Putman}}{2009}]{grc09a}
{Grcevich} J.,  {Putman} M.~E.,  2009, \mn@doi [\apj]
  {10.1088/0004-637X/696/1/385}, \href
  {http://adsabs.harvard.edu/abs/2009ApJ...696..385G} {696, 385}

\bibitem[\protect\citeauthoryear{{Gr{\o}nnow}, {Tepper-Garc{\'{\i}}a},
  {Bland-Hawthorn}  \& {McClure-Griffiths}}{{Gr{\o}nnow} et~al.}{2017}]{gro17a}
{Gr{\o}nnow} A.,  {Tepper-Garc{\'{\i}}a} T.,  {Bland-Hawthorn} J.,
  {McClure-Griffiths} N.~M.,  2017, \mn@doi [\apj] {10.3847/1538-4357/aa7ed2},
  \href {http://adsabs.harvard.edu/abs/2017ApJ...845...69G} {845, 69}

\bibitem[\protect\citeauthoryear{{Guglielmo}, {Lewis}  \&
  {Bland-Hawthorn}}{{Guglielmo} et~al.}{2014}]{gug14a}
{Guglielmo} M.,  {Lewis} G.~F.,   {Bland-Hawthorn} J.,  2014, \mn@doi [\mnras]
  {10.1093/mnras/stu1549}, \href
  {http://adsabs.harvard.edu/abs/2014MNRAS.444.1759G} {444, 1759}

\bibitem[\protect\citeauthoryear{{Hammer}}{{Hammer}}{1892}]{ham92a}
{Hammer} E.,  1892, Petermanns Mitteilungen, 38, 85

\bibitem[\protect\citeauthoryear{{Hernquist}}{{Hernquist}}{1990}]{her90a}
{Hernquist} L.,  1990, \mn@doi [\apj] {10.1086/168845}, \href
  {http://adsabs.harvard.edu/abs/1990ApJ...356..359H} {356, 359}

\bibitem[\protect\citeauthoryear{{Hernquist}}{{Hernquist}}{1993}]{her93b}
{Hernquist} L.,  1993, \mn@doi [\apjs] {10.1086/191784}, \href
  {http://adsabs.harvard.edu/abs/1993ApJS...86..389H} {86, 389}

\bibitem[\protect\citeauthoryear{{Hibbard} \& {Yun}}{{Hibbard} \&
  {Yun}}{1999}]{hib99a}
{Hibbard} J.~E.,  {Yun} M.~S.,  1999, \mn@doi [\aj] {10.1086/300928}, \href
  {http://adsabs.harvard.edu/abs/1999AJ....118..162H} {118, 162}

\bibitem[\protect\citeauthoryear{{Hibbard}, {Vacca}  \& {Yun}}{{Hibbard}
  et~al.}{2000}]{hib00a}
{Hibbard} J.~E.,  {Vacca} W.~D.,   {Yun} M.~S.,  2000, \mn@doi [\aj]
  {10.1086/301263}, \href {http://adsabs.harvard.edu/abs/2000AJ....119.1130H}
  {119, 1130}

\bibitem[\protect\citeauthoryear{{Hodges-Kluck}, {Miller}  \&
  {Bregman}}{{Hodges-Kluck} et~al.}{2016}]{hod16a}
{Hodges-Kluck} E.~J.,  {Miller} M.~J.,   {Bregman} J.~N.,  2016, \mn@doi [\apj]
  {10.3847/0004-637X/822/1/21}, \href
  {http://adsabs.harvard.edu/abs/2016ApJ...822...21H} {822, 21}

\bibitem[\protect\citeauthoryear{{Hou}, {Prantzos}  \& {Boissier}}{{Hou}
  et~al.}{2000}]{hou00a}
{Hou} J.~L.,  {Prantzos} N.,   {Boissier} S.,  2000, \aap, \href
  {http://adsabs.harvard.edu/abs/2000A%26A...362..921H} {362, 921}

\bibitem[\protect\citeauthoryear{{Ibata} \& {Lewis}}{{Ibata} \&
  {Lewis}}{1998}]{iba98a}
{Ibata} R.~A.,  {Lewis} G.~F.,  1998, \mn@doi [\apj] {10.1086/305773}, \href
  {http://adsabs.harvard.edu/abs/1998ApJ...500..575I} {500, 575}

\bibitem[\protect\citeauthoryear{{Ibata}, {Gilmore}  \& {Irwin}}{{Ibata}
  et~al.}{1994}]{iba94a}
{Ibata} R.~A.,  {Gilmore} G.,   {Irwin} M.~J.,  1994, \mn@doi [\nat]
  {10.1038/370194a0}, \href {http://adsabs.harvard.edu/abs/1994Natur.370..194I}
  {370, 194}

\bibitem[\protect\citeauthoryear{{Ibata}, {Wyse}, {Gilmore}, {Irwin}  \&
  {Suntzeff}}{{Ibata} et~al.}{1997}]{iba97a}
{Ibata} R.~A.,  {Wyse} R.~F.~G.,  {Gilmore} G.,  {Irwin} M.~J.,   {Suntzeff}
  N.~B.,  1997, \mn@doi [\aj] {10.1086/118283}, \href
  {http://adsabs.harvard.edu/abs/1997AJ....113..634I} {113, 634}

\bibitem[\protect\citeauthoryear{{Jethwa}, {Erkal}  \& {Belokurov}}{{Jethwa}
  et~al.}{2016}]{jet16a}
{Jethwa} P.,  {Erkal} D.,   {Belokurov} V.,  2016, \mn@doi [\mnras]
  {10.1093/mnras/stw1343}, \href
  {http://adsabs.harvard.edu/abs/2016MNRAS.461.2212J} {461, 2212}

\bibitem[\protect\citeauthoryear{{Jiang} \& {Binney}}{{Jiang} \&
  {Binney}}{2000}]{jia00a}
{Jiang} I.-G.,  {Binney} J.,  2000, \mn@doi [\mnras]
  {10.1046/j.1365-8711.2000.03311.x}, \href
  {http://adsabs.harvard.edu/abs/2000MNRAS.314..468J} {314, 468}

\bibitem[\protect\citeauthoryear{{Kazantzidis}, {Magorrian}  \&
  {Moore}}{{Kazantzidis} et~al.}{2004}]{kaz04a}
{Kazantzidis} S.,  {Magorrian} J.,   {Moore} B.,  2004, \mn@doi [\apj]
  {10.1086/380192}, \href {http://adsabs.harvard.edu/abs/2004ApJ...601...37K}
  {601, 37}

\bibitem[\protect\citeauthoryear{{King}}{{King}}{1962}]{kin62a}
{King} I.,  1962, \mn@doi [\aj]
  {file://localhost/Users/tepper/references/aaa.pdf}, \href
  {http://adsabs.harvard.edu/abs/1962AJ.....67..471K} {67, 471}

\bibitem[\protect\citeauthoryear{{Konz}, {Br{\"u}ns}  \& {Birk}}{{Konz}
  et~al.}{2002}]{kon02a}
{Konz} C.,  {Br{\"u}ns} C.,   {Birk} G.~T.,  2002, \mn@doi [\aap]
  {10.1051/0004-6361:20020863}, \href
  {http://adsabs.harvard.edu/abs/2002A%26A...391..713K} {391, 713}

\bibitem[\protect\citeauthoryear{{Koribalski}, {Johnston}  \&
  {Otrupcek}}{{Koribalski} et~al.}{1994}]{kor94a}
{Koribalski} B.,  {Johnston} S.,   {Otrupcek} R.,  1994, \mn@doi [\mnras]
  {10.1093/mnras/270.1.L43}, \href
  {http://adsabs.harvard.edu/abs/1994MNRAS.270L..43K} {270, L43}

\bibitem[\protect\citeauthoryear{{Kunder} \& {Chaboyer}}{{Kunder} \&
  {Chaboyer}}{2009}]{kun09a}
{Kunder} A.,  {Chaboyer} B.,  2009, \mn@doi [\aj]
  {10.1088/0004-6256/137/5/4478}, \href
  {http://adsabs.harvard.edu/abs/2009AJ....137.4478K} {137, 4478}

\bibitem[\protect\citeauthoryear{{Law} \& {Majewski}}{{Law} \&
  {Majewski}}{2010}]{law10a}
{Law} D.~R.,  {Majewski} S.~R.,  2010, \mn@doi [\apj]
  {10.1088/0004-637X/714/1/229}, \href
  {http://adsabs.harvard.edu/abs/2010ApJ...714..229L} {714, 229}

\bibitem[\protect\citeauthoryear{{Law} \& {Majewski}}{{Law} \&
  {Majewski}}{2016}]{law16a}
{Law} D.~R.,  {Majewski} S.~R.,  2016, \mn@doi [Tidal Streams in the Local
  Group and Beyond] {10.1007/978-3-319-19336-6_2}, \href
  {http://adsabs.harvard.edu/abs/2016ASSL..420...31L} {420, 31}

\bibitem[\protect\citeauthoryear{{{\L}okas}, {Kazantzidis}, {Majewski}, {Law},
  {Mayer}  \& {Frinchaboy}}{{{\L}okas} et~al.}{2010}]{lok10a}
{{\L}okas} E.~L.,  {Kazantzidis} S.,  {Majewski} S.~R.,  {Law} D.~R.,  {Mayer}
  L.,   {Frinchaboy} P.~M.,  2010, \mn@doi [\apj]
  {10.1088/0004-637X/725/2/1516}, \href
  {http://adsabs.harvard.edu/abs/2010ApJ...725.1516L} {725, 1516}

\bibitem[\protect\citeauthoryear{{Majewski}, {Skrutskie}, {Weinberg}  \&
  {Ostheimer}}{{Majewski} et~al.}{2003}]{maj03a}
{Majewski} S.~R.,  {Skrutskie} M.~F.,  {Weinberg} M.~D.,   {Ostheimer} J.~C.,
  2003, \mn@doi [\apj] {10.1086/379504}, \href
  {http://adsabs.harvard.edu/abs/2003ApJ...599.1082M} {599, 1082}

\bibitem[\protect\citeauthoryear{{Majewski} et~al.,}{{Majewski}
  et~al.}{2004}]{maj04a}
{Majewski} S.~R.,  et~al., 2004, \mn@doi [\aj] {10.1086/421372}, \href
  {http://adsabs.harvard.edu/abs/2004AJ....128..245M} {128, 245}

\bibitem[\protect\citeauthoryear{{Massari}, {Bellini}, {Ferraro}, {van der
  Marel}, {Anderson}, {Dalessandro}  \& {Lanzoni}}{{Massari}
  et~al.}{2013}]{mas13a}
{Massari} D.,  {Bellini} A.,  {Ferraro} F.~R.,  {van der Marel} R.~P.,
  {Anderson} J.,  {Dalessandro} E.,   {Lanzoni} B.,  2013, \mn@doi [\apj]
  {10.1088/0004-637X/779/1/81}, \href
  {http://adsabs.harvard.edu/abs/2013ApJ...779...81M} {779, 81}

\bibitem[\protect\citeauthoryear{{Mathewson}, {Cleary}  \&
  {Murray}}{{Mathewson} et~al.}{1974}]{mat74a}
{Mathewson} D.~S.,  {Cleary} M.~N.,   {Murray} J.~D.,  1974, \mn@doi [\apj]
  {10.1086/152875}, \href {http://adsabs.harvard.edu/abs/1974ApJ...190..291M}
  {190, 291}

\bibitem[\protect\citeauthoryear{{Mayer}, {Governato}, {Colpi}, {Moore},
  {Quinn}, {Wadsley}, {Stadel}  \& {Lake}}{{Mayer} et~al.}{2001}]{may01a}
{Mayer} L.,  {Governato} F.,  {Colpi} M.,  {Moore} B.,  {Quinn} T.,  {Wadsley}
  J.,  {Stadel} J.,   {Lake} G.,  2001, \mn@doi [\apjl] {10.1086/318898}, \href
  {http://adsabs.harvard.edu/abs/2001ApJ...547L.123M} {547, L123}

\bibitem[\protect\citeauthoryear{{Mayer}, {Mastropietro}, {Wadsley}, {Stadel}
  \& {Moore}}{{Mayer} et~al.}{2006}]{may06a}
{Mayer} L.,  {Mastropietro} C.,  {Wadsley} J.,  {Stadel} J.,   {Moore} B.,
  2006, \mn@doi [\mnras] {10.1111/j.1365-2966.2006.10403.x}, \href
  {http://adsabs.harvard.edu/abs/2006MNRAS.369.1021M} {369, 1021}

\bibitem[\protect\citeauthoryear{{McConnachie}}{{McConnachie}}{2012}]{mcc12a}
{McConnachie} A.~W.,  2012, \mn@doi [\aj] {10.1088/0004-6256/144/1/4}, \href
  {http://adsabs.harvard.edu/abs/2012AJ....144....4M} {144, 4}

\bibitem[\protect\citeauthoryear{{McGaugh}, {Schombert}, {de Blok}  \&
  {Zagursky}}{{McGaugh} et~al.}{2010}]{mcg10a}
{McGaugh} S.~S.,  {Schombert} J.~M.,  {de Blok} W.~J.~G.,   {Zagursky} M.~J.,
  2010, \mn@doi [\apjl] {10.1088/2041-8205/708/1/L14}, \href
  {http://adsabs.harvard.edu/abs/2010ApJ...708L..14M} {708, L14}

\bibitem[\protect\citeauthoryear{{Mihos}}{{Mihos}}{2001}]{mih01a}
{Mihos} J.~C.,  2001, \mn@doi [\apj] {10.1086/319721}, \href
  {http://adsabs.harvard.edu/abs/2001ApJ...550...94M} {550, 94}

\bibitem[\protect\citeauthoryear{{Miller} \& {Bregman}}{{Miller} \&
  {Bregman}}{2015}]{mil15a}
{Miller} M.~J.,  {Bregman} J.~N.,  2015, \mn@doi [\apj]
  {10.1088/0004-637X/800/1/14}, \href
  {http://adsabs.harvard.edu/abs/2015ApJ...800...14M} {800, 14}

\bibitem[\protect\citeauthoryear{{Miyamoto} \& {Nagai}}{{Miyamoto} \&
  {Nagai}}{1975}]{miy75a}
{Miyamoto} M.,  {Nagai} R.,  1975, \pasj, \href
  {http://adsabs.harvard.edu/abs/1975PASJ...27..533M} {27, 533}

\bibitem[\protect\citeauthoryear{{Moss}, {McClure-Griffiths}, {Murphy},
  {Pisano}, {Kummerfeld}  \& {Curran}}{{Moss} et~al.}{2013}]{mos13a}
{Moss} V.~A.,  {McClure-Griffiths} N.~M.,  {Murphy} T.,  {Pisano} D.~J.,
  {Kummerfeld} J.~K.,   {Curran} J.~R.,  2013, \mn@doi [\apjs]
  {10.1088/0067-0049/209/1/12}, \href
  {http://adsabs.harvard.edu/abs/2013ApJS..209...12M} {209, 12}

\bibitem[\protect\citeauthoryear{{Mucciarelli}, {Bellazzini}, {Ibata},
  {Romano}, {Chapman}  \& {Monaco}}{{Mucciarelli} et~al.}{2017}]{muc17a}
{Mucciarelli} A.,  {Bellazzini} M.,  {Ibata} R.,  {Romano} D.,  {Chapman}
  S.~C.,   {Monaco} L.,  2017, \mn@doi [\aap] {10.1051/0004-6361/201730707},
  \href {http://adsabs.harvard.edu/abs/2017A%26A...605A..46M} {605, A46}

\bibitem[\protect\citeauthoryear{{Myers}, {Snyder}, {Rusthoven}, {The}  \&
  {Hartmann}}{{Myers} et~al.}{2010}]{mye10a}
{Myers} J.~M.,  {Snyder} B.,  {Rusthoven} M.,  {The} L.-S.,   {Hartmann} D.~H.,
   2010, \mn@doi [\apj] {10.1088/0004-637X/723/2/1057}, \href
  {http://adsabs.harvard.edu/abs/2010ApJ...723.1057M} {723, 1057}

\bibitem[\protect\citeauthoryear{{Nichols} \& {Bland-Hawthorn}}{{Nichols} \&
  {Bland-Hawthorn}}{2011}]{nic11a}
{Nichols} M.,  {Bland-Hawthorn} J.,  2011, \mn@doi [\apj]
  {10.1088/0004-637X/732/1/17}, \href
  {http://adsabs.harvard.edu/abs/2011ApJ...732...17N} {732, 17}

\bibitem[\protect\citeauthoryear{{Nichols}, {Mirabal}, {Agertz}, {Lockman}  \&
  {Bland-Hawthorn}}{{Nichols} et~al.}{2014}]{nic14b}
{Nichols} M.,  {Mirabal} N.,  {Agertz} O.,  {Lockman} F.~J.,   {Bland-Hawthorn}
  J.,  2014, \mn@doi [\mnras] {10.1093/mnras/stu1028}, \href
  {http://adsabs.harvard.edu/abs/2014MNRAS.442.2883N} {442, 2883}

\bibitem[\protect\citeauthoryear{{Nichols}, {Revaz}  \& {Jablonka}}{{Nichols}
  et~al.}{2015}]{nic15a}
{Nichols} M.,  {Revaz} Y.,   {Jablonka} P.,  2015, \mn@doi [\aap]
  {10.1051/0004-6361/201526113}, \href
  {http://adsabs.harvard.edu/abs/2015A%26A...582A..23N} {582, A23}

\bibitem[\protect\citeauthoryear{{Nidever}, {Majewski}, {Butler Burton}  \&
  {Nigra}}{{Nidever} et~al.}{2010}]{nid10a}
{Nidever} D.~L.,  {Majewski} S.~R.,  {Butler Burton} W.,   {Nigra} L.,  2010,
  \mn@doi [\apj] {10.1088/0004-637X/723/2/1618}, \href
  {http://adsabs.harvard.edu/abs/2010ApJ...723.1618N} {723, 1618}

\bibitem[\protect\citeauthoryear{{Niederste-Ostholt}, {Belokurov}, {Evans}  \&
  {Pe{\~n}arrubia}}{{Niederste-Ostholt} et~al.}{2010}]{nie10a}
{Niederste-Ostholt} M.,  {Belokurov} V.,  {Evans} N.~W.,   {Pe{\~n}arrubia} J.,
   2010, \mn@doi [\apj] {10.1088/0004-637X/712/1/516}, \href
  {http://adsabs.harvard.edu/abs/2010ApJ...712..516N} {712, 516}

\bibitem[\protect\citeauthoryear{{Niederste-Ostholt}, {Belokurov}  \&
  {Evans}}{{Niederste-Ostholt} et~al.}{2012}]{nie12a}
{Niederste-Ostholt} M.,  {Belokurov} V.,   {Evans} N.~W.,  2012, \mn@doi
  [\mnras] {10.1111/j.1365-2966.2012.20602.x}, \href
  {http://adsabs.harvard.edu/abs/2012MNRAS.422..207N} {422, 207}

\bibitem[\protect\citeauthoryear{{Pe{\~n}arrubia}, {McConnachie}  \&
  {Navarro}}{{Pe{\~n}arrubia} et~al.}{2008}]{pen08a}
{Pe{\~n}arrubia} J.,  {McConnachie} A.~W.,   {Navarro} J.~F.,  2008, \mn@doi
  [\apj] {10.1086/521543}, \href
  {http://adsabs.harvard.edu/abs/2008ApJ...672..904P} {672, 904}

\bibitem[\protect\citeauthoryear{{Pe{\~n}arrubia}, {Belokurov}, {Evans},
  {Mart{\'{\i}}nez-Delgado}, {Gilmore}, {Irwin}, {Niederste-Ostholt}  \&
  {Zucker}}{{Pe{\~n}arrubia} et~al.}{2010}]{pen10b}
{Pe{\~n}arrubia} J.,  {Belokurov} V.,  {Evans} N.~W.,
  {Mart{\'{\i}}nez-Delgado} D.,  {Gilmore} G.,  {Irwin} M.,
  {Niederste-Ostholt} M.,   {Zucker} D.~B.,  2010, \mn@doi [\mnras]
  {10.1111/j.1745-3933.2010.00921.x}, \href
  {http://adsabs.harvard.edu/abs/2010MNRAS.408L..26P} {408, L26}

\bibitem[\protect\citeauthoryear{{Pe{\~n}arrubia} et~al.,}{{Pe{\~n}arrubia}
  et~al.}{2011}]{pen11b}
{Pe{\~n}arrubia} J.,  et~al., 2011, \mn@doi [\apjl]
  {10.1088/2041-8205/727/1/L2}, \href
  {http://adsabs.harvard.edu/abs/2011ApJ...727L...2P} {727, L2}

\bibitem[\protect\citeauthoryear{{Perret}, {Renaud}, {Epinat}, {Amram},
  {Bournaud}, {Contini}, {Teyssier}  \& {Lambert}}{{Perret}
  et~al.}{2014}]{per14c}
{Perret} V.,  {Renaud} F.,  {Epinat} B.,  {Amram} P.,  {Bournaud} F.,
  {Contini} T.,  {Teyssier} R.,   {Lambert} J.-C.,  2014, \mn@doi [\aap]
  {10.1051/0004-6361/201322395}, \href
  {http://adsabs.harvard.edu/abs/2014A%26A...562A...1P} {562, A1}

\bibitem[\protect\citeauthoryear{{Pettitt}, {Tasker}, {Wadsley}, {Keller}  \&
  {Benincasa}}{{Pettitt} et~al.}{2017}]{pet17a}
{Pettitt} A.~R.,  {Tasker} E.~J.,  {Wadsley} J.~W.,  {Keller} B.~W.,
  {Benincasa} S.~M.,  2017, \mn@doi [\mnras] {10.1093/mnras/stx736}, \href
  {http://adsabs.harvard.edu/abs/2017MNRAS.468.4189P} {468, 4189}

\bibitem[\protect\citeauthoryear{{Pietrzy{\'n}ski} et~al.,}{{Pietrzy{\'n}ski}
  et~al.}{2013}]{pie13b}
{Pietrzy{\'n}ski} G.,  et~al., 2013, \mn@doi [\nat] {10.1038/nature11878},
  \href {http://adsabs.harvard.edu/abs/2013Natur.495...76P} {495, 76}

\bibitem[\protect\citeauthoryear{{Planck Collaboration} et~al.,}{{Planck
  Collaboration} et~al.}{2014}]{pla14b}
{Planck Collaboration} et~al., 2014, \mn@doi [\aap]
  {10.1051/0004-6361/201321591}, \href
  {http://adsabs.harvard.edu/abs/2014A%26A...571A..16P} {571, A16}

\bibitem[\protect\citeauthoryear{{Pryor}, {Piatek}  \& {Olszewski}}{{Pryor}
  et~al.}{2010}]{pry10a}
{Pryor} C.,  {Piatek} S.,   {Olszewski} E.~W.,  2010, \mn@doi [\aj]
  {10.1088/0004-6256/139/3/839}, \href
  {http://adsabs.harvard.edu/abs/2010AJ....139..839P} {139, 839}

\bibitem[\protect\citeauthoryear{{Purcell}, {Bullock}, {Tollerud}, {Rocha}  \&
  {Chakrabarti}}{{Purcell} et~al.}{2011}]{pur11a}
{Purcell} C.~W.,  {Bullock} J.~S.,  {Tollerud} E.~J.,  {Rocha} M.,
  {Chakrabarti} S.,  2011, \mn@doi [\nat] {10.1038/nature10417}, \href
  {http://adsabs.harvard.edu/abs/2011Natur.477..301P} {477, 301}

\bibitem[\protect\citeauthoryear{{Putman}, {Thom}, {Gibson}  \&
  {Staveley-Smith}}{{Putman} et~al.}{2004}]{put04a}
{Putman} M.~E.,  {Thom} C.,  {Gibson} B.~K.,   {Staveley-Smith} L.,  2004,
  \mn@doi [\apjl] {10.1086/382728}, \href
  {http://adsabs.harvard.edu/abs/2004ApJ...603L..77P} {603, L77}

\bibitem[\protect\citeauthoryear{{Robitaille} \& {Whitney}}{{Robitaille} \&
  {Whitney}}{2010}]{rob10a}
{Robitaille} T.~P.,  {Whitney} B.~A.,  2010, \mn@doi [\apjl]
  {10.1088/2041-8205/710/1/L11}, \href
  {http://adsabs.harvard.edu/abs/2010ApJ...710L..11R} {710, L11}

\bibitem[\protect\citeauthoryear{{Rocha-Pinto}, {Scalo}, {Maciel}  \&
  {Flynn}}{{Rocha-Pinto} et~al.}{2000}]{roc00b}
{Rocha-Pinto} H.~J.,  {Scalo} J.,  {Maciel} W.~J.,   {Flynn} C.,  2000, \aap,
  \href {http://adsabs.harvard.edu/abs/2000A%26A...358..869R} {358, 869}

\bibitem[\protect\citeauthoryear{{Sawala}, {Scannapieco}  \& {White}}{{Sawala}
  et~al.}{2012}]{saw12a}
{Sawala} T.,  {Scannapieco} C.,   {White} S.,  2012, \mn@doi [\mnras]
  {10.1111/j.1365-2966.2011.20181.x}, \href
  {http://adsabs.harvard.edu/abs/2012MNRAS.420.1714S} {420, 1714}

\bibitem[\protect\citeauthoryear{{Sch{\"o}nrich}, {Binney}  \&
  {Dehnen}}{{Sch{\"o}nrich} et~al.}{2010}]{sch10b}
{Sch{\"o}nrich} R.,  {Binney} J.,   {Dehnen} W.,  2010, \mn@doi [\mnras]
  {10.1111/j.1365-2966.2010.16253.x}, \href
  {http://adsabs.harvard.edu/abs/2010MNRAS.403.1829S} {403, 1829}

\bibitem[\protect\citeauthoryear{{Shuter}}{{Shuter}}{1992}]{shu92a}
{Shuter} W.~L.~H.,  1992, \mn@doi [\apj] {10.1086/170996}, \href
  {http://adsabs.harvard.edu/abs/1992ApJ...386..101S} {386, 101}

\bibitem[\protect\citeauthoryear{{Siegel} et~al.,}{{Siegel}
  et~al.}{2007}]{sie07a}
{Siegel} M.~H.,  et~al., 2007, \mn@doi [\apjl] {10.1086/522003}, \href
  {http://adsabs.harvard.edu/abs/2007ApJ...667L..57S} {667, L57}

\bibitem[\protect\citeauthoryear{{Smith}}{{Smith}}{1963}]{smi63a}
{Smith} G.~P.,  1963, \bain, \href
  {http://adsabs.harvard.edu/abs/1963BAN....17..203S} {17, 203}

\bibitem[\protect\citeauthoryear{{Smith}, {Struck}  \& {Pogge}}{{Smith}
  et~al.}{1997}]{smi97a}
{Smith} B.~J.,  {Struck} C.,   {Pogge} R.~W.,  1997, \mn@doi [\apj]
  {10.1086/304286}, \href {http://adsabs.harvard.edu/abs/1997ApJ...483..754S}
  {483, 754}

\bibitem[\protect\citeauthoryear{{Spekkens}, {Urbancic}, {Mason}, {Willman}  \&
  {Aguirre}}{{Spekkens} et~al.}{2014}]{spe14b}
{Spekkens} K.,  {Urbancic} N.,  {Mason} B.~S.,  {Willman} B.,   {Aguirre}
  J.~E.,  2014, \mn@doi [\apjl] {10.1088/2041-8205/795/1/L5}, \href
  {http://adsabs.harvard.edu/abs/2014ApJ...795L...5S} {795, L5}

\bibitem[\protect\citeauthoryear{{Springel}, {Di Matteo}  \&
  {Hernquist}}{{Springel} et~al.}{2005}]{spr05c}
{Springel} V.,  {Di Matteo} T.,   {Hernquist} L.,  2005, \mn@doi [\mnras]
  {10.1111/j.1365-2966.2005.09238.x}, \href
  {http://adsabs.harvard.edu/abs/2005MNRAS.361..776S} {361, 776}

\bibitem[\protect\citeauthoryear{{Tepper-Garc{\'\i}a} \&
  {Bland-Hawthorn}}{{Tepper-Garc{\'\i}a} \& {Bland-Hawthorn}}{2018}]{tep18a}
{Tepper-Garc{\'\i}a} T.,  {Bland-Hawthorn} J.,  2018, \mn@doi [\mnras]
  {10.1093/mnras/stx2680}, \href
  {http://adsabs.harvard.edu/abs/2018MNRAS.473.5514T} {473, 5514}

\bibitem[\protect\citeauthoryear{Tepper-Garc{\'\i}a, Bland-Hawthorn  \&
  Sutherland}{Tepper-Garc{\'\i}a et~al.}{2015}]{tep15a}
Tepper-Garc{\'\i}a T.,  Bland-Hawthorn J.,   Sutherland R.~S.,  2015, The
  Astrophysical Journal, \href
  {http://adsabs.harvard.edu/abs/2015ApJ...813...94T} {813, 94}

\bibitem[\protect\citeauthoryear{{Teyssier}}{{Teyssier}}{2002}]{tey02a}
{Teyssier} R.,  2002, \mn@doi [\aap] {10.1051/0004-6361:20011817}, \href
  {http://adsabs.harvard.edu/abs/2002A%26A...385..337T} {385, 337}

\bibitem[\protect\citeauthoryear{{Tonnesen} \& {Bryan}}{{Tonnesen} \&
  {Bryan}}{2009}]{ton09a}
{Tonnesen} S.,  {Bryan} G.~L.,  2009, \mn@doi [\apj]
  {10.1088/0004-637X/694/2/789}, \href
  {http://adsabs.harvard.edu/abs/2009ApJ...694..789T} {694, 789}

\bibitem[\protect\citeauthoryear{{Wakker} \& {van Woerden}}{{Wakker} \& {van
  Woerden}}{1991}]{wak91b}
{Wakker} B.~P.,  {van Woerden} H.,  1991, \aap, \href
  {http://adsabs.harvard.edu/abs/1991A%26A...250..509W} {250, 509}

\bibitem[\protect\citeauthoryear{{Wetzel}, {Deason}  \&
  {Garrison-Kimmel}}{{Wetzel} et~al.}{2015}]{wet15b}
{Wetzel} A.~R.,  {Deason} A.~J.,   {Garrison-Kimmel} S.,  2015, \mn@doi [\apj]
  {10.1088/0004-637X/807/1/49}, \href
  {http://adsabs.harvard.edu/abs/2015ApJ...807...49W} {807, 49}

\bibitem[\protect\citeauthoryear{{de Boer}, {Belokurov}  \& {Koposov}}{{de
  Boer} et~al.}{2015}]{de-15c}
{de Boer} T.~J.~L.,  {Belokurov} V.,   {Koposov} S.,  2015, \mn@doi [\mnras]
  {10.1093/mnras/stv946}, \href
  {http://adsabs.harvard.edu/abs/2015MNRAS.451.3489D} {451, 3489}

\bibitem[\protect\citeauthoryear{{van den Bosch} \& {Ogiya}}{{van den Bosch} \&
  {Ogiya}}{2018}]{van18a}
{van den Bosch} F.~C.,  {Ogiya} G.,  2018, \mn@doi [\mnras]
  {10.1093/mnras/sty084}, \href
  {http://adsabs.harvard.edu/abs/2018MNRAS.tmp...67V} {}

\makeatother
\end{thebibliography}
\input{manuscript_mnras.bbl} 

\appendix

\section{A detailed view on gas accretion onto the Galactic plane} \label{sec:gacc2}

\begin{figure*}
\centering
\includegraphics[width=0.33\textwidth]{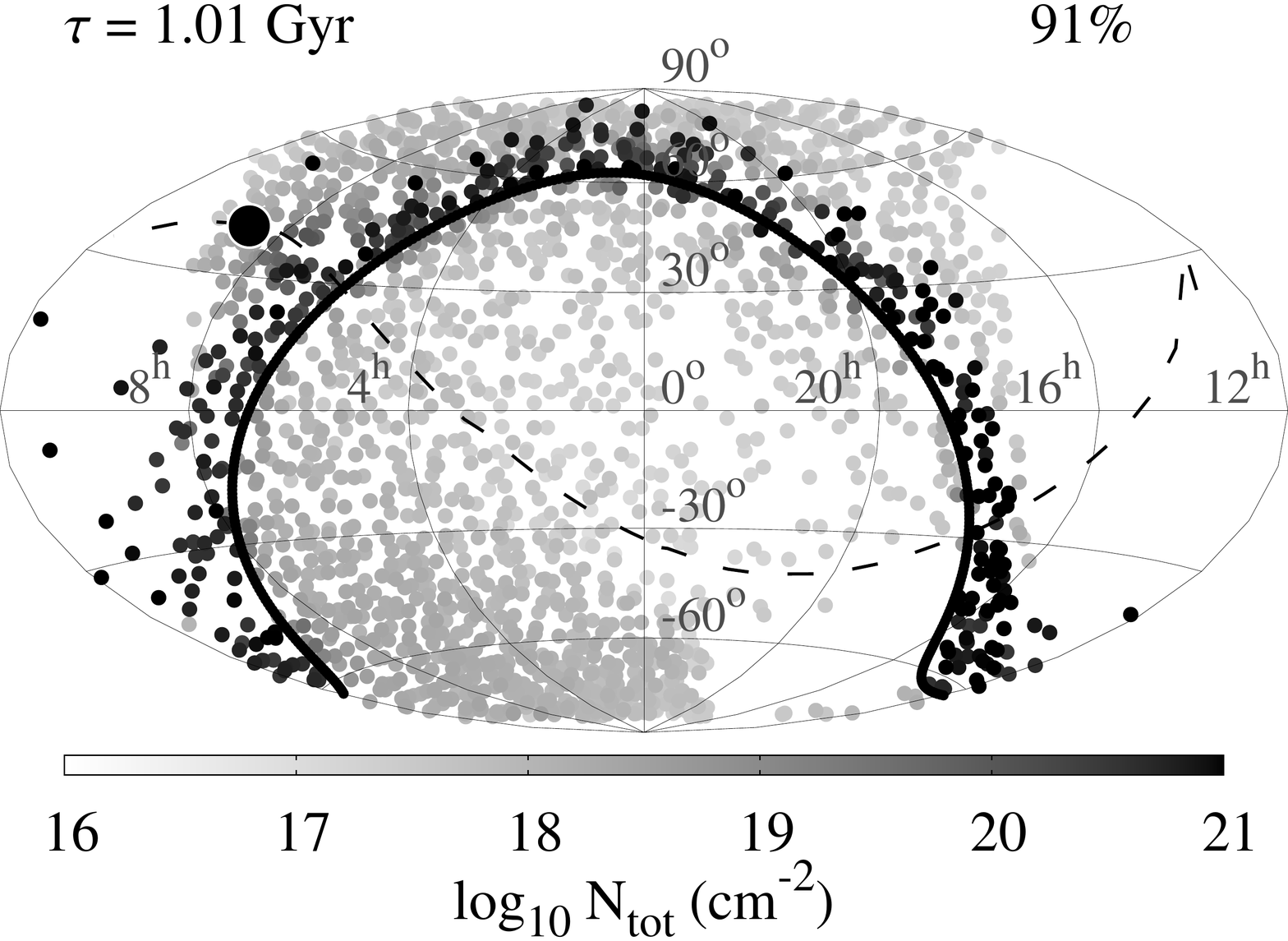}
\hfill
\includegraphics[width=0.33\textwidth]{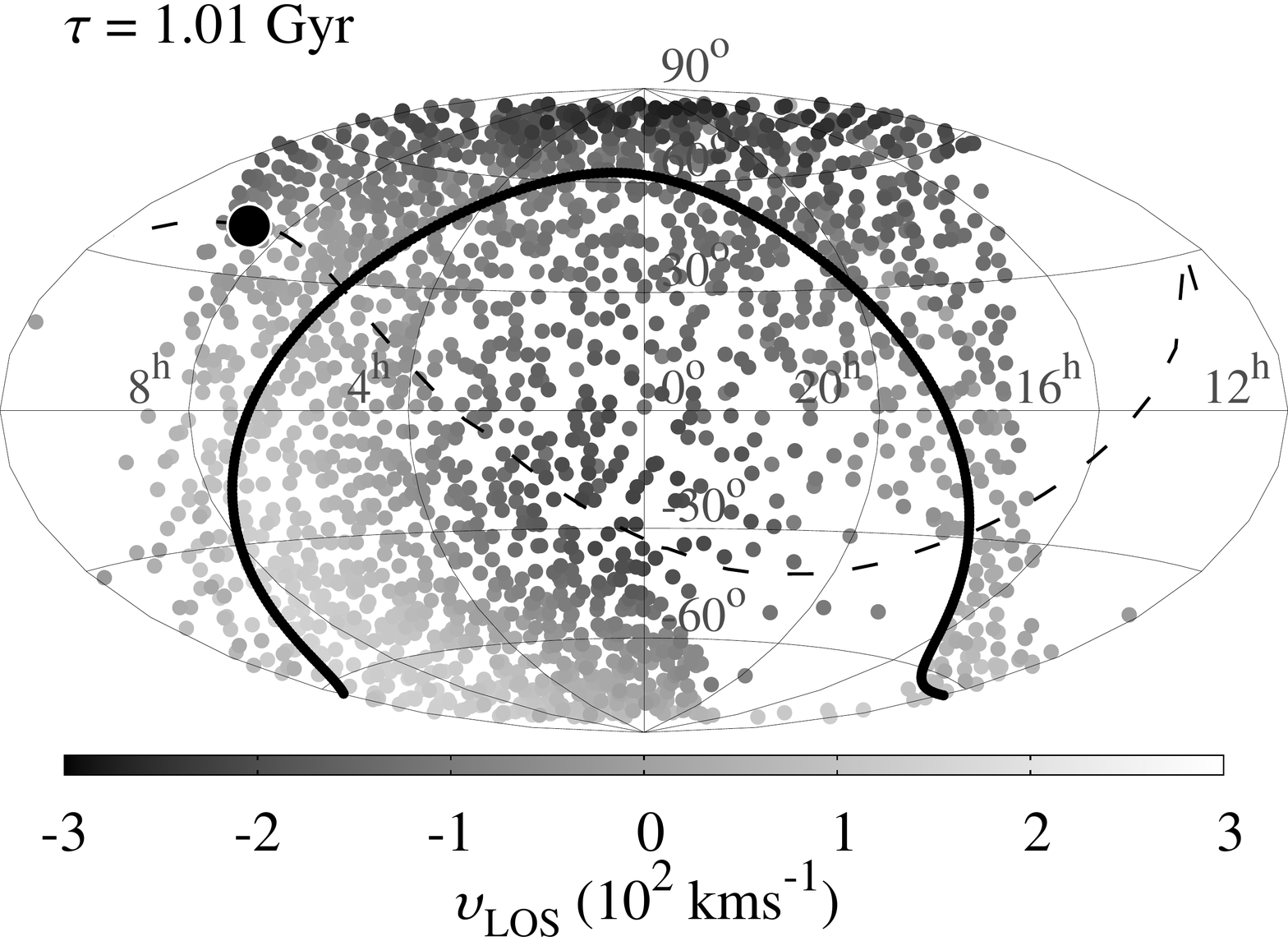}
\hfill
\includegraphics[width=0.33\textwidth]{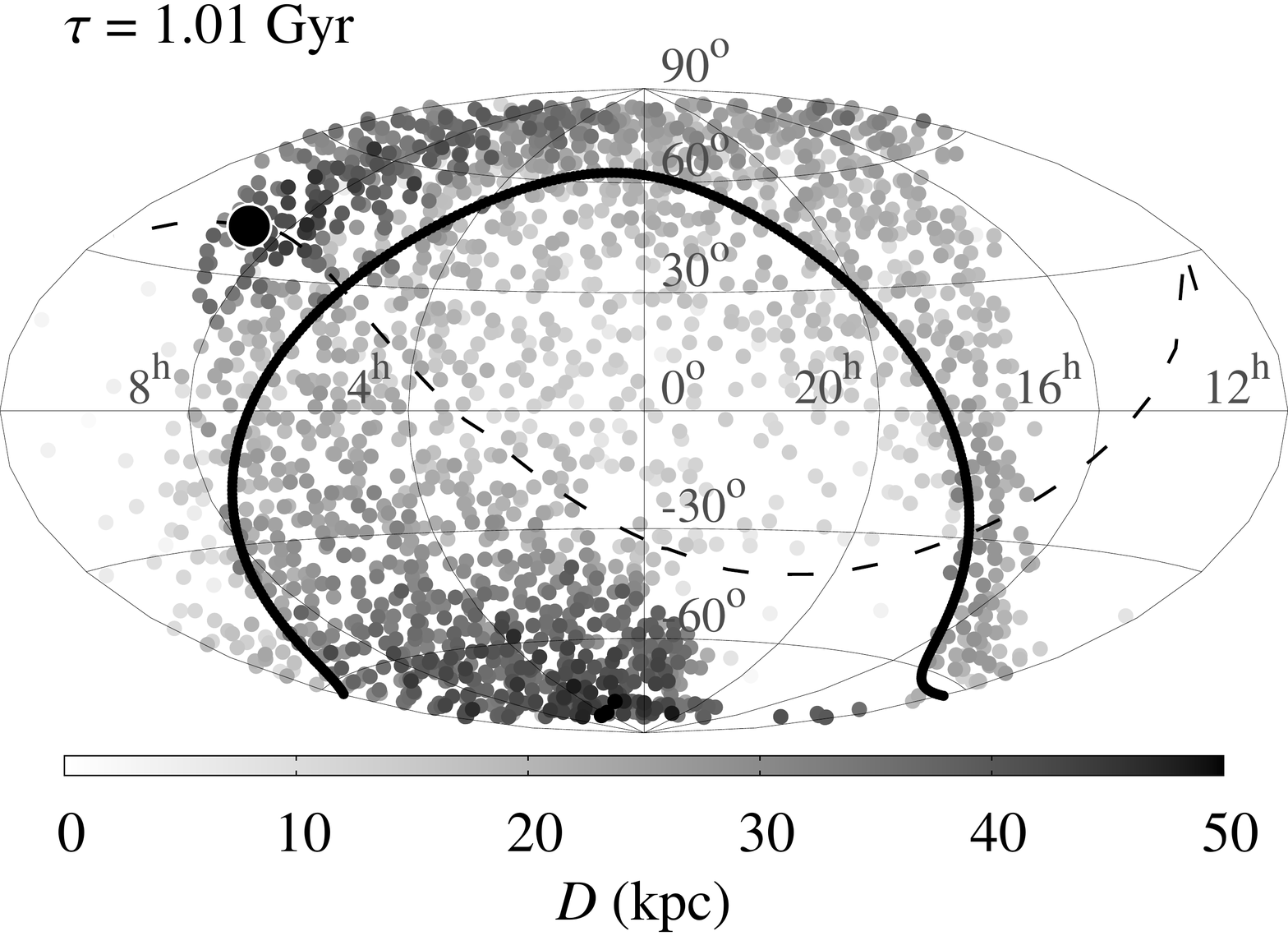}
\includegraphics[width=0.33\textwidth]{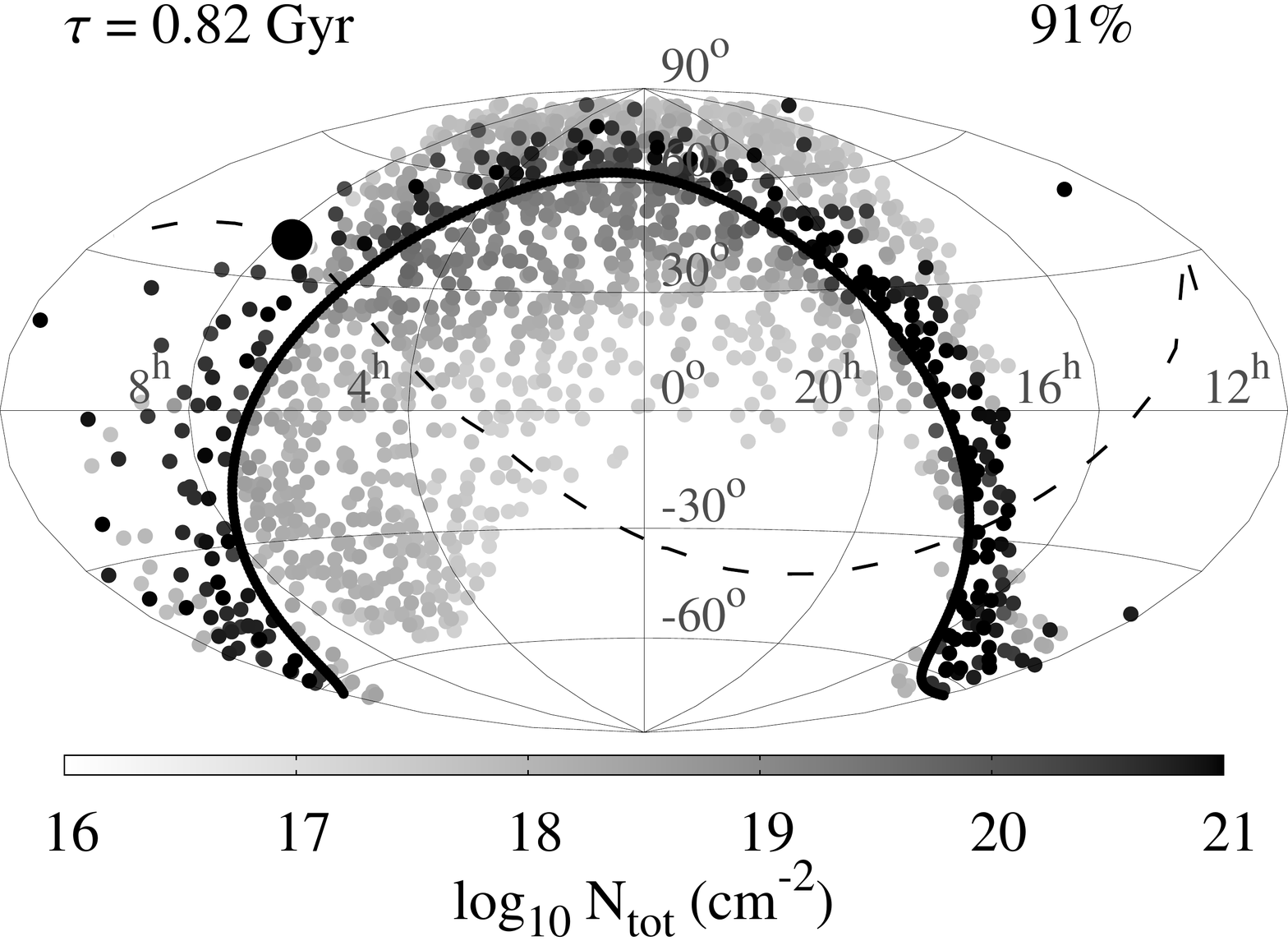}
\hfill
\includegraphics[width=0.33\textwidth]{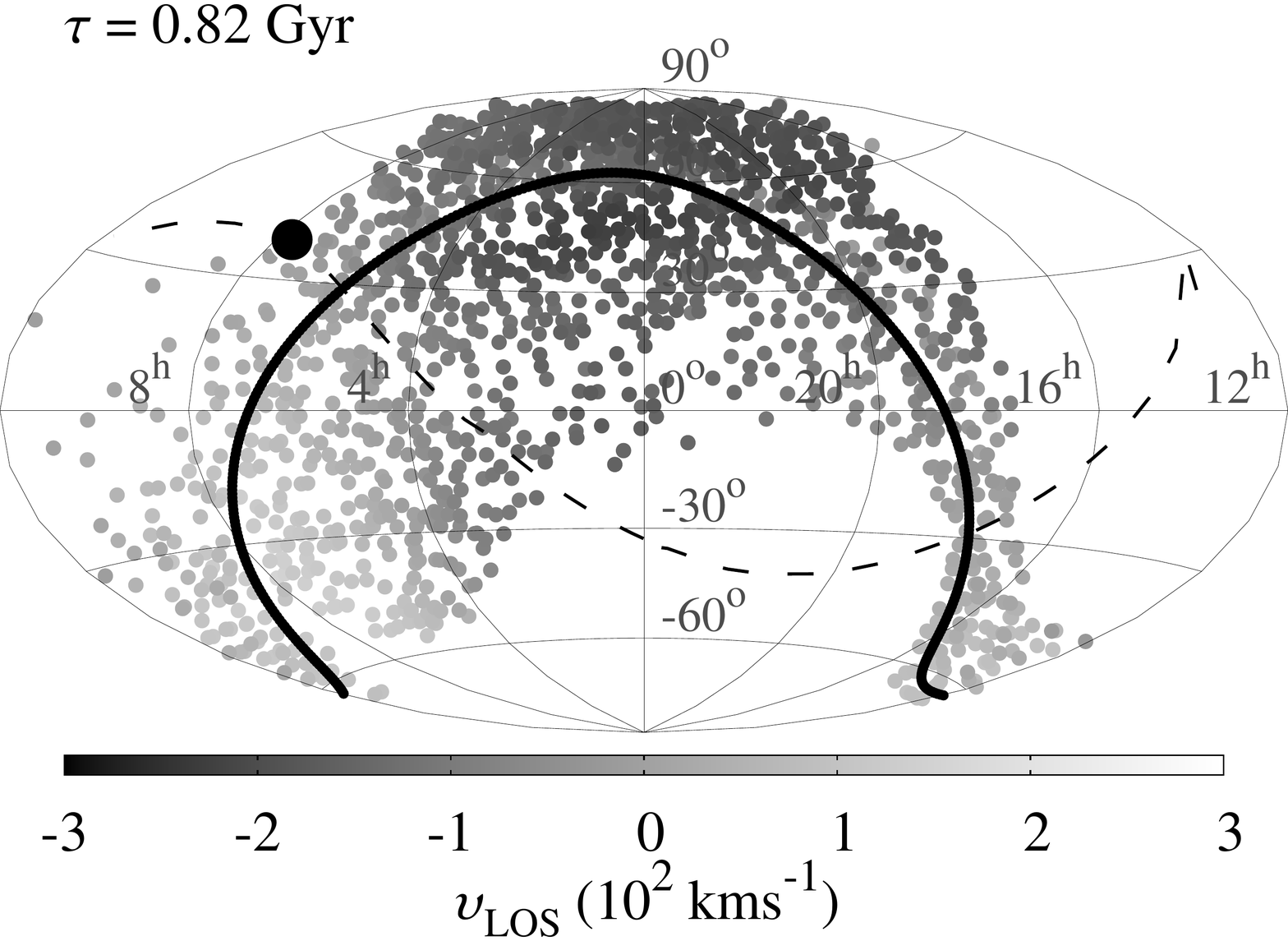}
\hfill
\includegraphics[width=0.33\textwidth]{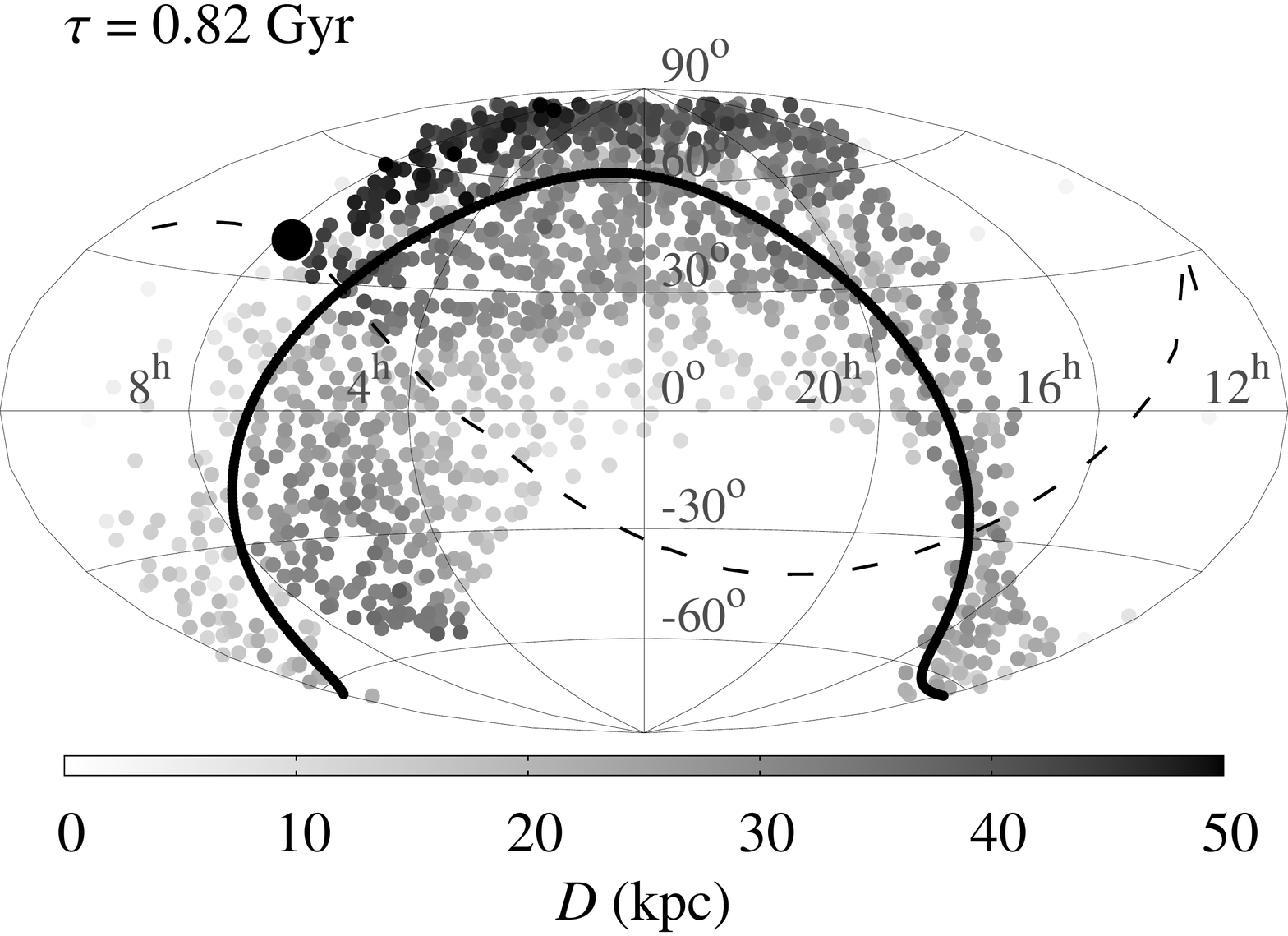}
\includegraphics[width=0.33\textwidth]{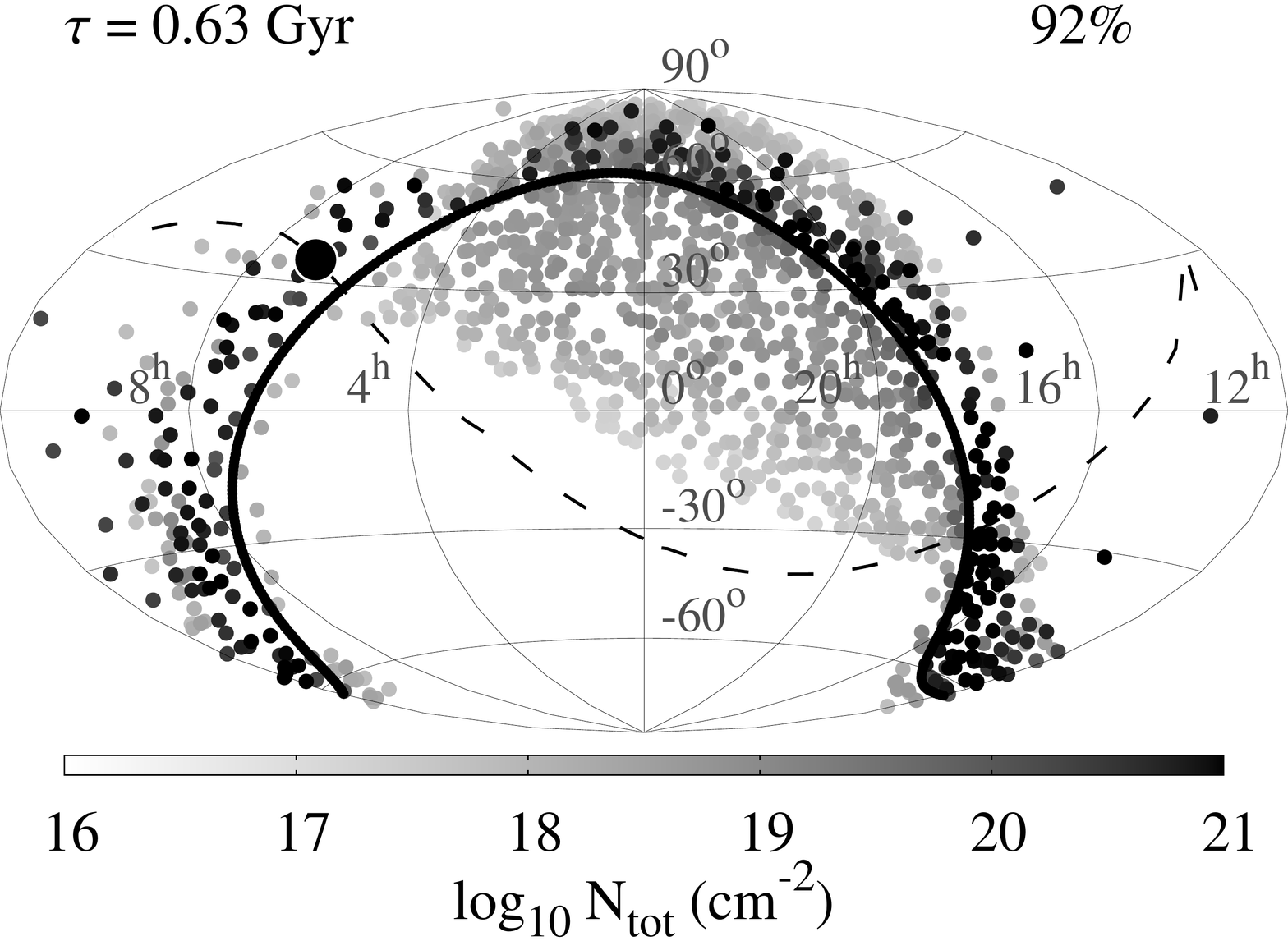}
\hfill
\includegraphics[width=0.33\textwidth]{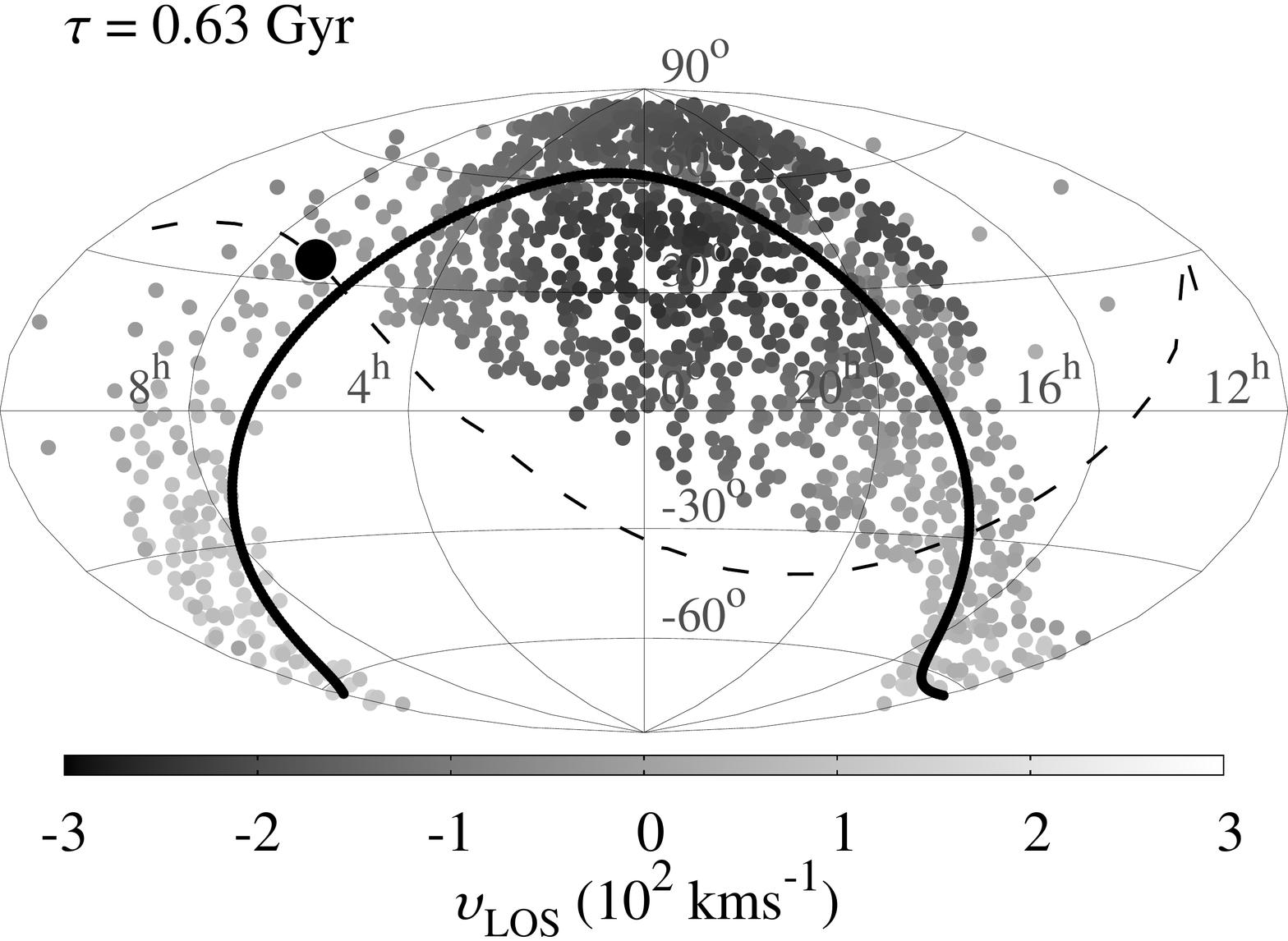}
\hfill
\includegraphics[width=0.33\textwidth]{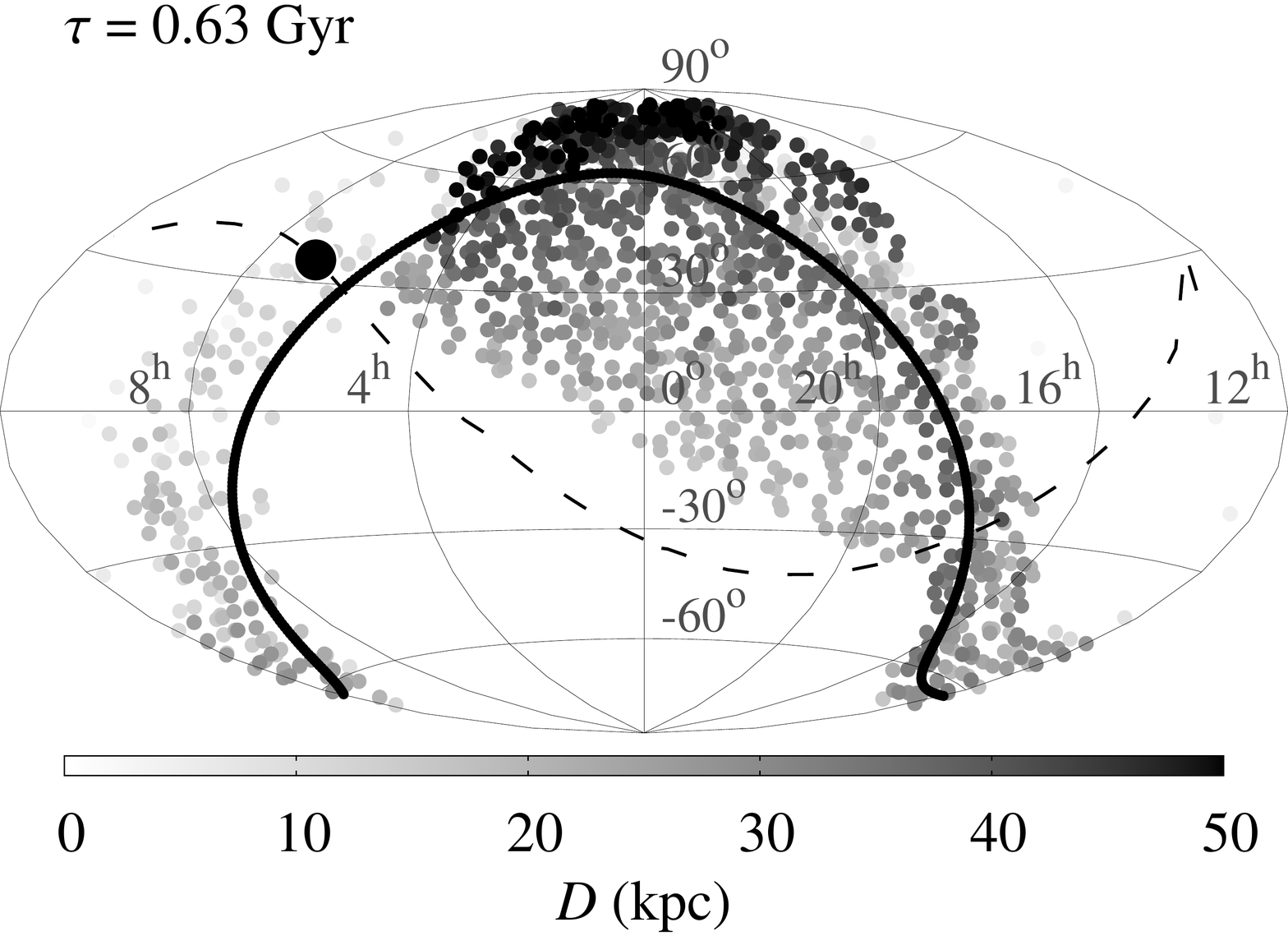}
\includegraphics[width=0.33\textwidth]{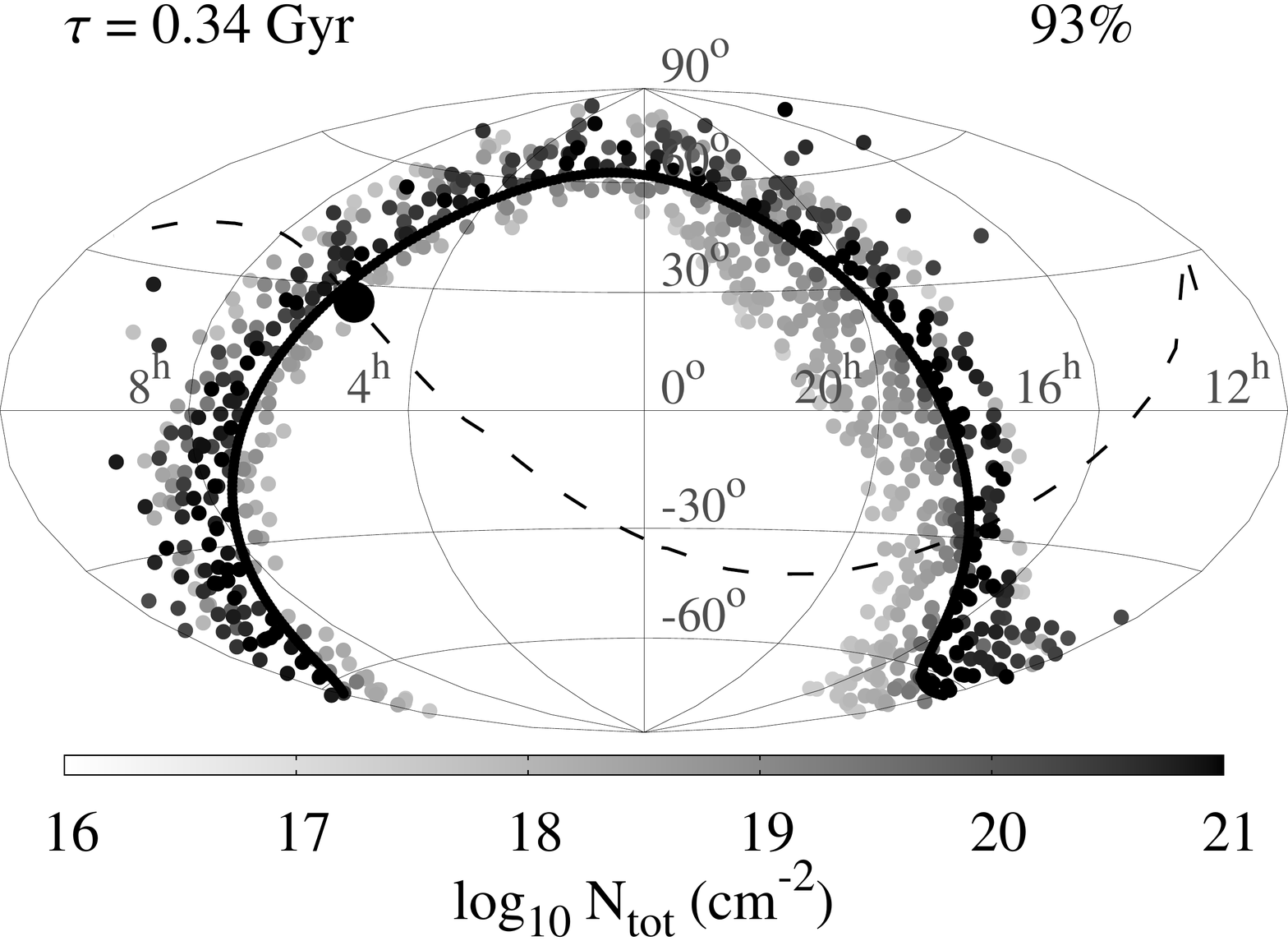}
\hfill
\includegraphics[width=0.33\textwidth]{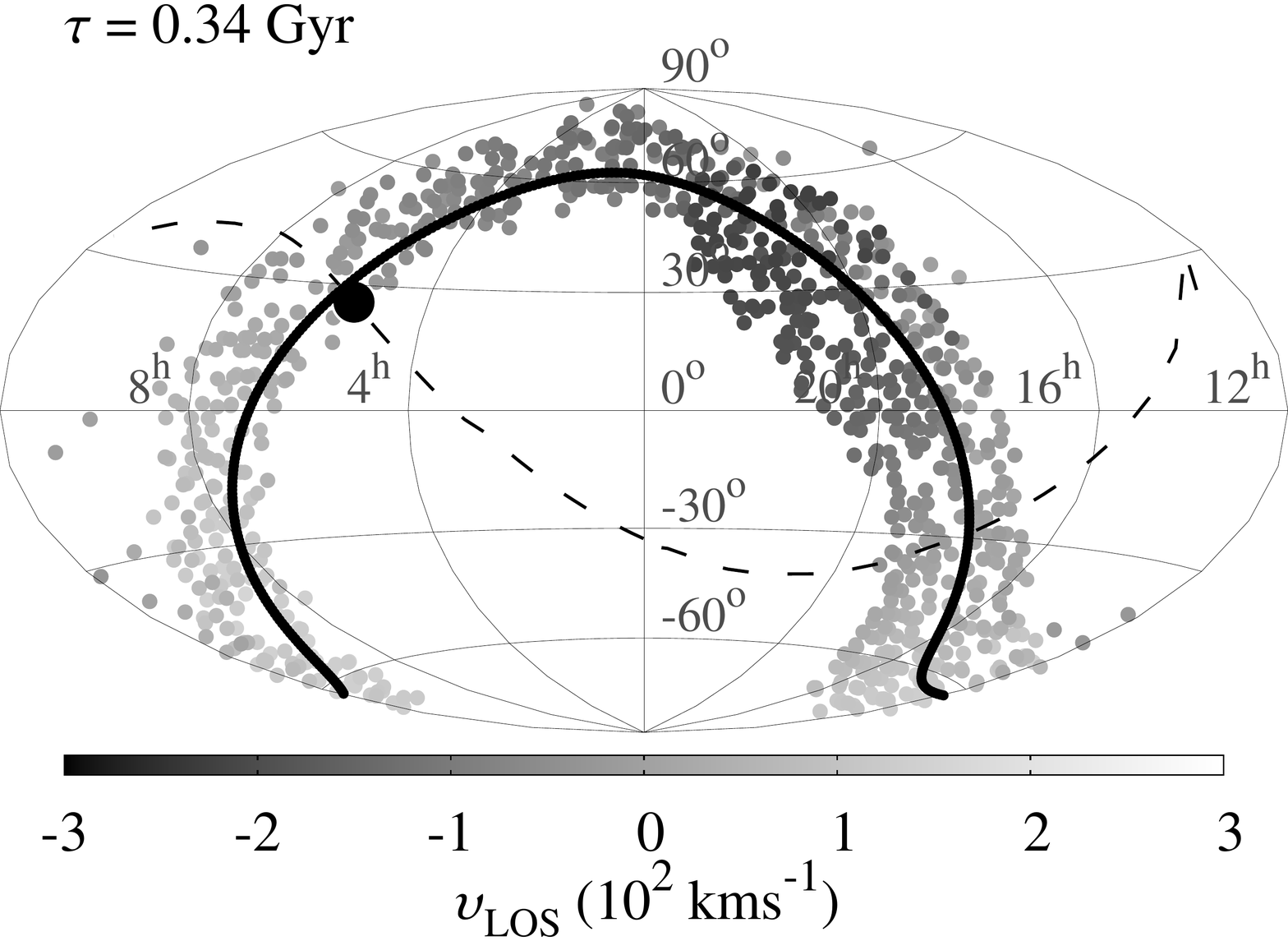}
\hfill
\includegraphics[width=0.33\textwidth]{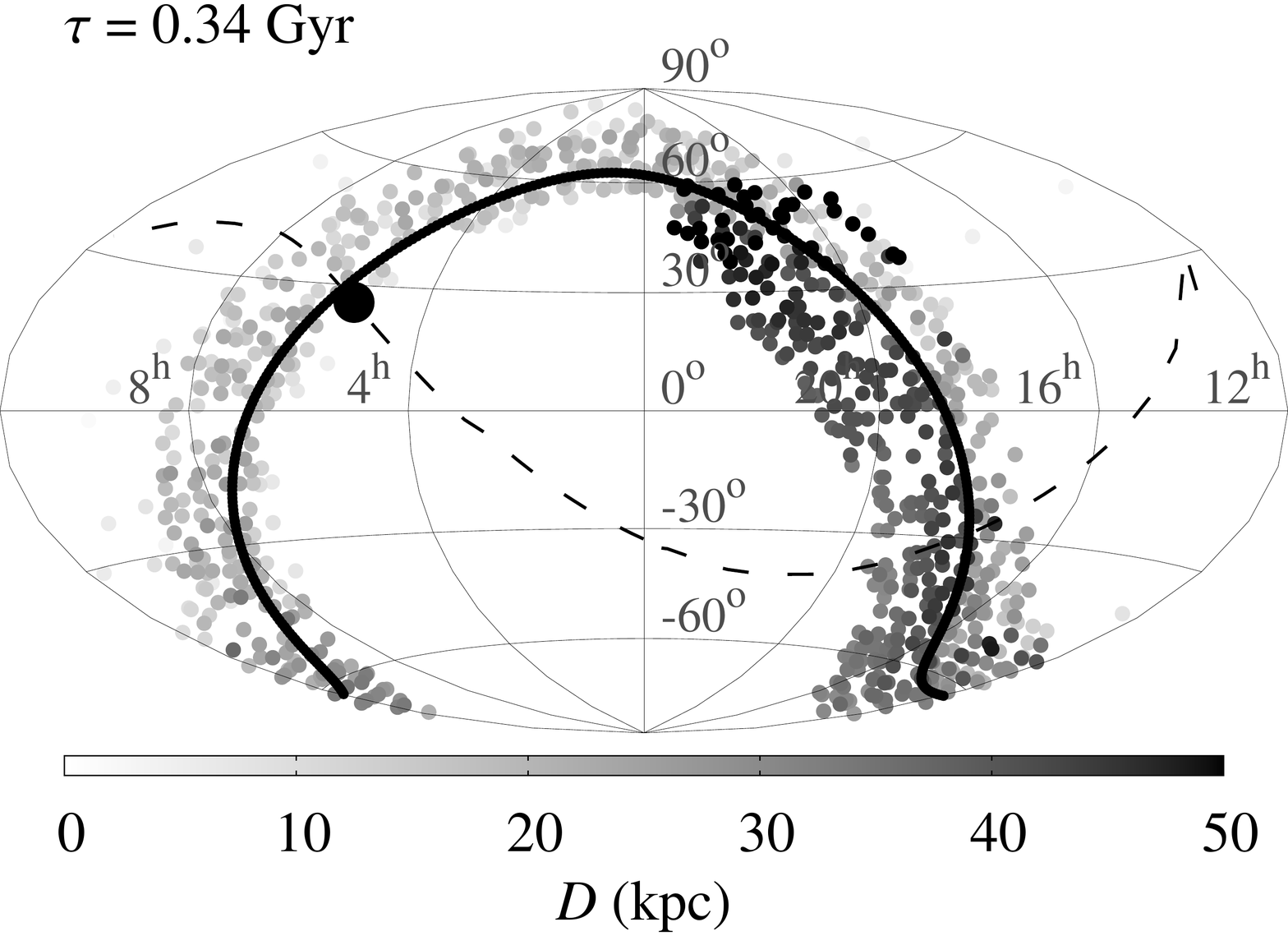}
\includegraphics[width=0.33\textwidth]{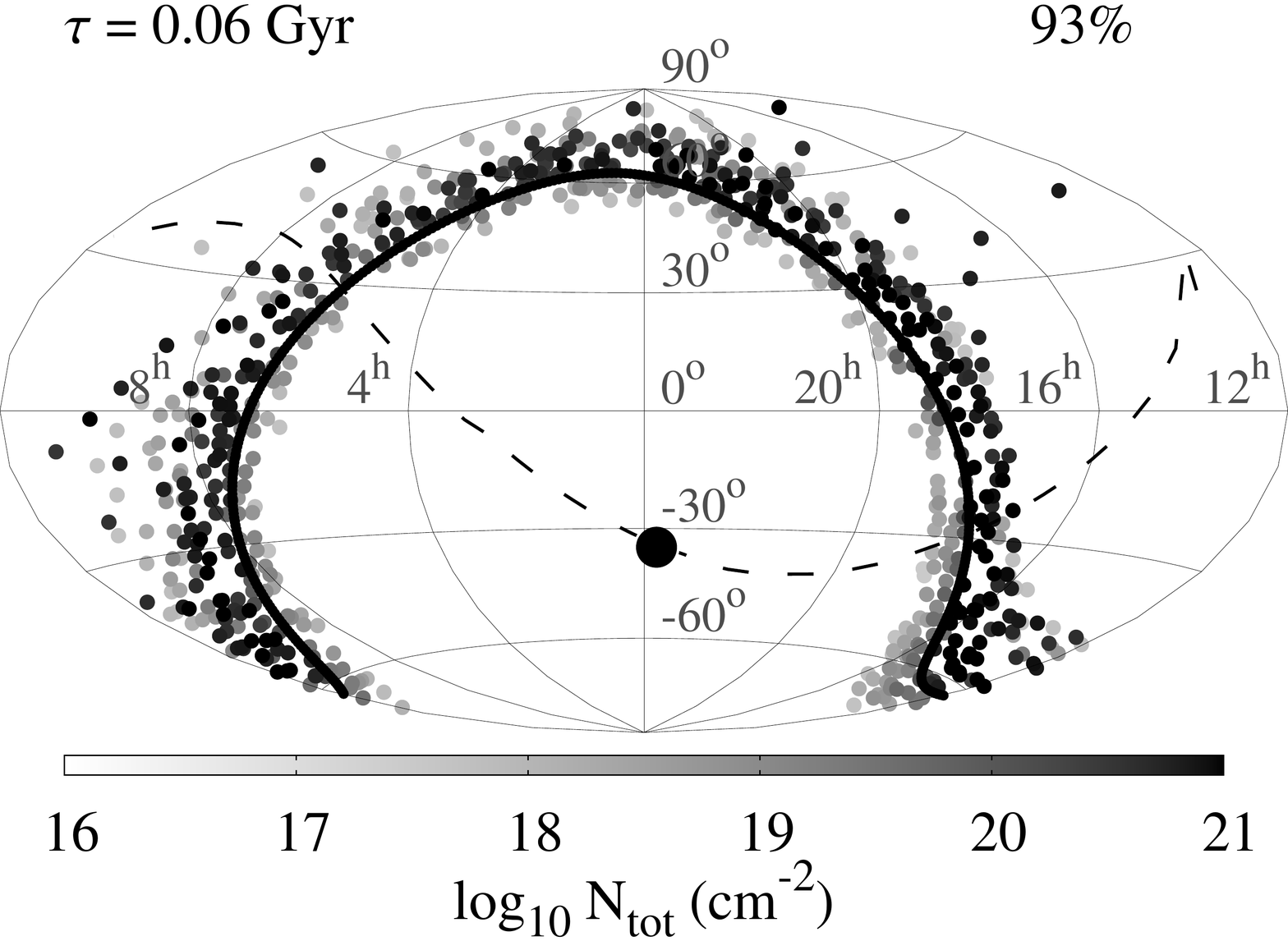}
\hfill
\includegraphics[width=0.33\textwidth]{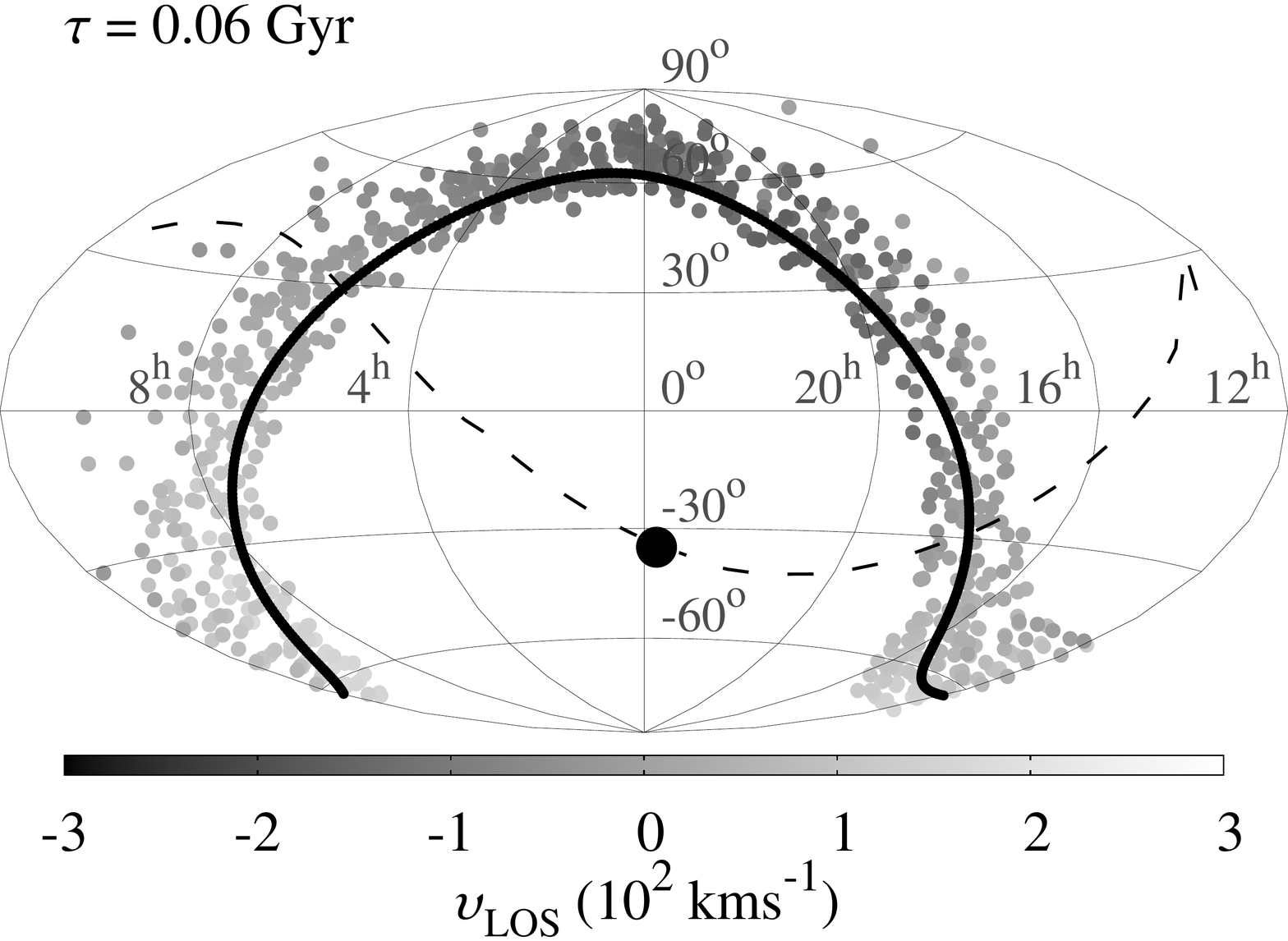}
\hfill
\includegraphics[width=0.33\textwidth]{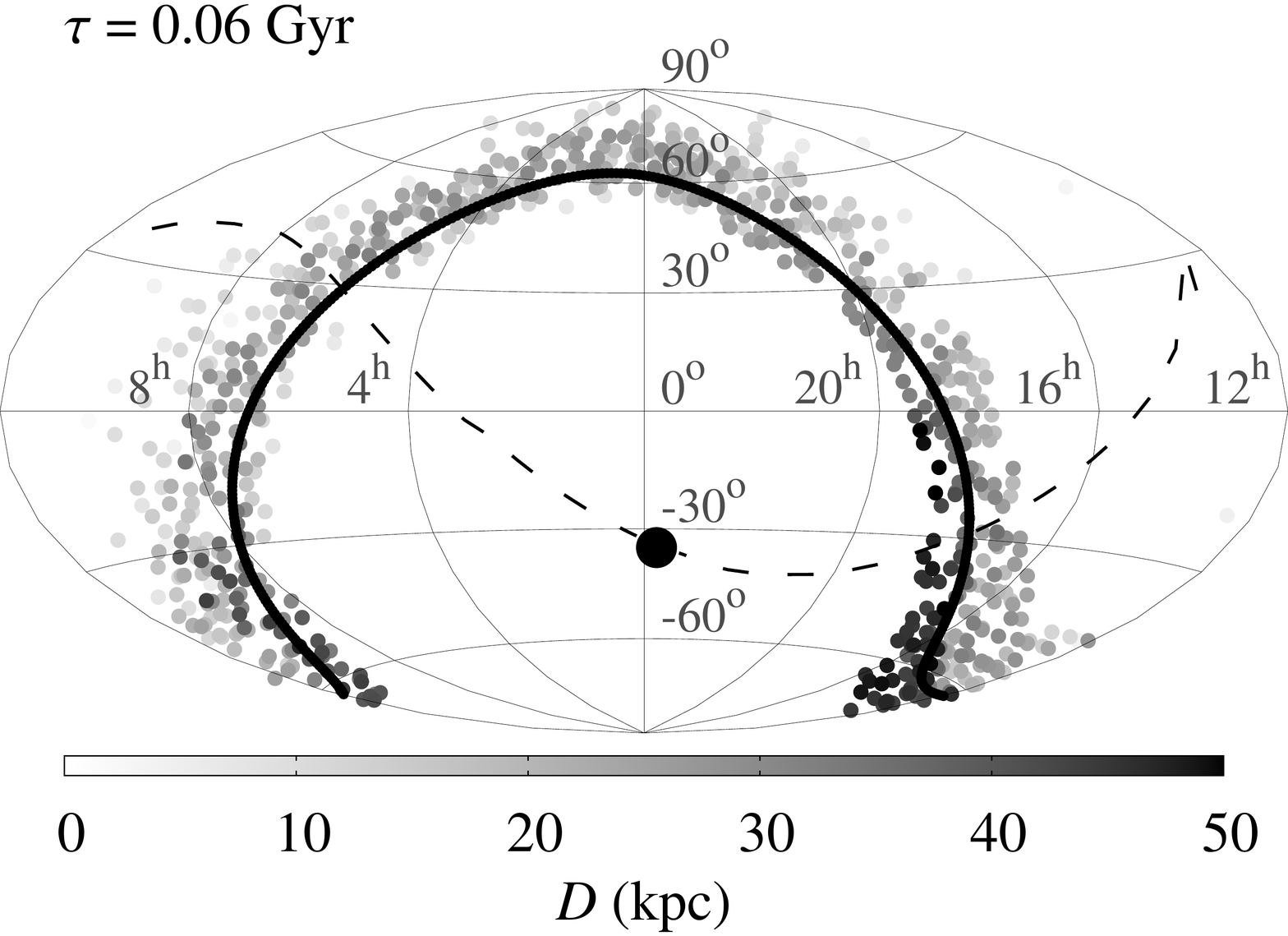}
\caption[ Model C All-sky distribution with gas col.dens., vel., and dist. ]{ Left, centre and right columns are similar to Figs.~\ref{fig:allskyC3_snpart}, \ref{fig:allskyC3_vel} and \ref{fig:allskyC3_dist}, respectively, but at slightly different epochs within the last Gyr of evolution of Sgr. The most relevant difference is that here we have adopted a gas tracer threshold of nearly 0 (rather than 1). Apparently, the gas stripped from Sgr settles onto the Galactic plane on a time scale of a few 100 Myr, and is found roughly evenly distributed at low latitude since $\gtrsim 300$ Myr ago. See text for further details. }
\label{fig:allskyC4}
\end{figure*}

In order to track in detail the location of the gas stripped away from Sgr, we repeat the exercise discussed in Sec.~\ref{sec:gacc}, with the following change: We set the gas tracer threshold to nearly zero. In principle, since initially only the gas associated to Sgr gas a non-vanishing tracer value, following this practice we should be able to trace nearly all of its gas, regardless of its location. The result is shown in Fig.~\ref{fig:allskyC4}. There, we show the total gas column density (left column), the line-of-sight velocity (central column), and the heliocentric distance (right column) of the gas with a tracer value above the threshold. Each row corresponds to a snapshot in decreasing order of lookback time (indicated in the top-left column of each panel), roughly spanning the last Gyr of Sgr's evolution. Note that the selected snapshots are slightly different from those shown in Figs.~\ref{fig:allskyC3_snpart} - \ref{fig:allskyC3_temp}. The percentage indicated in on the top-right corner of each of the panels in the left column indicates the total gas mass traced relative to the initial gas mass of the precursor. Clearly, virtually all the gas initially bound to Sgr is being accounted for. Naturally, more gas appears here than in the previous figures, covering a large fraction of the sky. In the previous sections we chose a threshold value for the gas tracer significantly higher than zero to avoid too cluttered images. Here we sacrifice somewhat graphical clarity in favour of a deeper insight into the fate of the gas.

The important result revealed here is that, apparently, the stripped gas slowly settles onto the disc, as can be judged by comparing the distribution of gas blobs relative to the locus of the Galactic plane (thick solid line). The process takes place over roughly the last Gyr. But the gas appears fairly settled, i.e. distributed evenly along the Galactic plane and at low latitude since, at least, $\sim 300$ Myr ago. Note that the increased column density of the gas initially associated to Sgr which appears now close to the Galactic plane is due to the fact that the it has effectively mixed with (and is part of) the dense Galactic gas disc.

\bsp	
\label{lastpage}
\end{document}